\documentclass[prl,
twocolumn, superscriptaddress,aps]{revtex4-1}

\usepackage[utf8]{inputenc}
\usepackage{graphicx}
\usepackage{xcolor}
\usepackage{hyperref}

\usepackage{amssymb} %\mathbb{R}
\usepackage{soul}
\usepackage{verbatim}
\usepackage{siunitx}
\usepackage{amsmath}
\usepackage{textcomp,gensymb}	
\usepackage{dsfont} %\mathds{1}
\usepackage{mathtools} % pmatrix, \coloneqq

\begin{document}

\title{Resonant Solitary States in Complex Networks}
\author{Jakob Niehues}
\affiliation{Corresponding author: jakob.niehues@pik-potsdam.de}
\affiliation{Potsdam Institute for Climate Impact Research (PIK), Member of the Leibniz Association, P.O. Box 60 12 03, D-14412 Potsdam, Germany}
\affiliation{Humboldt-Universit\"at zu Berlin, Department of Physics, Newtonstra\ss e 15, 12489 Berlin, Germany}
\affiliation{
Technische Universit\"at Berlin, ER 3-2, Hardenbergstrasse 36a, 10623 Berlin, Germany}
\author{Serhiy Yanchuk}
\affiliation{Potsdam Institute for Climate Impact Research (PIK), Member of the Leibniz Association, P.O. Box 60 12 03, D-14412 Potsdam, Germany}
\affiliation{University College Cork, School of Mathematical Sciences, Western Road, Cork, T12 XF62, Ireland}
\author{Rico Berner}
\affiliation{Humboldt-Universit\"at zu Berlin, Department of Physics, Newtonstra\ss e 15, 12489 Berlin, Germany}
\author{J\"urgen Kurths}
\affiliation{Potsdam Institute for Climate Impact Research (PIK), Member of the Leibniz Association, P.O. Box 60 12 03, D-14412 Potsdam, Germany}
\affiliation{Humboldt-Universit\"at zu Berlin, Department of Physics, Newtonstra\ss e 15, 12489 Berlin, Germany}
\author{Frank Hellmann}
\affiliation{Potsdam Institute for Climate Impact Research (PIK), Member of the Leibniz Association, P.O. Box 60 12 03, D-14412 Potsdam, Germany}
\author{Mehrnaz Anvari}
\affiliation{Potsdam Institute for Climate Impact Research (PIK), Member of the Leibniz Association, P.O. Box 60 12 03, D-14412 Potsdam, Germany}
\affiliation{Fraunhofer Institute for Algorithms and Scientific Computing, 53757 Sankt Augustin, Germany}

\begin{abstract}
Partially synchronized solitary states occur frequently when a synchronized system of networked oscillators with inertia is perturbed locally. Several asymptotic states of different frequencies can coexist at the same node.
Here, we reveal the mechanism behind this multistability: additional solitary frequencies arise from the coupling between network modes and the solitary oscillator's frequency, leading to significant energy transfer. This can cause the solitary node's frequency to resonate with a Laplacian eigenvalue. We analyze which network structures enable this resonance and explain longstanding numerical observations.
Another solitary state that is known in the literature is characterized by the effective decoupling of the synchronized network and the solitary node at the natural frequency. Our framework unifies the description of solitary states near and far from resonance, allowing to predict the behavior of complex networks from their topology.

\end{abstract}

\maketitle

\section{Introduction}
Many natural and human-made systems are characterized by various degrees of synchronization,  which is one of the most fundamental common aspects of their collective behavior \cite{pikovskySynchronizationUniversalConcept2001}.
Therefore, important extended systems as diverse as the heart, the brain, firefly populations, chemical reactions, and power grids, are ubiquitously modelled as networks of interconnected oscillators \cite{strogatzKuramotoCrawfordExploring2000, acebronSynchronizationPopulationsGlobally2000, acebronKuramotoModelSimple2005,pecora2014cluster,Rodrigues2016-ep}.
In some, such as power grids, global synchronization is essential for their proper functioning, in others, such as the brain, it can indicate severe dysfunction. The tendency to synchronize, and thus the function of the system, is strongly influenced by the underlying network's topology.

The paradigmatic models for understanding this relationship between structure and function in coupled oscillator networks are the Kuramoto model \cite{Kuramoto1984-wh} and its variants. These models feature extremely rich collective behavior, such as chimera states, frequency clusters, isolated desynchronization, and spatial chaos \cite{strogatzKuramotoCrawfordExploring2000,acebronSynchronizationPopulationsGlobally2000,acebronKuramotoModelSimple2005, pecora2014cluster, Rodrigues2016-ep, CyclopsStates, kovalenkoContrariansSynchronizeLimit2021, Kuramoto1984-wh}. The Kuramoto model with inertia \cite{tanakaFirstOrderPhase1997,tanakaSelfsynchronizationCoupledOscillators1997} has been developed independently to study synchronization properties in power grids \cite{BergenHill,Review} and biological systems \cite{ermentroutAdaptiveModelSynchrony1991}.

The transition from decoupling to synchrony in %inertial 
Kuramoto networks is characterized by spatial chaos, a form of extreme multistability \cite{Omelchenko2011-ql}. This regime typically has several states with similarly sized basins of attraction \cite{Gelbrecht_2020,halekotteTransientChaosEnforces2021}.
Among the most prominent are frequency clusters \cite{PhysRevE.103.042315,olmiHystereticTransitionsKuramoto2014}, the simplest of which are solitary states. In a solitary state, only one or a few independent oscillators are phase-shifted \cite{MaistrenkoSolitary14} or, in the presence of inertia, start to rotate at their own frequencies \cite{ChimeraJaros15}, while the rest of the network remains synchronized. Solitary states are especially prominent for the important situation of localized large perturbations \cite{Menck2014NatureComm,Hellmann2020-aj,Jaros2018-mu},  and they are particularly likely for perturbations at leaf nodes \cite{Menck2014NatureComm, Nitzbon_2017}. Localized perturbations are important in systems like power grids, where single component failures are common and can lead to desynchronization and blackouts \cite{UCTE_split2006}.

Interestingly, numerical studies have revealed that solitary nodes can exist not only at the natural frequency of the oscillators, but also at intermediate frequencies \cite{Nitzbon_2017,halekotteTransientChaosEnforces2021}.
In addition, the presence of multistability in these solitary nodes has been identified, allowing for the coexistence of both intermediate and natural solitary states \cite{Nitzbon_2017,halekotteTransientChaosEnforces2021} at the same node.

Previous work has discussed the role of a networks' topology in understanding its overall response to perturbations and the emergence of various stable states \cite{Rohden2012-iy,Menck2014NatureComm,Rodrigues2016-ep,Nitzbon_2017,Xiaozhu19}. In some contexts, the frequency and stability of the solitary states have been studied in terms of decoupling and entrainment arguments  \cite{Menck2014Diss,Hellmann2020-aj}. These explanations do not account for either intermediate frequency solitaries or the coexistence of several solitary states at the same node.

In this work, we use averaging theory \cite{Guckenheimer2002-rm} and linear response to develop a theory of the resonant coupling between the synchronized cluster and the solitary node. We uncover that the described phenomena arise from the cluster's resonantly excited complex network structure. This leads to a non-zero mean energy flow between the synchronized cluster and the solitary node, and a frequency shift (Fig.~\ref{fig:animation_summary}). While works like \cite{Self-consistent_method,yue2020model,StabilityRolSolInertia} have previously considered the interaction between individual oscillators and the mean field, such an interaction between collective network modes and individual oscillators has not been previously described.
Here, we uncover how the network's topology shapes the landscape of solitary states and illustrate our findings with a range of example systems, ranging from conceptual models to fully complex networks.

\section{Model}

\begin{figure}%[h]
    \centering    \includegraphics[width=\columnwidth]{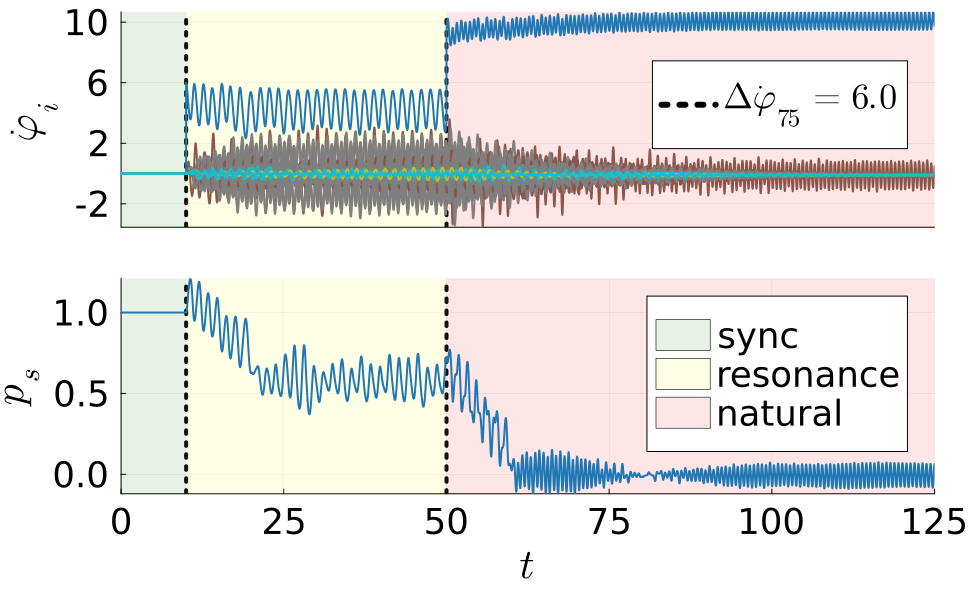}
    \caption{
    Example trajectories of solitary states of intermediate and natural frequencies.
    Frequencies ($\dot\varphi_i$) and time-averaged energy flow $p_s$ (moving average over 10 time units) between solitary and synchronous component are shown for a complex oscillator network, see \cite{Supplement} for an animation and the network topology (Supplementary Fig.~S2).
    The system is perturbed twice, with the timing indicated by vertical dashed lines.
    Both perturbations consist of an instantaneous step of $\Delta\dot\varphi_{75}= 6$ at node 75's frequency.
    The first perturbation brings the system from the synchronized regime into an intermediate solitary state in resonance with the synchronized cluster.
    The second perturbation causes the transition to a natural solitary state, where the solitary node is effectively decoupled and the oscillations of the synchronized cluster are much smaller.
    The distinct blue trajectory belongs to the solitary node 75 in Fig.~\ref{fig:asymtptotic_freq_histo}.
    }
    \label{fig:animation_summary}
\end{figure}

In the following, we will use the Kuramoto model with inertia on complex networks \cite{tanakaFirstOrderPhase1997,tanakaSelfsynchronizationCoupledOscillators1997,Hellmann2020-aj,Xiaozhu19,Menck2014NatureComm,halekotteMinimalFatalShocks2020,Manik2014-ye}. It is given by the coupled second-order equations
\begin{align}
    m_i \ddot\varphi_i &= P_i - \alpha_i \dot\varphi_i - \sum_{j=1}^N p_{ij}(\varphi_i,\varphi_j),
    \label{eq:swing_lossless}
\end{align}
where $\varphi_i$ are the phases of the $N$ oscillators, $P_i$ the driving powers, the $m_i$ are inertia constants, and the $\alpha_i > 0$ are damping coefficients. The coupling is given in terms of the weighted coupling matrix $\left\{\kappa_{ij}\right\}$ as
\begin{equation}
    p_{ij}(\varphi_i,\varphi_j) = \kappa_{ij} \sin(\varphi_i - \varphi_j).
\end{equation}
Note that the coupling matrix is symmetric, $\kappa_{ij} = \kappa_{ji}$, hence the coupling function is antisymmetric, $p_{ij} = - p_{ji}$.
We parametrize the model Eq.~(\ref{eq:swing_lossless}) such that the synchronous state coexists with various stable attractors of different degree of synchronization.
In particular, we set $P_i = 1$ for producers, $P_i = -1$ for consumers, $\alpha_i = 0.1$, and $\kappa_{ij} = 6$ for connected nodes $(i,j)$.
This is the parameter regime of, for example, real-world power grids \cite{Menck2014NatureComm,Nitzbon_2017,Hellmann2020-aj,halekotteMinimalFatalShocks2020,halekotteTransientChaosEnforces2021}.

To better understand the asymptotic behavior of the system, we recall a few properties of its attractors. In a synchronous state, we have $\varphi_i(t) = \varphi_i^* + \hat\omega t$, and constant frequencies $\dot\varphi_i^* = \hat\omega$, and the phases satisfy $P_i - \alpha_i \hat\omega = \sum_{j=1}^N p_{ij}(\varphi_i^*,\varphi_j^*)$. By going to a corotating frame via $\varphi_i \to \varphi_i - \hat\omega t$, we can always set $\hat \omega = 0$, and will assume so from now on. The $p_{ij}(\varphi_i^*,\varphi_j^*)$ can be interpreted as the energy flowing through the network to balance out the driving powers $P_i$, (this is the physical interpretation for power grids), and $\alpha_i$ determines the restoring force of the frequency.
The natural frequencies of the uncoupled oscillators (i.e., when all $\kappa_{ij}=0$) are determined by the condition that driving power and restoring force are in balance: $\Omega_i \coloneqq P_i/\alpha_i$.
Solitary states in systems of coupled oscillators can be quite similar to the natural solitary states of uncoupled oscillators, which can serve as an approximation \cite{Menck2014NatureComm}.

In this work we consider so-called \textit{frequency solitary states}, in which one or a few independent oscillators rotate at a distinct frequency, while the rest of the network forms a large synchronized cluster.
For simplicity, we restrict ourselves to 1-solitaries, which have only a single solitary oscillator at a node that we denote $z$.
However, the considerations in this work apply to the case of several solitary nodes as well, see Supplemental Material \cite{Supplement}, section I.F (SM I.F).
We denote the synchronized cluster $S = \{i : 1 \leq i \leq N, i \neq z \}$, and the long-term time average by $\langle\rangle$.
Then, for the purpose of this work, a solitary state is defined by $\langle\dot\varphi_z(t)\rangle = \omega_s$ and $\langle \varphi_i(t) \rangle = \omega_\text{sync}$, with $\omega_s \neq \omega_\text{sync}$.
Such solitary states only exist in the presence of inertia \cite{Jaros2018-mu}.

\begin{figure}%[h]
    \centering
    \includegraphics[width=\columnwidth]{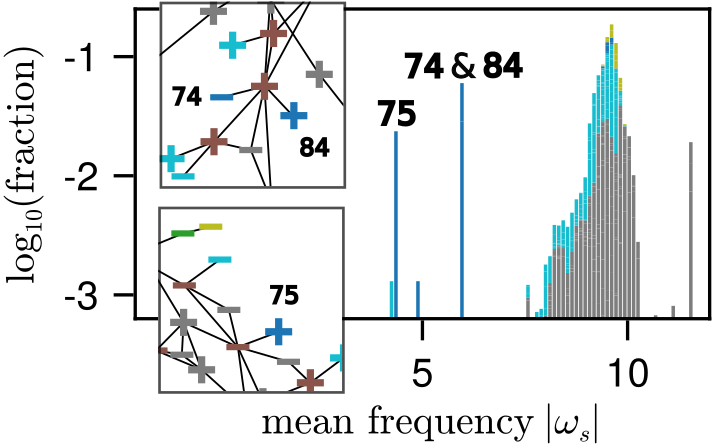}
    \caption{Local topological features determine the existence and relative occurrence of asymptotic solitary frequencies $\omega_s$ in a synthetic power grid observed after perturbations, see \cite{GitRepoReVelatioNs}.
    Three solitary states are clearly distinct from the natural frequencies at $\pm 10$;
    they occur at leaf nodes 74, 75 and 84 which have highly connected neighbors (cf. \cite{Nitzbon_2017}) as shown in the insets.
    Nodes shown as '$+$' ('$-$') have $P_i=1$ $(-1)$.
    See Fig.~S2 for the full network, and Fig.~\ref{fig:animation_summary} for a solitary state of node 75.
    }
    \label{fig:asymtptotic_freq_histo}
\end{figure}

To illustrate the phenomenon we describe and explain in this paper, we show simulation results in Fig.~\ref{fig:asymtptotic_freq_histo}. It displays the different asymptotic frequencies actually observed following single node perturbations in a complex network, the topology of which was generated with \cite{RandomGrowthModel}.
 The $|\omega_s|$, that coincide with $\omega_s \cdot \text{sgn}(P_z)$, are the absolute values of the mean asymptotic solitary frequencies $\omega_s$ of the trajectories $\dot\varphi_z(t)$ at the solitary nodes $z$. The trajectories are won from integrating Eq.~(\ref{eq:swing_lossless}) from many initial conditions generated by perturbing the synchronous state (see \cite{GitRepoReVelatioNs} for code and details).
    Here, asymptotic states are counted as 1-solitary if there is exactly one node with frequency $|\omega_s| > 1$, cf. Fig. S4.
Note that many solitary frequencies are close to the natural frequencies at $\Omega_i = \pm 10$.
We define such solitary states with $\omega_s \approx \Omega_z$ as \textit{natural}.
However, when perturbations occur at nodes  74, 75 and 84, they can also induce intermediate frequencies.
We define solitary states with such intermediate mean frequency $|\omega_s| < |\Omega_z|$ as \textit{resonant}, because (i) the fluctuation of the nodes in $S$ is more excited (see Fig.~\ref{fig:animation_summary}), and (ii) because we show that $\omega_s$ is in resonance with a network mode.
 The nodes 74, 75 and 84 share a specific topological property: they are leaf nodes (degree-1 nodes) with a high-degree neighbor. The intermediate resonant solitary states occur at all such nodes in this network (and other networks), and only at these nodes, discarding marginal effects.
We explore this strong link between the topology and the asymptotic behavior of the nodes in the following sections.

\section{Ansatz and self-consistency}
Here, we present our main result: the framework that allows us to make predictions of solitary states and relate them to the network's topology.
The framework predicts the existence and stability of both natural and resonant solitary states.
It consists of an ansatz and a self-consistent equation it has to fulfill.

An ansatz is a proposed form of a solution, often justified by prior knowledge about the system, and its sole purpose is to insert it and evaluate the result it produces. Here, the ansatz is a change of coordinates without loss of generality.

We begin by introducing suitable notation for describing a solitary leaf node $z$ and a synchronized cluster $S = \{i : 1 \leq i \leq N, i \neq z \}$.
For simplicity, we will assume that $\alpha_i$ and $m_i$ are constant throughout $S$. Node $z$ is connected to $S$ through the intermediary node $k \in S$. 
With all introduced notations, the considered system has the form
\begin{align}
    m \ddot{\varphi}_{i} & =P_{i}-\alpha \dot{\varphi}_{i}-\sum_{j=1}^{N}\kappa_{ij}\sin(\varphi_{i}-\varphi_{j}), 
    \quad i\in S,
    \label{eq:eom_S}
    \\
    m_{z}\ddot{\varphi}_{z} & =P_{z}-\alpha_{z}\dot{\varphi}_{z}-\kappa_{zk}\sin(\varphi_{z}-\varphi_{k}).
    \label{eq:eom_z}
\end{align}
The coupling between subsystems $S$ and $z$ is given by $p_{zk}$, which we decompose into a long-term time average $p_s \coloneqq \langle p_{zk} \rangle$ and a mean zero part $p_{zk}^\text{osc}(t)$:
\begin{equation}
    p_{zk}(t) = \kappa_{zk} \sin\left[\varphi_z(t) - \varphi_k(t)\right] \coloneqq p_s + p_{zk}^\text{osc}(t).
    \label{eq:p_zk}
\end{equation}
Note that we do not have an explicit expression for $p_s$ and $p_{zk}^\text{osc}(t)$ yet, but obtain them later.

Our ansatz for the trajectories, is to split the linear motion in time with the mean frequency from the mean zero part, similarly to $p_{zk}$ in Eq.~(\ref{eq:p_zk}),
\begin{align}
    \varphi_i(t) &\coloneqq \omega_\text{sync} t + \vartheta_i^* + \vartheta_i(t)
    \quad\text{for $i \in S$,}
    \label{eq:ansatz_S}
    \\
    \varphi_z(t) &\coloneqq \omega_\text{sync} t + \vartheta_k^* + \vartheta_z(t) \coloneqq\omega_s t + \vartheta_k^* + \varepsilon \psi_z(t),
    \label{eq:ansatz_z}
\end{align}
for some small $\varepsilon$ that we determine later.
We define $\omega_\text{sync}$ and $\omega_s$ as the long term average frequency of the synchronized cluster, and the solitary oscillator, respectively.
This means $\langle\vartheta_i(t)\rangle = 0 = \langle\psi_z\rangle(t)$ by construction.
The $\vartheta_i^*$ are a synchronous state of $S$ given a fixed power injection $p_s$ at node $k$.
Note that the ansatz is without loss of generality, and can be formally considered as a change of variables.
We use the form Eqs.~(\ref{eq:ansatz_S}-\ref{eq:ansatz_z}) because if a solitary state occurs, $\psi_z(t)$ and $\vartheta_i(t)$ are bounded, and the corotating solitary frequency, defined as $\omega_c \coloneqq \omega_s - \omega_\text{sync}$, is non-zero, cf. Supplementary Video~1 \cite{Supplement}.
We can use the ansatz Eqs.~(\ref{eq:p_zk}-\ref{eq:ansatz_z}) to determine the mean-zero part of the solitary trajectory $\psi_z(t)$, however, it is more informative to determine $\omega_s$ and $p_s$ from the ansatz as a characterization of the solitary state.

To determine $\omega_s$, we note that a condition for the existence of a solitary state at $\omega_s$ is that both $\omega_s$ and $p_s$ have to fulfill a self-consistent equation.
The argument goes as follows.
For $\psi_z(t)$ to stay bounded, and for $\omega_s$ to be the long-term average of the solitary node's frequency, the long-term average of $\ddot\vartheta_z = \varepsilon \ddot \psi_z$ has to be zero. Demanding this for Eq.~(\ref{eq:eom_z}) provides us with the condition $0 = P_z - \alpha_z \omega_s - p_s$.
We can interpret $p_s$ as a measure for the mean coupling of $S$ and $z$, while $\omega_s$ stands for their incoherence. Note that this relationship between $\omega_s$ and $p_s$ can be observed in Fig.~\ref{fig:animation_summary}: in synchrony ($\omega_s = 0$) there is a steady high energy flow $p_s=P_z$; in the natural solitary state, the mean energy flow $p_s\approx0$ and the system is effectively decoupled. For intermediate resonant solitary states, there is some intermediate amount of energy transfer $0 < |p_s| < |P_z|$.

If we now can express $p_s$ as a function of $\omega_s$, the above condition for a solitary state to have well-defined frequency becomes a self-consistent equation for $\omega_s$:
\begin{align}
0 = Z(\omega_s) \coloneqq P_z - \alpha_z \omega_s - p_s(\omega_s).
\label{eq:self_consistency_implicit}
\end{align}

The self-consistency function $Z(\omega_s)$ is defined as the average change of the solitary frequency as a function of the solitary mean frequency: $Z(\omega_s) \coloneqq \langle m_z \ddot\vartheta_z \rangle (\omega_s)$.
A potentially stable solitary state at $\omega_s$ requires the correct sign of the change in $Z$ with frequency: if a change in $\omega_s$ leads to a larger change in $\omega_s$ in the same direction, the solitary frequency is linearly unstable, and vice versa.
If we imagine $\omega_s$ as slowly varying, we have $m_z \dot \omega_s = Z(\omega_s)$. Thus, we expect the sign of the derivative
$\partial_{\omega_s} Z(\omega_s)$ to be informative about the stability of the solitary state, at least as a proxy.
In other words, by introducing the explicitly time-dependent ansatz Eqs.~(\ref{eq:ansatz_S}-\ref{eq:ansatz_z}), we get a non-autonomous system from Eqs.~(\ref{eq:eom_S}-\ref{eq:eom_z}). We are looking for limit cycles of this system that correspond to fixed points in the long-term averaged dynamics. These fixed points are solutions $\omega_s$ of Eq.~(\ref{eq:self_consistency_implicit}): mean frequencies that are allowed to persist by the dynamics, and that are stable under small perturbations.

To obtain the solutions, our strategy is as follows.
We assume that we can linearize Eqs.~(\ref{eq:eom_S}-\ref{eq:eom_z}) in $\vartheta_i(t)$ and $\varepsilon$ to obtain a system that quantitatively reflects the behavior of the trajectory of the solitary state. We will analyze the linearized system using averaging, and see that this is justified in many regimes. Together with the self-consistent equation, we obtain a proxy system for the description of solitary states. If the proxy system is stable, we conclude that a solitary state can (but is not guaranteed to) exist at $\omega_s$.
If the proxy system does not admit a stable solution for specific parameters, we interpret this as evidence that a solitary state can not exist with those parameters. Numerically, we see that if the proxy system admits a stable solution, we often do find solitary states with a large basin. We interpret this to mean that the mechanism revealed by the proxy system is indeed responsible for the formation and stabilization of solitary states.

To follow the strategy outlined above and evaluate Eq.~(\ref{eq:self_consistency_implicit}) to get predictions for solitary frequencies and their stability, we first need to determine $p_s$ as a function of $\omega_s$.
We present a derivation in the next section, and an interpretation and alternative derivation thereafter.

\section{Averaging}
Here, we outline the derivation of our main result, an explicit form for $p_s(\omega_s)$.
For a detailed step-by-step instruction, we refer to the Supplemental Material \cite{Supplement}, section I (SM I).

Both $p_s$ and $\omega_s$ depend on the trajectories $\varphi_z(t)$ and $\varphi_k(t)$ of the solution to Eqs.~(\ref{eq:eom_S}-\ref{eq:eom_z}). To obtain an explicit approximation of the solution, we will leverage a version of the averaging theorem \cite{Guckenheimer2002-rm} (and references therein).

We start by giving a brief summary of the averaging method following \cite{Guckenheimer2002-rm}.
Let $0 \leq \varepsilon \ll 1$, and $x\in U \subseteq \mathbb{R}^n$ for a bounded set $U$.
Let $f:\mathbb{R}^n \times \mathbb{R} \times \mathbb{R}^+ \to \mathbb{R}^n$ be a  $C^r$ function  with $r\ge 2$ (at least twice continuously differentiable) that is time-periodic with period $T>0$, and that determines the non-autonomous periodic dynamical system
\begin{equation}
    \dot x = \varepsilon f(x,t,\varepsilon).
    \label{eq:nonautonomous_system}
\end{equation}
Define the associated autonomous averaged system as
\begin{equation}
    \dot y = \varepsilon \frac{1}{T} \int_0^T f(y,t,0) dt \coloneqq\varepsilon \overline{f}(y).
    \label{eq:averaged_system}
\end{equation}

The averaging theorem states that a system of the form Eq.~(\ref{eq:nonautonomous_system}) can be cast into the form
\begin{equation}
    \dot y = \varepsilon \overline{f}(y) + \varepsilon^2 f_1(y,t,\varepsilon),
\end{equation}
where $f_1$ is also of period $T$ in $t$. This can be achieved by a coordinate change.
Moreover, solutions $x(t)$ of Eq.~(\ref{eq:nonautonomous_system}) and $y(t)$ of Eq.~(\ref{eq:averaged_system}) that are $\varepsilon$-close stay $\varepsilon$-close on a timescale $t\sim\varepsilon^{-1}$.
Further, hyperbolic fixed points of Eq.~(\ref{eq:averaged_system}) have corresponding unique $\varepsilon$-close hyperbolic periodic orbits of the same stability type.
We will leverage this theorem as follows: By using the ansatz Eqs.~(\ref{eq:ansatz_S}-\ref{eq:ansatz_z}) and appropriate approximations, we cast the system Eqs.~(\ref{eq:eom_S}-\ref{eq:eom_z}) into the form Eq.~(\ref{eq:nonautonomous_system}).
In the proper coordinate frame, the system is time-periodic and slowly varying.
We can then find approximate solutions by solving its corresponding averaged system for fixed points.
Those solutions, in particular $\vartheta_k(t)$, enable us to calculate $p_s$ as a function of $\omega_s$ and close the self-consistent equation, Eq.~(\ref{eq:self_consistency_implicit}).

First, we need to approximate the highly nonlinear dynamics in Eqs.~(\ref{eq:eom_S}-\ref{eq:eom_z}) appropriately.
Since the synchronized oscillators from the cluster $S$ are weakly perturbed by the solitary rotation, we assume smallness of $\vartheta_i(t)$, which is also justified by numerical observations, see Fig.~\ref{fig:animation_summary} and \cite{Supplement, GitRepoReVelatioNs}. Therefore, we linearize the system Eqs.~(\ref{eq:eom_S}-\ref{eq:eom_z}) with respect to $\vartheta_i(t) \in S$ around the origin.
It is crucial not to linearize with respect to $\vartheta_z(t)$, because it is not bounded.
For example, the coupling function $p_{zk}$ in Eqs.~(\ref{eq:eom_z}) and (\ref{eq:p_zk}) is approximated to linear order in $\vartheta_k$ as
\begin{align}
p_{zk}(t) &\approx \kappa_{zk} \left\{ \sin\left[\omega_c t + \varepsilon\psi_z(t)\right] - \vartheta_k(t) \cos\left[\omega_c t + \varepsilon\psi_z(t)\right] \right\}.
\label{eq:p_zk_linearized_theta}
\end{align}
From now on, $p_s$ and $p_{zk}^\text{osc}(t)$ represent the mean and oscillating part of the approximated $p_{zk}$.
To write the obtained system in vector form, we define the vectors $e^k_i := \delta_{ik}$, where $\delta_{ik} = 1$ if $i=k$, and $\delta_{ik} = 0$ else. Introducing $\vec{\vartheta}(t)$ with components $\vartheta_i(t)$ for all $i \in S$,
the linearized system is
\begin{align}
    m \ddot{\vec{\vartheta}} =& - \alpha \dot{\vec{\vartheta}} - L^* \vec{\vartheta} + e^k p_{zk}^\text{osc}(t),
    \label{eq:linear network}
    \\
    m_z \ddot\vartheta_z =& P_z - \alpha_z (\dot\vartheta_z + \omega_\text{sync}) - p_s - p_{zk}^\text{osc}(t),
    \label{eq:sol}
\end{align}
where $L^*$ is the effective coupling Laplacian on the synchronized cluster with weights $L^*_{ij} = \kappa_{ij}\cos(\vartheta^*_i - \vartheta^*_j)$ and eigenvectors $\vec{v}^{[\ell]}$. 
We have successfully transformed into a frame that accounts for the power imbalance due to $p_s$. This can be seen from the fact that $p_s$ does not appear in the equation for $S$, Eq.~(\ref{eq:linear network}), since the synchronized cluster steady state accounts for the average power injection $p_s$ from the solitary node with the $\omega_\text{sync}$ rotation.

Second, we need to properly take care of the fact that Eqs.~(\ref{eq:linear network}-\ref{eq:sol}) become a non-autonomous system when we insert the explicitly time-dependent ansatz Eqs.~(\ref{eq:ansatz_S}-\ref{eq:ansatz_z}) into $p_{zk}$.
We introduce a timescale $\varepsilon^{-1}$ that splits the autonomous and non-autonomous parts in Eqs.~(\ref{eq:linear network}-\ref{eq:sol}).
It allows us to write the system in a perturbative form, in which averaging can be applied.
Such a scaling can be achieved, if node $k$ has a sufficiently high degree (SM I.A).
In that case, there is a Laplacian eigenvalue $\lambda^{[r]}$ that is highly correlated to the node's degree $d_k$.
We will observe in the examples below that the solitary frequency tunes itself towards resonance with such eigenmodes: $\omega_s^2 \sim \lambda^{[r]}$.

The rescaling is given by
\begin{equation}
    t = \sqrt{\frac{m}{\lambda^{[r]}}} t',
    \qquad
    P_i = \frac{P'_i \alpha \sqrt{\lambda^{[r]}}}{\sqrt{m}},
    \qquad
    \kappa_{ij} = \frac{\kappa'_{ij}\alpha \sqrt{\lambda^{[r]}}}{\sqrt{m}},
    \label{eq:parameter_rescaling}
\end{equation}
dropping the prime from now on. Further, we set $\varepsilon = \alpha/\sqrt{m \lambda^{[r]}} \ll 1$.
In other words, by rescaling the time axis we can absorb $m$ in Eq.~(\ref{eq:linear network}), and set $\alpha$, $P$, and $\kappa$ to $\varepsilon$, $\varepsilon P$ and $\varepsilon \kappa$. We choose $\varepsilon$ such that, while the rescaled $\kappa$ remains of order one, the eigenvalues of the rescaled effective Laplacian $L^*$ for the eigenmodes localized at $k$ are of order $\varepsilon^{-1}$. As a result, we treat $L^r = \varepsilon L^*$ as being of order one. 
Eqs.~(\ref{eq:linear network}-\ref{eq:sol}) become
\begin{align}
    \ddot{\vec{\vartheta}} =&\varepsilon \left[ - \dot{\vec{\vartheta}} + e^k p_{zk}^\text{osc}(t)\right] - L^r \vec{\vartheta},
    \label{eq:eps linear network}
    \\
    \ddot\vartheta_z =& \frac{\varepsilon m}{m_z} \left[P_z - \frac{\alpha_z}{\alpha} \dot\varphi_z - p_{zk}(t) \right].
    \label{eq:eps sol}
\end{align}

Third, a crucial step for solvability is to approximate $\varepsilon p_{zk}(t)$, cf. Eq.~(\ref{eq:p_zk_linearized_theta}), to first order in $\varepsilon$, which finally allows us to identify an explicit expression for $p_s$  in terms of $\vartheta_k$ and $\omega_c$,
\begin{align}
    \varepsilon p_{zk}(t) &\approx \varepsilon \kappa_{zk} \left(\sin\omega_c t - \vartheta_k(t) \cos\omega_c t \right),
    \\
    p_s &\approx \kappa_{zk} \langle \vartheta_k(t) \sin\omega_c t \rangle.
    \label{eq:p_s_explicit}
\end{align}

Equation~(\ref{eq:eps linear network}) has the form of the perturbation of the Hamiltonian system $\ddot{\vec{\vartheta}} = - L^r \vec\vartheta$, where the perturbation is of order $\varepsilon$.
The perturbation is caused partly by the presence of $\vartheta_z(t)$, and by other small terms, such as damping.
The appropriate perturbative approach in this case is averaging \cite{Guckenheimer2002-rm}, where one should first write the system for the slowly varying amplitudes $\vec x(t)$ and $\vec z(t)$ of the network modes $\vec \xi(t)$, i.e., the amplitudes of the periodic solutions of the unperturbed system.
This is achieved by diagonalization of the Laplacian with the linear transformation $Q_{ij} \coloneqq v_i^{[j]}$,
\begin{align}
    \vec\xi(t) \coloneqq Q^T \vec \vartheta(t) \coloneqq \vec x(t) \cos\omega_c t - \vec z(t) \sin\omega_c t.
\end{align}
The amplitude dynamics for $\vec x(t)$ and $\vec z(t)$ can be inferred with the invertible van der Pol transformation (spelled out in Eq.~(S50)) into the frame rotating at the driving frequency $\omega_c$.
These amplitude dynamics is then decomposed into modes resonant with $\omega_c$ and non-resonant modes.
For the sake of brevity and clarity, we refer to SM I for details.
Most importantly, the contributions of both types of modes are small for different reasons.
On one hand, the resonant modes give small contributions due to a small prefactor $\varepsilon \Delta \sim \omega_c^2 - \lambda^{[r]}$.
On the other hand, the non-resonant modes, where $\varepsilon\Delta$ is of $O(1)$, give small contributions due to their weak localization, allowing us to average the dynamics of the slowly varying $\vec x(t)$ and $\vec z(t)$. We obtain an autonomous system that is a good approximation ($\varepsilon$-close) of the non-autonomous system for time scales up to $\varepsilon^{-1}$ \cite{Guckenheimer2002-rm}.
This averaged system (Eq.~(S55)) is linear  with a unique solution that gives us the average amplitudes $\vec x^*$ and $\vec z^*$ of all modes. From those, we obtain the average trajectory of $\vartheta_k(t)$, and insert it into Eq.~(\ref{eq:p_s_explicit}) to get an expression for $p_s(\omega_s)$.
This finally provides us with the explicit form of the self-consistent Eq.~(\ref{eq:self_consistency_implicit}):
\begin{align}
    0 = 
    Z(\omega_s) &= P_{z}-\alpha_z\omega_s - p_s
    \label{eq:self_consistency_explicit}
    \\
    &=P_{z}-\alpha_z\omega_s - \frac{ \kappa_{zk}^2 }{2} \sum_{\ell=1}^{N-1} \frac{\alpha \omega_c \left(v_k^{[\ell]}\right)^2}{\left(\lambda^{[\ell]} - m \omega_c^2\right)^2 + \alpha^2\omega_c^2}.
    \nonumber
\end{align}

In Eq.~(\ref{eq:self_consistency_explicit}), the sum runs over eigenmodes $\ell$ of $L^*$, with eigenvalue $\lambda^{[\ell]}$ and eigenvector $\vec{v}^{[\ell]}$. To efficiently evaluate Eq.~(\ref{eq:self_consistency_explicit}), we assume that the dependence of $L^*$ and its eigenmodes on $p_s$ is weak. This is a reasonable assumption if the synchronized cluster is significantly larger than the solitary cluster (node $z$), and the energy injected at node $k$ dissipates quickly into $S$.
Practically, the $\vartheta_i^*$, that depend on $p_s$, can be well approximated by the $\varphi_i^*$, that do not (SM I.B).

Lastly, to close Eq.~(\ref{eq:self_consistency_explicit}), we need to relate the solitary frequency $\omega_s$ to $\omega_c$, its relative value to $S$.
It can be shown by summing over Eq.~(\ref{eq:swing_lossless}) for all $i\in S$ and $i=z$, that the $\alpha$-weighted sum of frequencies decays towards zero from all initial conditions, hence $(N - 1) \alpha \omega_\text{sync} + \alpha_z \omega_s = 0$ (see SM I.A.). This gives us

\begin{align}
    \omega_c(\omega_s) = \left(1 + \frac{\alpha_z}{(N - 1) \alpha} \right) \omega_s
    \label{eq:omega_c}.
\end{align}
This linear relationship between $\omega_s$ and $\omega_c$ is useful in solving Eq.~(\ref{eq:self_consistency_explicit}).

Having obtained an explicit, closed, self-consistent equation for $\omega_s$ from our ansatz, we now turn to its interpretation and evaluation to make predictions of solitary attractor states and compare them to simulations.

\section{Interpretation: a linear response picture}
\label{sec:interpretation_a_linear_response_picture}
In this section, we provide an alternative derivation that is a shortcut to Eq.~(\ref{eq:self_consistency_explicit}), using linear response theory. For an extensive account of linear response in inertial Kuramoto networks, we refer to \cite{Xiaozhu19}.
As a motivation, we start by discussing the natural interpretation that the linear response offers, and that can also be deducted from the resulting Eq.~(\ref{eq:self_consistency_explicit}).

The obtained form of $p_s$ in Eq.~(\ref{eq:self_consistency_explicit}) has a natural interpretation. The solitary state, with frequency $\omega_s$, perturbs the synchronized cluster $S$, with relative frequency $\omega_c$. The response $\vec\vartheta(t)$ of the cluster $S$ oscillates at frequency $\omega_c$ with amplitudes $a_i$ and some well-defined phase shifts $\delta_i$, e.g., $\vartheta_k(t) \sim a_k \sin\left(\omega_c t + \delta_k\right)$. The phase shift $\delta_k$ enables a non-zero energy flow $p_s$ between $S$ and $z$, which is responsible for maintaining the relative frequency shift $\omega_c$.
The energy flow is proportional to the amplitude of the response, and especially the local amplitude of the corresponding eigenvector $v_k^{[\ell]}$ for each mode $\ell$.
Thus, large energy flows can occur if $\omega_c$ is close to resonance with an eigenmode of $S$ that is localized at $k$. As $\omega_c$ determines the magnitude of energy flow, but also shifts with the energy flow itself, the system can robustly tune itself towards such resonances if they are present. This resonant tuning mechanism predicts why certain topological features, namely the ones causing highly localized modes, enable the generation of additional solitary states, cf. section \ref{sec:evaluation}.

Following this interpretation, we can directly derive the result for $Z(\omega_s)$ in Eq.~(\ref{eq:self_consistency_explicit}) using linear response theory, see SM I.C.
We observe that the nonlinear part of $p_{zk}^\text{osc}(t)$ in Eq.~(\ref{eq:eps linear network}) is eventually averaged out in the averaging approach.
Linear response can be readily applied, if we neglect this nonlinearity from the outset. The assumption that this nonlinearity is sufficiently small can be justified, when it is only one of several contributions to the overall coupling of a well-connected node $k$.
In this case, the cluster $S$ is harmonically driven by $\sin\omega_c t$, and linear response similar to \cite{Xiaozhu19} can be used to calculate its response for each network mode $\ell$ separately, including the modal amplitudes $a_i^{(k)[\ell]}(\omega_c)$ at node $i$ and modal phase lags $\delta^{[\ell]}(\omega_c)$ given by
\begin{eqnarray}
    a_i^{(k)[\ell]}(\omega_c) &\coloneqq& \frac{ \kappa_{kz}  v_i^{[\ell]}  v_k^{[\ell]}  }{\sqrt{\left(\lambda^{[\ell]} - m \omega_c^2\right)^2 + \alpha^2 \omega_c^2 }},
    \label{eq:linear_response_resonance_amplitude}
    \\
    \sin\delta^{[\ell]}(\omega_c) &\coloneqq& -\frac{\alpha \omega_c}{\sqrt{\left(\lambda^{[\ell]} - m \omega_c^2\right)^2 + \alpha^2 \omega_c^2 }}.
\end{eqnarray}
Here, the superscript $^{(k)}$ indicates that the perturbation caused by $\varphi_z$ enters $S$ at node $k$.
For each mode $\ell$, the nodal amplitudes depend on the mismatch between the driving frequency $\omega_c$ and the eigenmodes of the network associated with the $\lambda^{[\ell]}$, as well as the localization of the corresponding eigenvectors at nodes $i$ and $k$.
The damping slightly shifts the location of the resonance peaks in the amplitude away from $\omega_c = \pm \sqrt{\lambda^{[\ell]}/m}$ and makes their height finite.

Combining the modal responses linearly provides us with a trajectory for the $\vartheta_i$,
\begin{equation}
    \kappa_{kz} \vartheta_i^{(k)}(t) =  \sum_{\ell=1}^{N-1} a_i^{(k)[\ell]} \sin\left(\omega_c t + \delta^{[\ell]} \right).
    \label{eq:linear_response_solution_with_amplitudes_and_phases}
\end{equation}
From the trajectory of $\vartheta_k$ we can determine the time averaged energy flow $p_s = \langle p_{zk}\rangle$ in terms of the amplitudes and phase lags:
\begin{eqnarray}
    p_s(\omega_c) &=&- \frac{\kappa_{kz}}{2} \sum_{\ell=1}^N a_k^{(k)[\ell]}(\omega_c) \sin\delta^{[\ell]}(\omega_c)
    \\
    &=& \frac{ \kappa_{zk}^2 }{2} \sum_{\ell=1}^{N-1} \frac{\alpha \omega_c \left(v_k^{[\ell]}\right)^2}{\left(\lambda^{[\ell]} - m \omega_c^2\right)^2 + \alpha^2\omega_c^2},
    \label{eq:self_consistency_power_flow_explicit}
\end{eqnarray}
which is the exact same as in Eq.~(\ref{eq:self_consistency_explicit}).
In fact, the terms in the denominators in the sum over all modes in Eq.~(\ref{eq:self_consistency_explicit}) resemble the resonance curve of linear oscillators.
It is known as Cauchy–Lorentz distribution, Lorentz(ian) function, or Breit–Wigner distribution.
Due to the generalization to the resonance of a network $S$, the Lorentzians are summed over all $N-1$ network modes associated to the $\lambda^{[\ell]}$.
Each contribution is weighted by the localization of the corresponding eigenvector $\vec{v}^{[\ell]}$ at node $k$, quantified by $\left(v_k^{[\ell]}\right)^2$.
This is how the topology affects the distribution of resonance in the network, see also \cite{Xiaozhu19}.

We remark that both approaches used to derive Eq.~(\ref{eq:self_consistency_explicit}) straightforwardly generalize to a solitary node connected to several nodes in $S$, see SM I.D. The back-reactions of the network combine linearly, and we simply sum over all neighbors of $z$ (Eq.~(S74)).

Furthermore, a treatment of heterogeneous system parameters within $S$ is also possible, see SM I.E.

\section{Evaluation}
\label{sec:evaluation}
Here, we make some general statements about the properties and solutions of our main result, Eq.~(\ref{eq:self_consistency_explicit}), and provide some examples.

First, note that the obtained expression for $p_s = \langle p_{zk}\rangle$ in Eq.~(\ref{eq:self_consistency_explicit}), has the same sign as $\omega_c$. Therefore, $p_s$ is antisymmetric in $\omega_c$ and hence in $\omega_s \sim \omega_c$, cf. Eq.~(\ref{eq:omega_c}).
It follows that Eq.~(\ref{eq:self_consistency_explicit}) is invariant under $\omega_s \to -\omega_s$ while $P_z \to -P_z$, i.e., generators and consumers behave similarly with their solitary frequencies having opposite signs.
In fact, it is easy to see from Eq.~(\ref{eq:self_consistency_explicit}) that all solutions lie between zero and the natural frequency, $0 \le |\omega_s| \le |\Omega_z| = |P_z|/\alpha_z$.
We see these bounds and symmetry properties confirmed in Fig.~\ref{fig:asymtptotic_freq_histo}, and the examples below.

As discussed above, the linear stability of solutions of $Z(\omega_s) = 0$ can be heuristically determined by evaluating the derivative: the orbit at $\omega_s$ is indicated to be linearly stable if $\partial_{\omega_s} Z(\omega_s) < 0$ and linearly unstable if $\partial_{\omega_s} Z(\omega_s) > 0$.
This consideration gains importance with the numerical observation that the instantaneous solitary frequency $\dot\varphi_z(t)$ oscillates with time around $\omega_s$, cf. Fig.~\ref{fig:animation_summary}.
The solution to $Z(\omega_s)=0$ is an intersection of $p_s(\omega_s)$ and the straight line $P_z - \alpha_z \omega_s$ with slope $-\alpha_z$.
Therefore, stability is given if $-\alpha_z < \partial_{\omega_s} p_s(\omega_s)$.
Graphically, this means that $p_s(\omega_s)$ intersects the straight line from below, counting in positive $\omega_s$ direction.

The existence of solitary frequencies $\omega_s$ as solutions to Eq.~(\ref{eq:self_consistency_explicit}) relies on sufficiently high eigenvector localization at the neighboring node $k$, quantified by $\left(v_k^{[\ell]}\right)^2$.
Such high localization is present in random networks of high degree heterogeneity \cite{Hata2017-jl}.
Indeed, we observe in Fig.~\ref{fig:asymtptotic_freq_histo} and in the examples below that resonant solitary states are most common at degree-one nodes with a high degree neighbor \cite{Nitzbon_2017}, and that the solitary frequency scales with the degree $d_k$ of node $k$.
For an estimate of the relation between the eigenvector localization, the corresponding eigenvalue and the node's degree, we refer to SM I.A.

The existence of solutions to Eq.~(\ref{eq:self_consistency_explicit}) also depends on system parameters and their topological distribution.
Whether and which solitary states exist in a complex system is a complex question with no easy answer, but we can provide some insights (see SM II.D for more details).
Generally, the parameters need to be chosen in an intermediate regime, such that the synchronous state is not globally attractive, but the coupling is also not too weak.
Relatively small damping $\alpha_i$ and strong coupling $\kappa_{ij}$, such as in power grids, are an indicator for this.
We remark that solving the self-consistent Eq.~(\ref{eq:self_consistency_explicit}) is an efficient way to determine which solitary states exist. Numerical simulations of the dynamics as in Fig.~\ref{fig:asymtptotic_freq_histo} are more precise but more expensive.

We now illustrate the power of Eq.~(\ref{eq:self_consistency_explicit}) using two examples. First we give a minimal effective model that exhibits the resonant tuning mechanism cleanly; then a full complex network with a complex resonant response. We compare the results with numerical simulations \cite{PowerDynamics2022}. Further examples can be found in the SM II.B.

\begin{figure*}%[h]
    \centering
    \includegraphics[width=\textwidth]{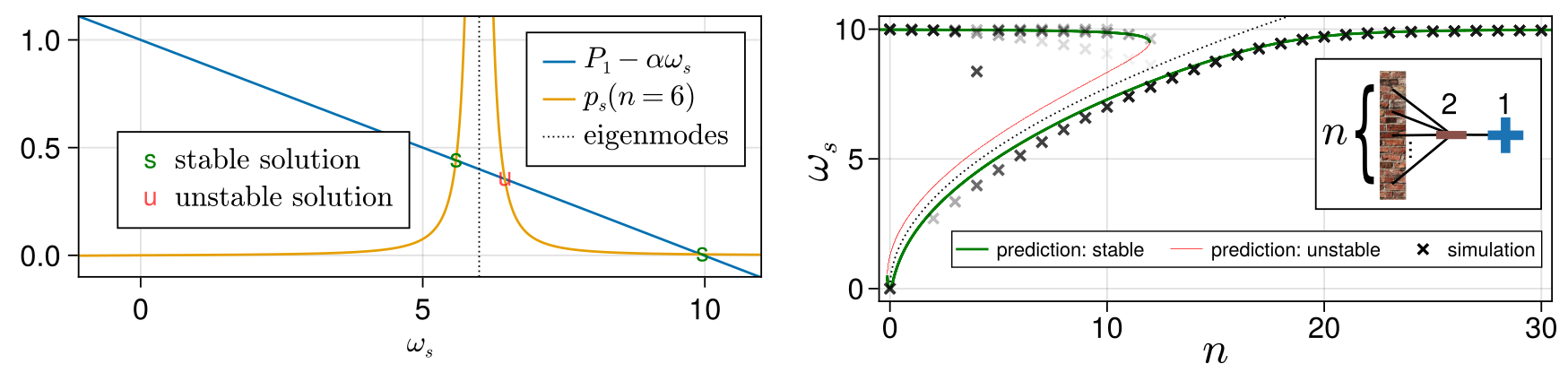}
    \caption{Network modes shape the landscape of solitary attractor states for the two-node model (inset).
    (left) Intersections between the straight line $P_1 - \alpha \omega_s$ and the mean energy flow $p_s(n=6)$ generate stable and unstable solutions according to Eq.~(\ref{eq:self_consistency_explicit}).
    We use $P_1 = 1$, $\alpha=0.1$.
    The single peak of $p_s$ is centered around the network mode close to $\sqrt{K n}$.
    (right) Bifurcation diagram with the bifurcation parameter $n$.
    The two stable branches are connected by an unstable branch and centered around the network mode.
    Numerical simulations \cite{GitRepoReVelatioNs} confirm predictions of both the location and stability of solitary states.
    }
    \label{fig:toy_model_intersection_and_bifurcation_diagram}
\end{figure*}

\subsection{Example I: Two-node model}
A minimal model that still shows tristability with two distinct solitary frequencies is a system that neglects the response of the network beyond the node $k$. We have a solitary node ($z=1$), its neighbor ($k=2$), and the rest of the network is assumed to have infinite inertia, and thus no dynamics. This model is shown in the inset in Fig.~\ref{fig:toy_model_intersection_and_bifurcation_diagram} (right). The neighbor has $n$ links into the infinitely inert part, each with coupling strength $K$.
See SM II.A for details.

The left panel in Fig.~\ref{fig:toy_model_intersection_and_bifurcation_diagram} shows the graphical solution of Eq.~(\ref{eq:self_consistency_explicit}) for this example system. We see the energy absorbed in node 1 by frequency adaptation, a straight line given by $P_1 - \alpha \omega_s$, and the energy flow due to the network's response, $p_s(\omega_s) = \langle p_{12}\rangle (\omega_s)$. The energy flow has a single peak centered around the only network mode, which is close to $\lambda \approx K n$. At the intersections of these two curves, we have zeros of Eq.~(\ref{eq:self_consistency_explicit}). However, the linear stability heuristic suggests that only intersections with $\partial_{\omega_s} Z < 0$ are candidates for stable solitary states.
Indeed, numerical simulations confirm the existence of the stable solutions, and that they correspond to the intersections, where $p_s$ comes from below.

Repeating this solution process for Eq.~(\ref{eq:self_consistency_explicit}) for a range of $n$, we can draw the bifurcation diagram in Fig.~\ref{fig:toy_model_intersection_and_bifurcation_diagram} (right).
Each vertical slice corresponds to the top view of an intersection plot like the one in the left panel.
For this minimal example system, we can obtain an explicit expression for several branches $n(\omega_s)$ of stable or unstable solutions (see SM II.A) that correspond to the intersections in the left panel.
 These branches predict that for $n \lesssim 12$, there is a pair of a stable and an unstable solution $\omega_s$ due to the resonance peak around $\lambda$, and a third, stable solution at $\omega_s \approx \Omega_1 = P_1/\alpha$. For larger values of $n$, the second and third solutions annihilate, while the first converges to $\omega_s \to \Omega_1$.
Numerical simulations with random initial conditions \cite{GitRepoReVelatioNs} closely correspond to the predicted frequency and stability of solitary states, confirming the existence of tristability between the synchronous state
and solitary states at multiple frequencies, and confirming the overall
accuracy of our results.

In summary, this minimal example system shows that the resonant (intermediate) solitary state and the natural (decoupled) solitary state are two ends of a spectrum of frequencies, and connected by an unstable branch of solutions. Further, the natural frequency is an upper bound for solitary frequencies and can be approached by tuning the resonant network mode higher.

\subsection{Example II: Complex network}
\begin{figure}%[h]
    \centering
    \includegraphics[width=\columnwidth]{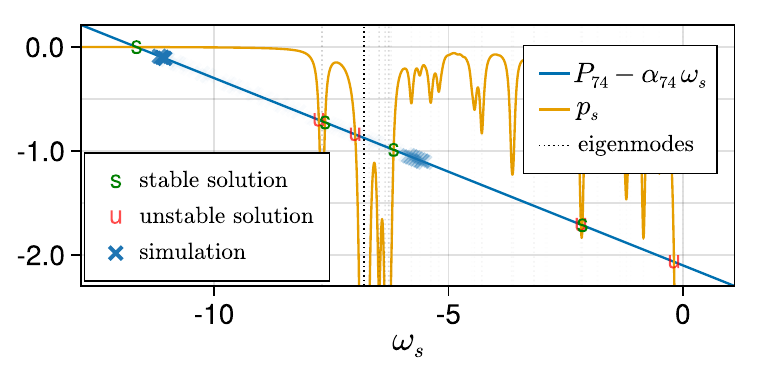}
    \caption{Solutions of Eq.~(\ref{eq:self_consistency_explicit}) for a dense sprout node ($i=74)$ in a synthetic power grid (Fig.~\ref{fig:asymtptotic_freq_histo}).
    For the sake of clarity, we slightly modified the node parameters to reduce the number of intersections in the figure (see SM II.C).}
    \label{fig:self_consistency_intersection_grid_1_node_74_P_-2.1_D_0.18}
\end{figure}

We now apply our result to the complex network shown in Figs.~\ref{fig:asymtptotic_freq_histo} and S2. For models of this complexity, the self-consistent equation has to be solved numerically for every node.
We observe that intermediate resonant solitary states are located at leaf nodes with neighbor degree $\geq$ 6, (cf. \cite{Nitzbon_2017}).
Figure~\ref{fig:self_consistency_intersection_grid_1_node_74_P_-2.1_D_0.18} shows the solutions of Eq.~(\ref{eq:self_consistency_explicit}) for $z=74$ for slightly heterogeneous parameters.

As many network modes are excited, the response curve $p_s(\omega_s)$ is considerably more complex than for the two-node model. However, simulation results reveal that there is still one dominant mode with high localization at the root node $k$. Solitary states obtained from random initial conditions are frequently found in close resonance with the dominant mode, or close to the natural frequency.

There are several tentative observations we can make by looking at complex networks, (see also SM II.C, Figs.~S5 and S6 for more examples):
(i)
If the self-consistent equation  yields more solitary states than the one at the natural frequency, there are two dominating solitary states: the one at the natural frequency, and one in resonance with a localized network mode.
There can also be many additional solutions with small basins of attraction, cf. Figs.~S4 and S5. 

(ii) Eq.~(\ref{eq:self_consistency_explicit}) tends to overestimate the magnitude of the most prominent solution for $\omega_s$ slightly, which can already be observed in the effective model (cf. Figs.~\ref{fig:toy_model_intersection_and_bifurcation_diagram} and \ref{fig:self_consistency_intersection_grid_1_node_74_P_-2.1_D_0.18}).
(ii) the most prominent natural solitary state lies at an intermediate frequency that has a high and broad peak in $p_s$ and some distance to the next solutions.
This can be understood by considering that every peak in $p_s$ produces a pair of solutions, a stable one and an unstable one, and that the solitary frequency $\varphi_z(t)$ is modulated around $\omega_s$, cf. Fig.~\ref{fig:animation_summary}.
Leaf nodes with high neighbor degree \cite{Nitzbon_2017}, typically produce such situations (cf. SM I.A) with a highly localized network mode that results in a high peak. Thus, these findings explain why such nodes feature resonant solitary states with intermediate frequencies.
(iii) Besides the main peak at the most localized network mode, there can be several minor peaks that result in solitary states with much smaller basins of attraction due to three reasons: First, the peaks might be narrow, such that the stable and unstable solution are close, second, the peak might not exceed the intersection with the straight line by much, hence the pair of solutions is already close to their annihilation bifurcation, and third, the peak might be close to other peaks and their solutions.
This can account for the observation that parts of the phase space can be dominated by basin boundaries of many, different, unlikely states \cite{Gelbrecht_2020}, inducing transient chaos \cite{halekotteTransientChaosEnforces2021}.

As a closing remark, the derivation of the minimal model, and the explicit relationship between network mode and neighbor degree (SM I.A and II.A), suggest the following. Even in complex networks, the neighbor degree $d_k$ can serve as an upper bound for the (squared) intermediate solitary frequency, when weighted with the coupling $\kappa_{ij}$. This is also observed numerically (Fig. S8).

\section{Summary and discussion} 
We present an analytical description of the effective coupling between a solitary oscillator and the synchronized cluster in a complex network. To approximate the trajectories, we utilize partial linearization and an averaging theorem. This way, we discover that a resonant excitation of the linear modes of the synchronized cluster can couple coherently to the shifted frequency of the solitary state, resulting in a large energy flow between the synchronized cluster and the solitary oscillator, and effectively shifting the frequency.

We have mainly found that intermediate solitary states, first observed in \cite{Nitzbon_2017}, are actually resonant solitary states, and we have uncovered how the network's topology shapes their properties. Furthermore, we have unified their analytical description with the one of natural solitary states \cite{Menck2014NatureComm}, and shown that there is a spectrum connecting those extremes.

Our self-consistent framework predicts solitary states in reduced models, where it is solvable, and is also in excellent agreement with what we observe numerically in the original complex networks.
With this structure-function relationship, we can explain the observation that solitary states appear mostly in specific parts of the network \cite{PhysRevE.103.042315}, more specifically leaf nodes and tree-shaped structures, which act as weak points in perturbation scenarios \cite{Menck2014NatureComm, Nitzbon_2017, halekotteMinimalFatalShocks2020,halekotteTransientChaosEnforces2021}. These results have important implications for improving grid stability through modifications of parameters \cite{Motter2013-yo,wassmerTargetedSuppressionFailure2021,Menara2022-bg} and topology \cite{Witthaut_2012,Menck2014NatureComm,Schultz16Tweak,kaiserNetworkIsolatorsInhibit2021}:
Our framework can be used to evaluate how these modifications affect the presence and stability of potentially harmful solitary attractors and their proximity to the synchronous state.
Furthermore, we can efficiently identify the troublemaker nodes that tend to desynchronize the easiest due to their intermediate frequencies: resonant solitary states at certain leaf nodes (dense sprout nodes, see \cite{Supplement,Nitzbon_2017}).
We can identify them from the network topology only, and without expensive simulations.
It has been shown that targeted control of such troublemaker nodes can improve grid stability drastically \cite{Olmi_Delayed_Feedback}.

\section{Conclusion and Outlook}

While the presented mechanism can explain the potential existence and stability of resonant solitary states, there are still open questions that require further investigation. A key practical question is the estimation of the basin boundaries of these states, which would provide deeper insights into the dynamics and robustness of synchronous systems against perturbations \cite{Manik2014-ye,klinshovStabilityThresholdApproach2015,halekotteMinimalFatalShocks2020,halekotteTransientChaosEnforces2021,StabilityRolSolInertia}.
It is known that the basin boundaries depend on the phase and frequency \cite{Menck2014NatureComm,Hellmann2020-aj,halekotteTransientChaosEnforces2021}. Averaging the phase dependence out, an estimate of the basin boundaries in $\omega$-direction could be estimated by the turning points of $Z(\omega)$ around a stable solution.

Detailed numerical studies suggest that additional classes of attractive states exist, and the interaction of losses on the lines and resonant energy flow remains unexplored \cite{Hellmann2020-aj}. Furthermore, a more detailed study of the interaction of multiple solitary nodes with distinct frequencies or in frequency clusters is needed (see SM I.F for an outline).

Finally, while this paper focuses on explaining a known phenomenon of Kuramoto oscillators, which exhibit a purely sinusoidal coupling, our derivation generally only demands a $2\pi$-periodic coupling function. This includes coupling functions with phase shifts and harmonics.
It would also be interesting to treat phase-amplitude oscillators that play a central role in networks such as power grids \cite{PRXEnergy.1.013008}.
We expect that the overall analytical approach we introduced  is a promising tool in describing the synchronization of rotation and oscillation, and leave it to future work to develop and adapt our theory to address a wide range of models.

\paragraph{Code Availability}
The numerical experiments were conducted with the \texttt{PowerDynamics.jl} package \cite{PowerDynamics2022}.
Further details can be found in SM III.
The code for reproducing our simulations and figures can be found in \cite{GitRepoReVelatioNs}.
The repository can also be applied to find solitary frequencies in complex networks by solving Eq.~(\ref{eq:self_consistency_explicit}).

\begin{acknowledgments}
\paragraph{Acknowledgments}
Initial investigations of this phenomenon grew out of work by Anton Plietzsch and discussions with Paul Schultz, Chris Bick, Jeroen Lamb, Deniz Eroglu, Tiago Pereira, and Carsten Grabow during a stay at Nesim Mathematics Village.

Part of this work was funded by the DFG grant \mbox{CoCoHype} (DFG KU 837/39-2; J.N., J.K., F.H., M.A.) and BMWK grant OpPoDyn (03EI1071A; J.N., F.H.).

J.N. gratefully acknowledges support by Studien\-stiftung des Deutschen Volkes scholarship foundation and Berlin International Graduate School in Model and Simulation based Research (BIMoS) at Technische Universit\"at Berlin.
S.Y. and J.K. acknowledge funding by DFG, German Research Foundation, Project No. 411803875.
\end{acknowledgments}

\bibliography{main}

\begin{thebibliography}{49}%
\makeatletter
\providecommand \@ifxundefined [1]{%
 \@ifx{#1\undefined}
}%
\providecommand \@ifnum [1]{%
 \ifnum #1\expandafter \@firstoftwo
 \else \expandafter \@secondoftwo
 \fi
}%
\providecommand \@ifx [1]{%
 \ifx #1\expandafter \@firstoftwo
 \else \expandafter \@secondoftwo
 \fi
}%
\providecommand \natexlab [1]{#1}%
\providecommand \enquote  [1]{``#1''}%
\providecommand \bibnamefont  [1]{#1}%
\providecommand \bibfnamefont [1]{#1}%
\providecommand \citenamefont [1]{#1}%
\providecommand \href@noop [0]{\@secondoftwo}%
\providecommand \href [0]{\begingroup \@sanitize@url \@href}%
\providecommand \@href[1]{\@@startlink{#1}\@@href}%
\providecommand \@@href[1]{\endgroup#1\@@endlink}%
\providecommand \@sanitize@url [0]{\catcode `\\12\catcode `\$12\catcode `\&12\catcode `\#12\catcode `\^12\catcode `\_12\catcode `\%12\relax}%
\providecommand \@@startlink[1]{}%
\providecommand \@@endlink[0]{}%
\providecommand \url  [0]{\begingroup\@sanitize@url \@url }%
\providecommand \@url [1]{\endgroup\@href {#1}{\urlprefix }}%
\providecommand \urlprefix  [0]{URL }%
\providecommand \Eprint [0]{\href }%
\providecommand \doibase [0]{http://dx.doi.org/}%
\providecommand \selectlanguage [0]{\@gobble}%
\providecommand \bibinfo  [0]{\@secondoftwo}%
\providecommand \bibfield  [0]{\@secondoftwo}%
\providecommand \translation [1]{[#1]}%
\providecommand \BibitemOpen [0]{}%
\providecommand \bibitemStop [0]{}%
\providecommand \bibitemNoStop [0]{.\EOS\space}%
\providecommand \EOS [0]{\spacefactor3000\relax}%
\providecommand \BibitemShut  [1]{\csname bibitem#1\endcsname}%
\let\auto@bib@innerbib\@empty
%</preamble>
\bibitem [{\citenamefont {Pikovsky}\ \emph {et~al.}(2001)\citenamefont {Pikovsky}, \citenamefont {Rosenblum},\ and\ \citenamefont {Kurths}}]{pikovskySynchronizationUniversalConcept2001}%
  \BibitemOpen
  \bibfield  {author} {\bibinfo {author} {\bibfnamefont {A.}~\bibnamefont {Pikovsky}}, \bibinfo {author} {\bibfnamefont {M.}~\bibnamefont {Rosenblum}}, \ and\ \bibinfo {author} {\bibfnamefont {J.}~\bibnamefont {Kurths}},\ }\href {\doibase 10.1017/CBO9780511755743} {\emph {\bibinfo {title} {Synchronization: {{A Universal Concept}} in {{Nonlinear Sciences}}}}},\ Cambridge {{Nonlinear Science Series}}\ (\bibinfo  {publisher} {{Cambridge University Press}},\ \bibinfo {address} {{Cambridge}},\ \bibinfo {year} {2001})\BibitemShut {NoStop}%
\bibitem [{\citenamefont {Strogatz}(2000)}]{strogatzKuramotoCrawfordExploring2000}%
  \BibitemOpen
  \bibfield  {author} {\bibinfo {author} {\bibfnamefont {S.~H.}\ \bibnamefont {Strogatz}},\ }\href {\doibase 10.1016/S0167-2789(00)00094-4} {\bibfield  {journal} {\bibinfo  {journal} {Physica D: Nonlinear Phenomena}\ }\textbf {\bibinfo {volume} {143}},\ \bibinfo {pages} {1} (\bibinfo {year} {2000})}\BibitemShut {NoStop}%
\bibitem [{\citenamefont {Acebr{\'o}n}\ \emph {et~al.}(2000)\citenamefont {Acebr{\'o}n}, \citenamefont {Bonilla},\ and\ \citenamefont {Spigler}}]{acebronSynchronizationPopulationsGlobally2000}%
  \BibitemOpen
  \bibfield  {author} {\bibinfo {author} {\bibfnamefont {J.~A.}\ \bibnamefont {Acebr{\'o}n}}, \bibinfo {author} {\bibfnamefont {L.~L.}\ \bibnamefont {Bonilla}}, \ and\ \bibinfo {author} {\bibfnamefont {R.}~\bibnamefont {Spigler}},\ }\href {\doibase 10.1103/PhysRevE.62.3437} {\bibfield  {journal} {\bibinfo  {journal} {Physical Review E}\ }\textbf {\bibinfo {volume} {62}},\ \bibinfo {pages} {3437} (\bibinfo {year} {2000})}\BibitemShut {NoStop}%
\bibitem [{\citenamefont {Acebr{\'o}n}\ \emph {et~al.}(2005)\citenamefont {Acebr{\'o}n}, \citenamefont {Bonilla}, \citenamefont {P{\'e}rez~Vicente}, \citenamefont {Ritort},\ and\ \citenamefont {Spigler}}]{acebronKuramotoModelSimple2005}%
  \BibitemOpen
  \bibfield  {author} {\bibinfo {author} {\bibfnamefont {J.~A.}\ \bibnamefont {Acebr{\'o}n}}, \bibinfo {author} {\bibfnamefont {L.~L.}\ \bibnamefont {Bonilla}}, \bibinfo {author} {\bibfnamefont {C.~J.}\ \bibnamefont {P{\'e}rez~Vicente}}, \bibinfo {author} {\bibfnamefont {F.}~\bibnamefont {Ritort}}, \ and\ \bibinfo {author} {\bibfnamefont {R.}~\bibnamefont {Spigler}},\ }\href {\doibase 10.1103/RevModPhys.77.137} {\bibfield  {journal} {\bibinfo  {journal} {Reviews of Modern Physics}\ }\textbf {\bibinfo {volume} {77}},\ \bibinfo {pages} {137} (\bibinfo {year} {2005})}\BibitemShut {NoStop}%
\bibitem [{\citenamefont {Pecora}\ \emph {et~al.}(2014)\citenamefont {Pecora}, \citenamefont {Sorrentino}, \citenamefont {Hagerstrom}, \citenamefont {Murphy},\ and\ \citenamefont {Roy}}]{pecora2014cluster}%
  \BibitemOpen
  \bibfield  {author} {\bibinfo {author} {\bibfnamefont {L.~M.}\ \bibnamefont {Pecora}}, \bibinfo {author} {\bibfnamefont {F.}~\bibnamefont {Sorrentino}}, \bibinfo {author} {\bibfnamefont {A.~M.}\ \bibnamefont {Hagerstrom}}, \bibinfo {author} {\bibfnamefont {T.~E.}\ \bibnamefont {Murphy}}, \ and\ \bibinfo {author} {\bibfnamefont {R.}~\bibnamefont {Roy}},\ }\href@noop {} {\bibfield  {journal} {\bibinfo  {journal} {Nature communications}\ }\textbf {\bibinfo {volume} {5}},\ \bibinfo {pages} {4079} (\bibinfo {year} {2014})}\BibitemShut {NoStop}%
\bibitem [{\citenamefont {Rodrigues}\ \emph {et~al.}(2016)\citenamefont {Rodrigues}, \citenamefont {Peron}, \citenamefont {Ji},\ and\ \citenamefont {Kurths}}]{Rodrigues2016-ep}%
  \BibitemOpen
  \bibfield  {author} {\bibinfo {author} {\bibfnamefont {F.~A.}\ \bibnamefont {Rodrigues}}, \bibinfo {author} {\bibfnamefont {T.~K. D.~M.}\ \bibnamefont {Peron}}, \bibinfo {author} {\bibfnamefont {P.}~\bibnamefont {Ji}}, \ and\ \bibinfo {author} {\bibfnamefont {J.}~\bibnamefont {Kurths}},\ }\href@noop {} {\bibfield  {journal} {\bibinfo  {journal} {Phys. Rep.}\ }\textbf {\bibinfo {volume} {610}},\ \bibinfo {pages} {1} (\bibinfo {year} {2016})}\BibitemShut {NoStop}%
\bibitem [{\citenamefont {Kuramoto}(1984)}]{Kuramoto1984-wh}%
  \BibitemOpen
  \bibfield  {author} {\bibinfo {author} {\bibfnamefont {Y.}~\bibnamefont {Kuramoto}},\ }\href@noop {} {\emph {\bibinfo {title} {Chemical oscillations, waves, and turbulence}}},\ \bibinfo {edition} {1st}\ ed.,\ Springer series in synergetics\ (\bibinfo  {publisher} {Springer},\ \bibinfo {address} {Berlin, Germany},\ \bibinfo {year} {1984})\BibitemShut {NoStop}%
\bibitem [{\citenamefont {Munyayev}\ \emph {et~al.}(2023)\citenamefont {Munyayev}, \citenamefont {Bolotov}, \citenamefont {Smirnov}, \citenamefont {Osipov},\ and\ \citenamefont {Belykh}}]{CyclopsStates}%
  \BibitemOpen
  \bibfield  {author} {\bibinfo {author} {\bibfnamefont {V.~O.}\ \bibnamefont {Munyayev}}, \bibinfo {author} {\bibfnamefont {M.~I.}\ \bibnamefont {Bolotov}}, \bibinfo {author} {\bibfnamefont {L.~A.}\ \bibnamefont {Smirnov}}, \bibinfo {author} {\bibfnamefont {G.~V.}\ \bibnamefont {Osipov}}, \ and\ \bibinfo {author} {\bibfnamefont {I.}~\bibnamefont {Belykh}},\ }\href {\doibase 10.1103/PhysRevLett.130.107201} {\bibfield  {journal} {\bibinfo  {journal} {Phys. Rev. Lett.}\ }\textbf {\bibinfo {volume} {130}},\ \bibinfo {pages} {107201} (\bibinfo {year} {2023})}\BibitemShut {NoStop}%
\bibitem [{\citenamefont {Kovalenko}\ \emph {et~al.}(2021)\citenamefont {Kovalenko}, \citenamefont {Dai}, \citenamefont {{Alfaro-Bittner}}, \citenamefont {Raigorodskii}, \citenamefont {Perc},\ and\ \citenamefont {Boccaletti}}]{kovalenkoContrariansSynchronizeLimit2021}%
  \BibitemOpen
  \bibfield  {author} {\bibinfo {author} {\bibfnamefont {K.}~\bibnamefont {Kovalenko}}, \bibinfo {author} {\bibfnamefont {X.}~\bibnamefont {Dai}}, \bibinfo {author} {\bibfnamefont {K.}~\bibnamefont {{Alfaro-Bittner}}}, \bibinfo {author} {\bibfnamefont {A.~M.}\ \bibnamefont {Raigorodskii}}, \bibinfo {author} {\bibfnamefont {M.}~\bibnamefont {Perc}}, \ and\ \bibinfo {author} {\bibfnamefont {S.}~\bibnamefont {Boccaletti}},\ }\href {\doibase 10.1103/PhysRevLett.127.258301} {\bibfield  {journal} {\bibinfo  {journal} {Physical Review Letters}\ }\textbf {\bibinfo {volume} {127}},\ \bibinfo {pages} {258301} (\bibinfo {year} {2021})}\BibitemShut {NoStop}%
\bibitem [{\citenamefont {Tanaka}\ \emph {et~al.}(1997{\natexlab{a}})\citenamefont {Tanaka}, \citenamefont {Lichtenberg},\ and\ \citenamefont {Oishi}}]{tanakaFirstOrderPhase1997}%
  \BibitemOpen
  \bibfield  {author} {\bibinfo {author} {\bibfnamefont {H.-A.}\ \bibnamefont {Tanaka}}, \bibinfo {author} {\bibfnamefont {A.~J.}\ \bibnamefont {Lichtenberg}}, \ and\ \bibinfo {author} {\bibfnamefont {S.}~\bibnamefont {Oishi}},\ }\href {\doibase 10.1103/PhysRevLett.78.2104} {\bibfield  {journal} {\bibinfo  {journal} {Physical Review Letters}\ }\textbf {\bibinfo {volume} {78}},\ \bibinfo {pages} {2104} (\bibinfo {year} {1997}{\natexlab{a}})}\BibitemShut {NoStop}%
\bibitem [{\citenamefont {Tanaka}\ \emph {et~al.}(1997{\natexlab{b}})\citenamefont {Tanaka}, \citenamefont {Lichtenberg},\ and\ \citenamefont {Oishi}}]{tanakaSelfsynchronizationCoupledOscillators1997}%
  \BibitemOpen
  \bibfield  {author} {\bibinfo {author} {\bibfnamefont {H.-A.}\ \bibnamefont {Tanaka}}, \bibinfo {author} {\bibfnamefont {A.~J.}\ \bibnamefont {Lichtenberg}}, \ and\ \bibinfo {author} {\bibfnamefont {S.}~\bibnamefont {Oishi}},\ }\href {\doibase 10.1016/S0167-2789(96)00193-5} {\bibfield  {journal} {\bibinfo  {journal} {Physica D: Nonlinear Phenomena}\ }\textbf {\bibinfo {volume} {100}},\ \bibinfo {pages} {279} (\bibinfo {year} {1997}{\natexlab{b}})}\BibitemShut {NoStop}%
\bibitem [{\citenamefont {Bergen}\ and\ \citenamefont {Hill}(1981)}]{BergenHill}%
  \BibitemOpen
  \bibfield  {author} {\bibinfo {author} {\bibfnamefont {A.}~\bibnamefont {Bergen}}\ and\ \bibinfo {author} {\bibfnamefont {D.}~\bibnamefont {Hill}},\ }\href {\doibase 10.1109/TPAS.1981.316883} {\bibfield  {journal} {\bibinfo  {journal} {IEEE Transactions on Power Apparatus and Systems}\ }\textbf {\bibinfo {volume} {PAS-100}},\ \bibinfo {pages} {25} (\bibinfo {year} {1981})}\BibitemShut {NoStop}%
\bibitem [{\citenamefont {Witthaut}\ \emph {et~al.}(2022)\citenamefont {Witthaut}, \citenamefont {Hellmann}, \citenamefont {Kurths}, \citenamefont {Kettemann}, \citenamefont {Meyer-Ortmanns},\ and\ \citenamefont {Timme}}]{Review}%
  \BibitemOpen
  \bibfield  {author} {\bibinfo {author} {\bibfnamefont {D.}~\bibnamefont {Witthaut}}, \bibinfo {author} {\bibfnamefont {F.}~\bibnamefont {Hellmann}}, \bibinfo {author} {\bibfnamefont {J.}~\bibnamefont {Kurths}}, \bibinfo {author} {\bibfnamefont {S.}~\bibnamefont {Kettemann}}, \bibinfo {author} {\bibfnamefont {H.}~\bibnamefont {Meyer-Ortmanns}}, \ and\ \bibinfo {author} {\bibfnamefont {M.}~\bibnamefont {Timme}},\ }\href@noop {} {\bibfield  {journal} {\bibinfo  {journal} {Rev. Mod. Phys.}\ }\textbf {\bibinfo {volume} {94}} (\bibinfo {year} {2022})}\BibitemShut {NoStop}%
\bibitem [{\citenamefont {Ermentrout}(1991)}]{ermentroutAdaptiveModelSynchrony1991}%
  \BibitemOpen
  \bibfield  {author} {\bibinfo {author} {\bibfnamefont {B.}~\bibnamefont {Ermentrout}},\ }\href {\doibase 10.1007/BF00164052} {\bibfield  {journal} {\bibinfo  {journal} {Journal of Mathematical Biology}\ }\textbf {\bibinfo {volume} {29}},\ \bibinfo {pages} {571} (\bibinfo {year} {1991})}\BibitemShut {NoStop}%
\bibitem [{\citenamefont {Omelchenko}\ \emph {et~al.}(2011)\citenamefont {Omelchenko}, \citenamefont {Maistrenko}, \citenamefont {H{\"o}vel},\ and\ \citenamefont {Sch{\"o}ll}}]{Omelchenko2011-ql}%
  \BibitemOpen
  \bibfield  {author} {\bibinfo {author} {\bibfnamefont {I.}~\bibnamefont {Omelchenko}}, \bibinfo {author} {\bibfnamefont {Y.}~\bibnamefont {Maistrenko}}, \bibinfo {author} {\bibfnamefont {P.}~\bibnamefont {H{\"o}vel}}, \ and\ \bibinfo {author} {\bibfnamefont {E.}~\bibnamefont {Sch{\"o}ll}},\ }\href@noop {} {\bibfield  {journal} {\bibinfo  {journal} {Phys. Rev. Lett.}\ }\textbf {\bibinfo {volume} {106}},\ \bibinfo {pages} {234102} (\bibinfo {year} {2011})}\BibitemShut {NoStop}%
\bibitem [{\citenamefont {Gelbrecht}\ \emph {et~al.}(2020)\citenamefont {Gelbrecht}, \citenamefont {Kurths},\ and\ \citenamefont {Hellmann}}]{Gelbrecht_2020}%
  \BibitemOpen
  \bibfield  {author} {\bibinfo {author} {\bibfnamefont {M.}~\bibnamefont {Gelbrecht}}, \bibinfo {author} {\bibfnamefont {J.}~\bibnamefont {Kurths}}, \ and\ \bibinfo {author} {\bibfnamefont {F.}~\bibnamefont {Hellmann}},\ }\href {\doibase 10.1088/1367-2630/ab7a05} {\bibfield  {journal} {\bibinfo  {journal} {New Journal of Physics}\ }\textbf {\bibinfo {volume} {22}},\ \bibinfo {pages} {033032} (\bibinfo {year} {2020})}\BibitemShut {NoStop}%
\bibitem [{\citenamefont {Halekotte}\ \emph {et~al.}(2021)\citenamefont {Halekotte}, \citenamefont {Vanselow},\ and\ \citenamefont {Feudel}}]{halekotteTransientChaosEnforces2021}%
  \BibitemOpen
  \bibfield  {author} {\bibinfo {author} {\bibfnamefont {L.}~\bibnamefont {Halekotte}}, \bibinfo {author} {\bibfnamefont {A.}~\bibnamefont {Vanselow}}, \ and\ \bibinfo {author} {\bibfnamefont {U.}~\bibnamefont {Feudel}},\ }\href {\doibase 10.1088/2632-072X/ac080f} {\bibfield  {journal} {\bibinfo  {journal} {Journal of Physics: Complexity}\ }\textbf {\bibinfo {volume} {2}},\ \bibinfo {pages} {035015} (\bibinfo {year} {2021})}\BibitemShut {NoStop}%
\bibitem [{\citenamefont {Berner}\ \emph {et~al.}(2021)\citenamefont {Berner}, \citenamefont {Yanchuk},\ and\ \citenamefont {Sch\"oll}}]{PhysRevE.103.042315}%
  \BibitemOpen
  \bibfield  {author} {\bibinfo {author} {\bibfnamefont {R.}~\bibnamefont {Berner}}, \bibinfo {author} {\bibfnamefont {S.}~\bibnamefont {Yanchuk}}, \ and\ \bibinfo {author} {\bibfnamefont {E.}~\bibnamefont {Sch\"oll}},\ }\href {\doibase 10.1103/PhysRevE.103.042315} {\bibfield  {journal} {\bibinfo  {journal} {Phys. Rev. E}\ }\textbf {\bibinfo {volume} {103}},\ \bibinfo {pages} {042315} (\bibinfo {year} {2021})}\BibitemShut {NoStop}%
\bibitem [{\citenamefont {Olmi}\ \emph {et~al.}(2014)\citenamefont {Olmi}, \citenamefont {Navas}, \citenamefont {Boccaletti},\ and\ \citenamefont {Torcini}}]{olmiHystereticTransitionsKuramoto2014}%
  \BibitemOpen
  \bibfield  {author} {\bibinfo {author} {\bibfnamefont {S.}~\bibnamefont {Olmi}}, \bibinfo {author} {\bibfnamefont {A.}~\bibnamefont {Navas}}, \bibinfo {author} {\bibfnamefont {S.}~\bibnamefont {Boccaletti}}, \ and\ \bibinfo {author} {\bibfnamefont {A.}~\bibnamefont {Torcini}},\ }\href {\doibase 10.1103/PhysRevE.90.042905} {\bibfield  {journal} {\bibinfo  {journal} {Physical Review E}\ }\textbf {\bibinfo {volume} {90}},\ \bibinfo {pages} {042905} (\bibinfo {year} {2014})}\BibitemShut {NoStop}%
\bibitem [{\citenamefont {Maistrenko}\ \emph {et~al.}(2014)\citenamefont {Maistrenko}, \citenamefont {Penkovsky},\ and\ \citenamefont {Rosenblum}}]{MaistrenkoSolitary14}%
  \BibitemOpen
  \bibfield  {author} {\bibinfo {author} {\bibfnamefont {Y.}~\bibnamefont {Maistrenko}}, \bibinfo {author} {\bibfnamefont {B.}~\bibnamefont {Penkovsky}}, \ and\ \bibinfo {author} {\bibfnamefont {M.}~\bibnamefont {Rosenblum}},\ }\href {\doibase 10.1103/PhysRevE.89.060901} {\bibfield  {journal} {\bibinfo  {journal} {Phys. Rev. E}\ }\textbf {\bibinfo {volume} {89}},\ \bibinfo {pages} {060901} (\bibinfo {year} {2014})}\BibitemShut {NoStop}%
\bibitem [{\citenamefont {Jaros}\ \emph {et~al.}(2015)\citenamefont {Jaros}, \citenamefont {Maistrenko},\ and\ \citenamefont {Kapitaniak}}]{ChimeraJaros15}%
  \BibitemOpen
  \bibfield  {author} {\bibinfo {author} {\bibfnamefont {P.}~\bibnamefont {Jaros}}, \bibinfo {author} {\bibfnamefont {Y.}~\bibnamefont {Maistrenko}}, \ and\ \bibinfo {author} {\bibfnamefont {T.}~\bibnamefont {Kapitaniak}},\ }\href {\doibase 10.1103/PhysRevE.91.022907} {\bibfield  {journal} {\bibinfo  {journal} {Phys. Rev. E}\ }\textbf {\bibinfo {volume} {91}},\ \bibinfo {pages} {022907} (\bibinfo {year} {2015})}\BibitemShut {NoStop}%
\bibitem [{\citenamefont {Menck}\ \emph {et~al.}(2014)\citenamefont {Menck}, \citenamefont {Heitzig}, \citenamefont {Kurths},\ and\ \citenamefont {Schellnhuber}}]{Menck2014NatureComm}%
  \BibitemOpen
  \bibfield  {author} {\bibinfo {author} {\bibfnamefont {P.~J.}\ \bibnamefont {Menck}}, \bibinfo {author} {\bibfnamefont {J.}~\bibnamefont {Heitzig}}, \bibinfo {author} {\bibfnamefont {J.}~\bibnamefont {Kurths}}, \ and\ \bibinfo {author} {\bibfnamefont {H.}~\bibnamefont {Schellnhuber}},\ }\href {https://doi.org/10.1038/ncomms4969} {\bibfield  {journal} {\bibinfo  {journal} {Nature Communications}\ } (\bibinfo {year} {2014})}\BibitemShut {NoStop}%
\bibitem [{\citenamefont {Hellmann}\ \emph {et~al.}(2020)\citenamefont {Hellmann}, \citenamefont {Schultz}, \citenamefont {Jaros}, \citenamefont {Levchenko}, \citenamefont {Kapitaniak}, \citenamefont {Kurths},\ and\ \citenamefont {Maistrenko}}]{Hellmann2020-aj}%
  \BibitemOpen
  \bibfield  {author} {\bibinfo {author} {\bibfnamefont {F.}~\bibnamefont {Hellmann}}, \bibinfo {author} {\bibfnamefont {P.}~\bibnamefont {Schultz}}, \bibinfo {author} {\bibfnamefont {P.}~\bibnamefont {Jaros}}, \bibinfo {author} {\bibfnamefont {R.}~\bibnamefont {Levchenko}}, \bibinfo {author} {\bibfnamefont {T.}~\bibnamefont {Kapitaniak}}, \bibinfo {author} {\bibfnamefont {J.}~\bibnamefont {Kurths}}, \ and\ \bibinfo {author} {\bibfnamefont {Y.}~\bibnamefont {Maistrenko}},\ }\href@noop {} {\bibfield  {journal} {\bibinfo  {journal} {Nat. Commun.}\ }\textbf {\bibinfo {volume} {11}},\ \bibinfo {pages} {592} (\bibinfo {year} {2020})}\BibitemShut {NoStop}%
\bibitem [{\citenamefont {Jaros}\ \emph {et~al.}(2018)\citenamefont {Jaros}, \citenamefont {Brezetsky}, \citenamefont {Levchenko}, \citenamefont {Dudkowski}, \citenamefont {Kapitaniak},\ and\ \citenamefont {Maistrenko}}]{Jaros2018-mu}%
  \BibitemOpen
  \bibfield  {author} {\bibinfo {author} {\bibfnamefont {P.}~\bibnamefont {Jaros}}, \bibinfo {author} {\bibfnamefont {S.}~\bibnamefont {Brezetsky}}, \bibinfo {author} {\bibfnamefont {R.}~\bibnamefont {Levchenko}}, \bibinfo {author} {\bibfnamefont {D.}~\bibnamefont {Dudkowski}}, \bibinfo {author} {\bibfnamefont {T.}~\bibnamefont {Kapitaniak}}, \ and\ \bibinfo {author} {\bibfnamefont {Y.}~\bibnamefont {Maistrenko}},\ }\href@noop {} {\bibfield  {journal} {\bibinfo  {journal} {Chaos}\ }\textbf {\bibinfo {volume} {28}},\ \bibinfo {pages} {011103} (\bibinfo {year} {2018})}\BibitemShut {NoStop}%
\bibitem [{\citenamefont {Nitzbon}\ \emph {et~al.}(2017)\citenamefont {Nitzbon}, \citenamefont {Schultz}, \citenamefont {Heitzig}, \citenamefont {Kurths},\ and\ \citenamefont {Hellmann}}]{Nitzbon_2017}%
  \BibitemOpen
  \bibfield  {author} {\bibinfo {author} {\bibfnamefont {J.}~\bibnamefont {Nitzbon}}, \bibinfo {author} {\bibfnamefont {P.}~\bibnamefont {Schultz}}, \bibinfo {author} {\bibfnamefont {J.}~\bibnamefont {Heitzig}}, \bibinfo {author} {\bibfnamefont {J.}~\bibnamefont {Kurths}}, \ and\ \bibinfo {author} {\bibfnamefont {F.}~\bibnamefont {Hellmann}},\ }\href {\doibase 10.1088/1367-2630/aa6321} {\bibfield  {journal} {\bibinfo  {journal} {New Journal of Physics}\ }\textbf {\bibinfo {volume} {19}},\ \bibinfo {pages} {033029} (\bibinfo {year} {2017})}\BibitemShut {NoStop}%
\bibitem [{\citenamefont {UCTE}()}]{UCTE_split2006}%
  \BibitemOpen
  \bibfield  {author} {\bibinfo {author} {\bibnamefont {UCTE}},\ }\href@noop {} {\enquote {\bibinfo {title} {Final report system disturbance on 4 november 2006},}\ }\bibinfo {howpublished} {\url{https://eepublicdownloads.entsoe.eu/clean-documents/pre2015/publications/ce/otherreports/Final-Report-20070130.pdf}},\ \bibinfo {note} {accessed: 2022-12-19 17:25}\BibitemShut {NoStop}%
\bibitem [{\citenamefont {Rohden}\ \emph {et~al.}(2012)\citenamefont {Rohden}, \citenamefont {Sorge}, \citenamefont {Timme},\ and\ \citenamefont {Witthaut}}]{Rohden2012-iy}%
  \BibitemOpen
  \bibfield  {author} {\bibinfo {author} {\bibfnamefont {M.}~\bibnamefont {Rohden}}, \bibinfo {author} {\bibfnamefont {A.}~\bibnamefont {Sorge}}, \bibinfo {author} {\bibfnamefont {M.}~\bibnamefont {Timme}}, \ and\ \bibinfo {author} {\bibfnamefont {D.}~\bibnamefont {Witthaut}},\ }\href@noop {} {\bibfield  {journal} {\bibinfo  {journal} {Phys. Rev. Lett.}\ }\textbf {\bibinfo {volume} {109}},\ \bibinfo {pages} {064101} (\bibinfo {year} {2012})}\BibitemShut {NoStop}%
\bibitem [{\citenamefont {Zhang}\ \emph {et~al.}(2019)\citenamefont {Zhang}, \citenamefont {Hallerberg}, \citenamefont {Matthiae}, \citenamefont {Witthaut},\ and\ \citenamefont {Timme}}]{Xiaozhu19}%
  \BibitemOpen
  \bibfield  {author} {\bibinfo {author} {\bibfnamefont {X.}~\bibnamefont {Zhang}}, \bibinfo {author} {\bibfnamefont {S.}~\bibnamefont {Hallerberg}}, \bibinfo {author} {\bibfnamefont {M.}~\bibnamefont {Matthiae}}, \bibinfo {author} {\bibfnamefont {D.}~\bibnamefont {Witthaut}}, \ and\ \bibinfo {author} {\bibfnamefont {M.}~\bibnamefont {Timme}},\ }\href {\doibase 10.1126/sciadv.aav1027} {\bibfield  {journal} {\bibinfo  {journal} {Science Advances}\ }\textbf {\bibinfo {volume} {5}},\ \bibinfo {pages} {eaav1027} (\bibinfo {year} {2019})}\BibitemShut {NoStop}%
\bibitem [{\citenamefont {Menck}(2014)}]{Menck2014Diss}%
  \BibitemOpen
  \bibfield  {author} {\bibinfo {author} {\bibfnamefont {P.-J.}\ \bibnamefont {Menck}},\ }\emph {\bibinfo {title} {How wires shape volumes}},\ \href {\doibase http://dx.doi.org/10.18452/16930} {Ph.D. thesis},\ \bibinfo  {school} {Humboldt-Universität zu Berlin, Mathematisch-Naturwissenschaftliche Fakultät I} (\bibinfo {year} {2014})\BibitemShut {NoStop}%
\bibitem [{\citenamefont {Guckenheimer}\ and\ \citenamefont {Holmes}(2002)}]{Guckenheimer2002-rm}%
  \BibitemOpen
  \bibfield  {author} {\bibinfo {author} {\bibfnamefont {J.}~\bibnamefont {Guckenheimer}}\ and\ \bibinfo {author} {\bibfnamefont {P.}~\bibnamefont {Holmes}},\ }\href@noop {} {\emph {\bibinfo {title} {Nonlinear oscillations, dynamical systems, and bifurcations of vector fields}}},\ \bibinfo {edition} {1st}\ ed.,\ Applied mathematical sciences\ (\bibinfo  {publisher} {Springer},\ \bibinfo {address} {New York, NY},\ \bibinfo {year} {2002})\BibitemShut {NoStop}%
\bibitem [{\citenamefont {Gao}\ and\ \citenamefont {Efstathiou}(2018)}]{Self-consistent_method}%
  \BibitemOpen
  \bibfield  {author} {\bibinfo {author} {\bibfnamefont {J.}~\bibnamefont {Gao}}\ and\ \bibinfo {author} {\bibfnamefont {K.}~\bibnamefont {Efstathiou}},\ }\href {\doibase 10.1103/PhysRevE.98.042201} {\bibfield  {journal} {\bibinfo  {journal} {Phys. Rev. E}\ }\textbf {\bibinfo {volume} {98}},\ \bibinfo {pages} {042201} (\bibinfo {year} {2018})}\BibitemShut {NoStop}%
\bibitem [{\citenamefont {Yue}\ \emph {et~al.}(2020)\citenamefont {Yue}, \citenamefont {Smith},\ and\ \citenamefont {Gottwald}}]{yue2020model}%
  \BibitemOpen
  \bibfield  {author} {\bibinfo {author} {\bibfnamefont {W.}~\bibnamefont {Yue}}, \bibinfo {author} {\bibfnamefont {L.~D.}\ \bibnamefont {Smith}}, \ and\ \bibinfo {author} {\bibfnamefont {G.~A.}\ \bibnamefont {Gottwald}},\ }\href@noop {} {\bibfield  {journal} {\bibinfo  {journal} {Physical Review E}\ }\textbf {\bibinfo {volume} {101}},\ \bibinfo {pages} {062213} (\bibinfo {year} {2020})}\BibitemShut {NoStop}%
\bibitem [{\citenamefont {Munyayev}\ \emph {et~al.}(2022)\citenamefont {Munyayev}, \citenamefont {Bolotov}, \citenamefont {Smirnov}, \citenamefont {Osipov},\ and\ \citenamefont {Belykh}}]{StabilityRolSolInertia}%
  \BibitemOpen
  \bibfield  {author} {\bibinfo {author} {\bibfnamefont {V.~O.}\ \bibnamefont {Munyayev}}, \bibinfo {author} {\bibfnamefont {M.~I.}\ \bibnamefont {Bolotov}}, \bibinfo {author} {\bibfnamefont {L.~A.}\ \bibnamefont {Smirnov}}, \bibinfo {author} {\bibfnamefont {G.~V.}\ \bibnamefont {Osipov}}, \ and\ \bibinfo {author} {\bibfnamefont {I.~V.}\ \bibnamefont {Belykh}},\ }\href {\doibase 10.1103/PhysRevE.105.024203} {\bibfield  {journal} {\bibinfo  {journal} {Phys. Rev. E}\ }\textbf {\bibinfo {volume} {105}},\ \bibinfo {pages} {024203} (\bibinfo {year} {2022})}\BibitemShut {NoStop}%
\bibitem [{Sup()}]{Supplement}%
  \BibitemOpen
  \href@noop {} {}\bibinfo {note} {See Supplemental Material (attached) for animations of the dynamics of an exemplary complex network; an in-depth, self-contained step-by-step derivation of our main results; more numerical studies; an outline of several possible generalizations; more examples and their comparison; and details of the networks used and numerical experiments.}\BibitemShut {Stop}%
\bibitem [{\citenamefont {Halekotte}\ and\ \citenamefont {Feudel}(2020)}]{halekotteMinimalFatalShocks2020}%
  \BibitemOpen
  \bibfield  {author} {\bibinfo {author} {\bibfnamefont {L.}~\bibnamefont {Halekotte}}\ and\ \bibinfo {author} {\bibfnamefont {U.}~\bibnamefont {Feudel}},\ }\href {\doibase 10.1038/s41598-020-68805-6} {\bibfield  {journal} {\bibinfo  {journal} {Scientific Reports}\ }\textbf {\bibinfo {volume} {10}},\ \bibinfo {pages} {11783} (\bibinfo {year} {2020})}\BibitemShut {NoStop}%
\bibitem [{\citenamefont {Manik}\ \emph {et~al.}(2014)\citenamefont {Manik}, \citenamefont {Witthaut}, \citenamefont {Sch{\"a}fer}, \citenamefont {Matthiae}, \citenamefont {Sorge}, \citenamefont {Rohden}, \citenamefont {Katifori},\ and\ \citenamefont {Timme}}]{Manik2014-ye}%
  \BibitemOpen
  \bibfield  {author} {\bibinfo {author} {\bibfnamefont {D.}~\bibnamefont {Manik}}, \bibinfo {author} {\bibfnamefont {D.}~\bibnamefont {Witthaut}}, \bibinfo {author} {\bibfnamefont {B.}~\bibnamefont {Sch{\"a}fer}}, \bibinfo {author} {\bibfnamefont {M.}~\bibnamefont {Matthiae}}, \bibinfo {author} {\bibfnamefont {A.}~\bibnamefont {Sorge}}, \bibinfo {author} {\bibfnamefont {M.}~\bibnamefont {Rohden}}, \bibinfo {author} {\bibfnamefont {E.}~\bibnamefont {Katifori}}, \ and\ \bibinfo {author} {\bibfnamefont {M.}~\bibnamefont {Timme}},\ }\href@noop {} {\bibfield  {journal} {\bibinfo  {journal} {Eur. Phys. J. Spec. Top.}\ }\textbf {\bibinfo {volume} {223}},\ \bibinfo {pages} {2527} (\bibinfo {year} {2014})}\BibitemShut {NoStop}%
\bibitem [{\citenamefont {Schultz}\ \emph {et~al.}(2014)\citenamefont {Schultz}, \citenamefont {Heitzig},\ and\ \citenamefont {Kurths}}]{RandomGrowthModel}%
  \BibitemOpen
  \bibfield  {author} {\bibinfo {author} {\bibfnamefont {P.}~\bibnamefont {Schultz}}, \bibinfo {author} {\bibfnamefont {J.}~\bibnamefont {Heitzig}}, \ and\ \bibinfo {author} {\bibfnamefont {J.}~\bibnamefont {Kurths}},\ }\href {https://doi.org/10.1140/epjst/e2014-02279-6} {\bibfield  {journal} {\bibinfo  {journal} {The European Physical Journal Special Topics}\ } (\bibinfo {year} {2014})}\BibitemShut {NoStop}%
\bibitem [{\citenamefont {Niehues}(2023)}]{GitRepoReVelatioNs}%
  \BibitemOpen
  \bibfield  {author} {\bibinfo {author} {\bibfnamefont {J.}~\bibnamefont {Niehues}},\ }\href@noop {} {\enquote {\bibinfo {title} {Revelations.jl},}\ }\bibinfo {howpublished} {\url{https://doi.org/10.5281/zenodo.12636090}} (\bibinfo {year} {2023})\BibitemShut {NoStop}%
\bibitem [{\citenamefont {Hata}\ and\ \citenamefont {Nakao}(2017)}]{Hata2017-jl}%
  \BibitemOpen
  \bibfield  {author} {\bibinfo {author} {\bibfnamefont {S.}~\bibnamefont {Hata}}\ and\ \bibinfo {author} {\bibfnamefont {H.}~\bibnamefont {Nakao}},\ }\href@noop {} {\bibfield  {journal} {\bibinfo  {journal} {Sci. Rep.}\ }\textbf {\bibinfo {volume} {7}},\ \bibinfo {pages} {1121} (\bibinfo {year} {2017})}\BibitemShut {NoStop}%
\bibitem [{\citenamefont {Plietzsch}\ \emph {et~al.}(2022)\citenamefont {Plietzsch}, \citenamefont {Kogler}, \citenamefont {Auer}, \citenamefont {Merino}, \citenamefont {Gil-de Muro}, \citenamefont {Li{\ss}e}, \citenamefont {Vogel},\ and\ \citenamefont {Hellmann}}]{PowerDynamics2022}%
  \BibitemOpen
  \bibfield  {author} {\bibinfo {author} {\bibfnamefont {A.}~\bibnamefont {Plietzsch}}, \bibinfo {author} {\bibfnamefont {R.}~\bibnamefont {Kogler}}, \bibinfo {author} {\bibfnamefont {S.}~\bibnamefont {Auer}}, \bibinfo {author} {\bibfnamefont {J.}~\bibnamefont {Merino}}, \bibinfo {author} {\bibfnamefont {A.}~\bibnamefont {Gil-de Muro}}, \bibinfo {author} {\bibfnamefont {J.}~\bibnamefont {Li{\ss}e}}, \bibinfo {author} {\bibfnamefont {C.}~\bibnamefont {Vogel}}, \ and\ \bibinfo {author} {\bibfnamefont {F.}~\bibnamefont {Hellmann}},\ }\href@noop {} {\bibfield  {journal} {\bibinfo  {journal} {SoftwareX}\ }\textbf {\bibinfo {volume} {17}},\ \bibinfo {pages} {100861} (\bibinfo {year} {2022})}\BibitemShut {NoStop}%
\bibitem [{\citenamefont {Motter}\ \emph {et~al.}(2013)\citenamefont {Motter}, \citenamefont {Myers}, \citenamefont {Anghel},\ and\ \citenamefont {Nishikawa}}]{Motter2013-yo}%
  \BibitemOpen
  \bibfield  {author} {\bibinfo {author} {\bibfnamefont {A.~E.}\ \bibnamefont {Motter}}, \bibinfo {author} {\bibfnamefont {S.~A.}\ \bibnamefont {Myers}}, \bibinfo {author} {\bibfnamefont {M.}~\bibnamefont {Anghel}}, \ and\ \bibinfo {author} {\bibfnamefont {T.}~\bibnamefont {Nishikawa}},\ }\href@noop {} {\bibfield  {journal} {\bibinfo  {journal} {Nat. Phys.}\ }\textbf {\bibinfo {volume} {9}},\ \bibinfo {pages} {191} (\bibinfo {year} {2013})}\BibitemShut {NoStop}%
\bibitem [{\citenamefont {Wassmer}\ \emph {et~al.}(2021)\citenamefont {Wassmer}, \citenamefont {Witthaut},\ and\ \citenamefont {Kaiser}}]{wassmerTargetedSuppressionFailure2021}%
  \BibitemOpen
  \bibfield  {author} {\bibinfo {author} {\bibfnamefont {J.}~\bibnamefont {Wassmer}}, \bibinfo {author} {\bibfnamefont {D.}~\bibnamefont {Witthaut}}, \ and\ \bibinfo {author} {\bibfnamefont {F.}~\bibnamefont {Kaiser}},\ }\href {\doibase 10.1088/2632-072X/abf090} {\bibfield  {journal} {\bibinfo  {journal} {Journal of Physics: Complexity}\ }\textbf {\bibinfo {volume} {2}},\ \bibinfo {pages} {035003} (\bibinfo {year} {2021})}\BibitemShut {NoStop}%
\bibitem [{\citenamefont {Menara}\ \emph {et~al.}(2022)\citenamefont {Menara}, \citenamefont {Baggio}, \citenamefont {Bassett},\ and\ \citenamefont {Pasqualetti}}]{Menara2022-bg}%
  \BibitemOpen
  \bibfield  {author} {\bibinfo {author} {\bibfnamefont {T.}~\bibnamefont {Menara}}, \bibinfo {author} {\bibfnamefont {G.}~\bibnamefont {Baggio}}, \bibinfo {author} {\bibfnamefont {D.}~\bibnamefont {Bassett}}, \ and\ \bibinfo {author} {\bibfnamefont {F.}~\bibnamefont {Pasqualetti}},\ }\href@noop {} {\bibfield  {journal} {\bibinfo  {journal} {Nat. Commun.}\ }\textbf {\bibinfo {volume} {13}},\ \bibinfo {pages} {4721} (\bibinfo {year} {2022})}\BibitemShut {NoStop}%
\bibitem [{\citenamefont {Witthaut}\ and\ \citenamefont {Timme}(2012)}]{Witthaut_2012}%
  \BibitemOpen
  \bibfield  {author} {\bibinfo {author} {\bibfnamefont {D.}~\bibnamefont {Witthaut}}\ and\ \bibinfo {author} {\bibfnamefont {M.}~\bibnamefont {Timme}},\ }\href {\doibase 10.1088/1367-2630/14/8/083036} {\bibfield  {journal} {\bibinfo  {journal} {New Journal of Physics}\ }\textbf {\bibinfo {volume} {14}},\ \bibinfo {pages} {083036} (\bibinfo {year} {2012})}\BibitemShut {NoStop}%
\bibitem [{\citenamefont {Schultz}\ \emph {et~al.}(2016)\citenamefont {Schultz}, \citenamefont {Peron}, \citenamefont {Eroglu}, \citenamefont {Stemler}, \citenamefont {Ram\'{\i}rez~\'Avila}, \citenamefont {Rodrigues},\ and\ \citenamefont {Kurths}}]{Schultz16Tweak}%
  \BibitemOpen
  \bibfield  {author} {\bibinfo {author} {\bibfnamefont {P.}~\bibnamefont {Schultz}}, \bibinfo {author} {\bibfnamefont {T.}~\bibnamefont {Peron}}, \bibinfo {author} {\bibfnamefont {D.}~\bibnamefont {Eroglu}}, \bibinfo {author} {\bibfnamefont {T.}~\bibnamefont {Stemler}}, \bibinfo {author} {\bibfnamefont {G.~M.}\ \bibnamefont {Ram\'{\i}rez~\'Avila}}, \bibinfo {author} {\bibfnamefont {F.~A.}\ \bibnamefont {Rodrigues}}, \ and\ \bibinfo {author} {\bibfnamefont {J.}~\bibnamefont {Kurths}},\ }\href {\doibase 10.1103/PhysRevE.93.062211} {\bibfield  {journal} {\bibinfo  {journal} {Phys. Rev. E}\ }\textbf {\bibinfo {volume} {93}},\ \bibinfo {pages} {062211} (\bibinfo {year} {2016})}\BibitemShut {NoStop}%
\bibitem [{\citenamefont {Kaiser}\ \emph {et~al.}(2021)\citenamefont {Kaiser}, \citenamefont {Latora},\ and\ \citenamefont {Witthaut}}]{kaiserNetworkIsolatorsInhibit2021}%
  \BibitemOpen
  \bibfield  {author} {\bibinfo {author} {\bibfnamefont {F.}~\bibnamefont {Kaiser}}, \bibinfo {author} {\bibfnamefont {V.}~\bibnamefont {Latora}}, \ and\ \bibinfo {author} {\bibfnamefont {D.}~\bibnamefont {Witthaut}},\ }\href {\doibase 10.1038/s41467-021-23292-9} {\bibfield  {journal} {\bibinfo  {journal} {Nature Communications}\ }\textbf {\bibinfo {volume} {12}},\ \bibinfo {pages} {3143} (\bibinfo {year} {2021})}\BibitemShut {NoStop}%
\bibitem [{\citenamefont {Taher}\ \emph {et~al.}(2019)\citenamefont {Taher}, \citenamefont {Olmi},\ and\ \citenamefont {Sch\"oll}}]{Olmi_Delayed_Feedback}%
  \BibitemOpen
  \bibfield  {author} {\bibinfo {author} {\bibfnamefont {H.}~\bibnamefont {Taher}}, \bibinfo {author} {\bibfnamefont {S.}~\bibnamefont {Olmi}}, \ and\ \bibinfo {author} {\bibfnamefont {E.}~\bibnamefont {Sch\"oll}},\ }\href {\doibase 10.1103/PhysRevE.100.062306} {\bibfield  {journal} {\bibinfo  {journal} {Phys. Rev. E}\ }\textbf {\bibinfo {volume} {100}},\ \bibinfo {pages} {062306} (\bibinfo {year} {2019})}\BibitemShut {NoStop}%
\bibitem [{\citenamefont {Klinshov}\ \emph {et~al.}(2015)\citenamefont {Klinshov}, \citenamefont {Nekorkin},\ and\ \citenamefont {Kurths}}]{klinshovStabilityThresholdApproach2015}%
  \BibitemOpen
  \bibfield  {author} {\bibinfo {author} {\bibfnamefont {V.~V.}\ \bibnamefont {Klinshov}}, \bibinfo {author} {\bibfnamefont {V.~I.}\ \bibnamefont {Nekorkin}}, \ and\ \bibinfo {author} {\bibfnamefont {J.}~\bibnamefont {Kurths}},\ }\href {\doibase 10.1088/1367-2630/18/1/013004} {\bibfield  {journal} {\bibinfo  {journal} {New Journal of Physics}\ }\textbf {\bibinfo {volume} {18}},\ \bibinfo {pages} {013004} (\bibinfo {year} {2015})}\BibitemShut {NoStop}%
\bibitem [{\citenamefont {Kogler}\ \emph {et~al.}(2022)\citenamefont {Kogler}, \citenamefont {Plietzsch}, \citenamefont {Schultz},\ and\ \citenamefont {Hellmann}}]{PRXEnergy.1.013008}%
  \BibitemOpen
  \bibfield  {author} {\bibinfo {author} {\bibfnamefont {R.}~\bibnamefont {Kogler}}, \bibinfo {author} {\bibfnamefont {A.}~\bibnamefont {Plietzsch}}, \bibinfo {author} {\bibfnamefont {P.}~\bibnamefont {Schultz}}, \ and\ \bibinfo {author} {\bibfnamefont {F.}~\bibnamefont {Hellmann}},\ }\href {\doibase 10.1103/PRXEnergy.1.013008} {\bibfield  {journal} {\bibinfo  {journal} {PRX Energy}\ }\textbf {\bibinfo {volume} {1}},\ \bibinfo {pages} {013008} (\bibinfo {year} {2022})}\BibitemShut {NoStop}%
\end{thebibliography}%

\end{document}

\title{Supplementary Material to:\\
Resonant Solitary States in Complex Networks}

\author{Jakob Niehues}
\affiliation{Corresponding author: jakob.niehues@pik-potsdam.de}
\affiliation{Potsdam Institute for Climate Impact Research, Telegrafenberg A56, 14473 Potsdam, Germany}
\affiliation{Humboldt-Universit\"at zu Berlin, Department of Physics, Newtonstra\ss e 15, 12489 Berlin, Germany}
\affiliation{Technische Universit\"at Berlin, ER 3-2, Hardenbergstrasse 36a, 10623 Berlin, Germany}
\author{S. Yanchuk}
\affiliation{Potsdam Institute for Climate Impact Research, Telegrafenberg A56, 14473 Potsdam, Germany}
\affiliation{University College Cork, School of Mathematical Sciences, Western Road, Cork, T12 XF62, Ireland}
\author{R. Berner}
\affiliation{Humboldt-Universit\"at zu Berlin, Department of Physics, Newtonstra\ss e 15, 12489 Berlin, Germany}
\author{J. Kurths}
\affiliation{Potsdam Institute for Climate Impact Research, Telegrafenberg A56, 14473 Potsdam, Germany}
\affiliation{Humboldt-Universi\"at zu Berlin, Department of Physics, Newtonstra\ss e 15, 12489 Berlin, Germany}
\author{F. Hellmann}
\affiliation{Potsdam Institute for Climate Impact Research, Telegrafenberg A56, 14473 Potsdam, Germany}
\author{M. Anvari}
\affiliation{Potsdam Institute for Climate Impact Research, Telegrafenberg A56, 14473 Potsdam, Germany}
\affiliation{Fraunhofer Institute for Algorithms and Scientific Computing, 53757 Sankt Augustin, Germany}

\maketitle
This supplementary material contains a detailed derivation of the central self-consistent equation in the main article.
Moreover, we provide details on the examples studied.

\tableofcontents

\section{Derivation of the self-consistent equation}
\label{sec:derivation_of_the_self_consistency_equation}
\subsection{Preliminaries}
\label{sub:preliminaries}
We start with a network of $N$ Kuramoto oscillators with inertia (swing equations),
\begin{equation}
    m_i \ddot\varphi_i = P_i - \alpha_i \dot\varphi_i - \sum_{j=1}^N \kappa_{ij} \sin(\varphi_i - \varphi_j).
    \label{eq:swing_lossless}
\end{equation}
A necessary condition for the existence of a steady state is that the driving powers are balanced, that is, $\sum_{i=1}^N P_i = 0$. As long as the sum of damping coefficients is not zero, this can always be achieved by shifting to a rotating frame of reference with $\omega_\text{ref} = \frac{\sum_{i=1}^N P_i}{\sum_{i=1}^N \alpha_i}$. Without loss of generality, we will assume $\sum_{i=1}^N P_i = 0$ from here on.
Alternatively, Eq.~(\ref{eq:swing_lossless}) can be written as a set of $2 N$ first-order equations,
\begin{eqnarray}
    \dot\varphi_i &=& \omega_i,
    \label{eq:swing_lossless_phi}
    \\
    m_i \dot\omega_i &=& P_i - \alpha_i \omega_i - \sum_{j=1}^N \kappa_{ij} \sin(\varphi_i - \varphi_j).
    \label{eq:swing_lossless_omega}
\end{eqnarray}
The steady state $\varphi_i^*$, i.e., $\dot\varphi_i = 0$,  is given by the power flow equations,
\begin{equation}
    P_i =  \sum_{j=1}^N \kappa_{ij} \sin(\varphi_i^* - \varphi_j^*).
    \label{eq:power_flow_equations}
\end{equation}

\subsubsection{Network modes}
The steady state induces a weighted steady-state graph Laplacian
\begin{equation}
    \mathcal{L}_{ij}
    =
    \begin{cases}
        \begin{aligned}
            \sum_{j, j \neq i}^N \kappa_{ij} \cos\left(\varphi_i^* - \varphi_j^*\right) &\quad \text{iff $i = j$},\\
            - \kappa_{ij} \cos\left(\varphi_i^* - \varphi_j^*\right) &\quad \text{iff $i\neq j$.}
        \end{aligned}
    \end{cases}
    \label{eq:steady_state_Laplacian}
\end{equation}
It appears in the linearized dynamics with Jacobian $\mathcal{J}$,
\begin{equation}
    \frac{\D}{\D t}
    \begin{bmatrix}
        \mathds{1} & 0\\
        0 & \mathcal{M}
    \end{bmatrix}
    \begin{bmatrix}
        \delta \varphi\\
        \delta \omega
    \end{bmatrix}
    =
    \mathcal{J}
    \begin{bmatrix}
        \delta \varphi\\
        \delta \omega
    \end{bmatrix}
    =
    \begin{bmatrix}
        0 & \mathds{1}\\
        - \mathcal{L} & -\mathcal{A}
    \end{bmatrix}
    \begin{bmatrix}
        \delta \varphi\\
        \delta \omega
    \end{bmatrix}.
    \label{eq:linearized_dynamics}
\end{equation}
where $\delta\varphi$ is a vector with entries $\varphi_i - \varphi_i^*$, analogously $\delta\omega$ is a vector with entries $\omega_i - \omega_i^* = \omega_i$, and $\mathcal{M} \coloneqq \text{diag}(m_i)$, and $\mathcal{A} \coloneqq \text{diag}(\alpha_i)$.
$\mathcal{L}$ is positive semi-definite, see reference [43].
If the graph is connected, which we assume, $\mathcal{L}$ has exactly one zero eigenvalue.
The other eigenvalues are real and positive, and can be ordered $0 = \Lambda^{\left[1\right]} < \Lambda^{\left[2\right]} < \ldots < \Lambda^{\left[N\right]}$.
For $\mathcal{M} = \mathds{1}$ and $\mathcal{A} = \alpha \mathds{1}$,\footnote{Note that for homogeneous $m_i$ and $\alpha_i$, one of the two parameters can always be rescaled to 1 by rescaling time.} the eigenvalues of $\mathcal{J}$ are given by
\begin{equation}
    \sigma_\pm^{\left[\ell\right]} = -\frac{\alpha}{2} \pm \sqrt{\frac{\alpha^2}{4} - \Lambda^{\left[\ell\right]}},
    \label{eq:jacobian_eigenvalues}
\end{equation}
see reference [28].
Since their real parts are negative by assumption,\footnote{We only treat networks with linearly stable steady states.} except for one zero eigenvalue, the synchronous state is linearly stable as small perturbations are damped exponentially with time.
The imaginary parts of $\sigma_\pm^{\left[\ell\right]}$ are the \textit{eigenmodes} of the network, or \textit{network modes}, $\pm\omega^{\left[\ell\right]}$, that describe oscillations around the fixed point.
They are given by $\omega^{[1]}=0$ and
\begin{equation}
    \omega^{\left[\ell\right]} = \sqrt{\Lambda^{\left[\ell\right]} - \frac{\alpha^2}{4}}.
    \label{eq:network_modes}
\end{equation}
for $\ell = 2,\ldots,N$.
We assume that $\Lambda^{\left[\ell\right]} \gg \alpha^2/4$ for $\ell \geq \ell_\text{th}$, where $\ell_\text{th} \ge 2$ is some threshold.
Therefore, $\omega^{[\ell]} \approx \sqrt{\Lambda^{[\ell]} }$ for $\ell \geq \ell_\text{th}$.

The components $V_i^{[\ell]}$ of the corresponding eigenvectors $\vec V^{[\ell]}$ of $\mathcal{L}$ give us the nodal excitation of $\varphi_i$ of network mode $\ell$ at node location $i$.
The zero eigenmode $\omega^{[1]}=0$ corresponds to the uniform eigenvector $V_i^{[1]} = 1/\sqrt{N} \quad \forall i$.
It represents the symmetry of Eq.~(\ref{eq:swing_lossless}) under a constant global phase shift.
It does not describe an oscillation, but the uniform movement of all phases.
Therefore, it is also called \textit{bulk mode}.

\subsubsection{Eigenvector localization}
\label{subsub:eigenvector_localization}
Here, we provide an estimate for the eigenvalue of a highly concentrated steady-state Laplacian eigenvector.
It is a typical feature of the networks at hand and will become important in the derivation of the self-consistent equation.
Let us assume that a network mode is localized at node $i$, or as one could say, its Laplacian eigenvector $\vec V^{[\ell]}$ has a strong support there.
We quantify this by
\begin{equation}
    \left| \frac{V_j^{[\ell]}}{V_i^{[\ell]}} \right| < \varepsilon_i \quad \forall j \in \mathcal{N}_i,
\end{equation}
where $\varepsilon_i$ is a small parameter, and we denote the neighbors of node $i$ with $\mathcal{N}_i$.
When evaluated explicitly for component $i$, the eigenvector equation yields
\begin{equation}
    \Lambda^{[\ell]} V_i^{[\ell]}
    = \mathcal{L}_{ii} V_i^{[\ell]} + \sum_{j\in \mathcal{N}_i} \mathcal{L}_{ij} V_j^{[\ell]}
    = \mathcal{L}_{ii} V_i^{[\ell]} - \sum_{j\in \mathcal{N}_i} \kappa_{ij} \cos\left(\varphi_i^* - \varphi_j^*\right) V_j^{[\ell]}.
\end{equation}
This gives us tight bounds for the value of $\Lambda^{[\ell]}$,
\begin{equation}
    \left|\Lambda^{[\ell]} - \mathcal{L}_{ii}\right|
    = \left| \sum_{j\in \mathcal{N}_i} \kappa_{ij} \cos\left(\varphi_i^* - \varphi_j^*\right) \frac{V_j^{[\ell]}}{V_i^{[\ell]}} \right|
    \leq \mathcal{L}_{ii} \varepsilon_i.
\end{equation}
For $\cos\left(\varphi_i^* - \varphi_j^*\right) \approx 1$, this reduces to
\begin{equation}
    \left|\Lambda^{[\ell]} - K_i^* d_i \right|
    \leq  K_i^* d_i \varepsilon_i,
\end{equation}
where $K_i^*$ is the average coupling strength of the links $\{(i,j)|j\in\mathcal{N}_i\}$ with coupling strengths $\{\kappa_{ij}\}$, and $d_i = |\mathcal{N}_i|$ is the degree  of node $i$.
We conclude that a highly localized network mode comes with a Laplacian eigenvalue that is closely connected to the degree of the node where the network mode is localized.

\subsubsection{Partition of the system}
For the analysis of solitary states, we view the original system as two interacting subsystems, namely, the solitary oscillator $z$ and the rest, which we denote 
$S = \{1,\ldots,N\}\setminus z$, and which forms a synchronized cluster.
For now, we assume that the solitary oscillator has only one neighbor, $k$, without loss of generality.
As we show later on in section \ref{sub:several_solitary_nodes}, the results can be generalized to several neighbors in a straightforward way due to the additive nature of the coupling via the links from $z$ to $k\in\mathcal{N}_z$.

Node $z$ has a different mean frequency than the synchronized cluster, i.e.,
\begin{equation}
    \varphi_z(t) = \omega_s t + \varepsilon \psi_z(t) + \text{const.},
    \label{eq:ansatz_solitary_node}
\end{equation}
where $\omega_s$ is the time-averaged frequency of node $z$, $\omega_s \coloneqq \langle\dot \varphi_z \rangle$, $\varepsilon$ a small parameter, and $\langle \psi_z(t)\rangle = 0$ by assumption.
Note that separating the linear motion in time with $\omega_s t$ is without loss of generality.
A constant offset in $\varphi_z$ can be set to zero by time shifts.
We will later insert the ansatz in Eq.~(\ref{eq:ansatz_solitary_node}) into the dynamics to get an approximate solution, from which we can determine $\omega_s$.

For simplicity, we assume homogeneous inertia $m$ and damping constants $\alpha$ for the synchronized cluster $S$ for now.
We show how to generalize to heterogeneous parameters later on in section \ref{sub:heterogeneous_system_parameters}.

We denote the energy flow (coupling function) from node $i$ to node $j$ as
\begin{equation}
    p_{ij} \coloneqq \kappa_{ij} \sin(\varphi_i - \varphi_j) = -p_{ji}.
\end{equation}

For the solitary oscillator, we have
\begin{equation}
    m_z \ddot\varphi_z = P_z - \alpha_z \dot\varphi_z - p_{zk}.
    \label{eq:swing_eq_solitary_node}
\end{equation}
Regarding the coupling of the two subsystems $z$ and $S$, we separate the mean energy flow and the oscillating, mean-zero energy flow without loss of generality
\begin{equation}
    p_{zk}(t) \coloneqq \langle p_{zk} \rangle + p_{zk}^{\text{osc}}(t).
    \label{eq:def_mean_and_time_dependent_power_flow_Nk}
\end{equation}
With the ansatz from Eq.~(\ref{eq:ansatz_solitary_node}), time averaging of Eq.~(\ref{eq:swing_eq_solitary_node}) yields a self-consistent equation for the solitary mean frequency,
\begin{equation}
    0 = Z(\omega_s) \coloneqq P_z - \alpha_z \omega_s - \langle p_{zk} \rangle.
    \label{eq:self_consistency_implicit}
\end{equation}
In order to evaluate Eq.~(\ref{eq:self_consistency_implicit}), we need an approximation for $\langle p_{zk}\rangle$ as a function of $\omega_s$, thus closing the equation.
By assuming $\omega_s$ to be generally slowly varying, we can provide the following heuristic dynamical interpretation.
The self-consistency function $Z(\omega_s)$ is the average change in $m_z\dot\varphi_z$ over a period, which needs to be zero for a stable, periodic solitary state.
In other words, solutions $\omega_s$ to $Z(\omega_s)=0$ are fixed points $\omega_s = v^* = \text{const.}$ of the dynamical system $\dot v = Z(v)$ with state variable $v$.
Hence, the sign of $\mathrm{d} Z/\mathrm{d} \omega_s$ gives us the linear stability of the zeros of $Z(\omega_s)$, providing us with a heuristic for the stability of the corresponding periodic orbit with $\langle\dot\varphi_z\rangle = \omega_s$.

To determine $\langle p_{zk} \rangle$ as a function of $\omega_s$, system parameters and topology, we first need to find an explicit approximation for the trajectory of $\varphi_k(t)$, as Eq.~(\ref{eq:swing_lossless}) can not be solved analytically.

We start by considering $\langle p_{zk} \rangle$ and $p_{zk}^{\text{osc}}(t)$ one at a time.
First, we need to account for the power-imbalance in $S$ that is induced by the difference between $P_z$ and $\langle p_{zk} \rangle$.
Eq.~(\ref{eq:self_consistency_implicit}) shows that non-zero $\omega_s$ causes a deviation in $p_{zk}$ from its value in the synchronous state, $P_z$.
Furthermore, Fig.~1 in the main article shows that the mean coupling between the solitary node and $S$ is maximal in the synchronous state, intermediate for solitary states with intermediate solitary frequency, and approximately zero for natural solitary states, which have $\omega_s \approx P_z/\alpha_z$.
Hence, in a solitary state, where the power in the synchronized cluster is unbalanced, the oscillators in $S$ perform a slow collective counter-rotation to the solitary node, which can be observed numerically (see Supplementary Video 1 [32]).
To determine the frequency of the collective counter-rotation of $S$, we add up Eq.~(\ref{eq:swing_lossless}) for the synchronized cluster,
\begin{equation}
    m \frac{\mathrm{d}}{\mathrm{d} t}\sum_{i\in S}\dot\varphi_i 
    = - \alpha \sum_{i\in S} \dot\varphi_i - P_z + p_{zk}.
    \label{eq:reduced_network_sum_swing_eqs}
\end{equation}

We start by considering only the mean energy flow $\langle p_{zk} \rangle$ between node $z$ and the synchronized cluster.
Then, the power in the synchronized cluster is unbalanced, and the solution of Eq.~(\ref{eq:reduced_network_sum_swing_eqs}) exponentially converges towards
\begin{equation}
    \sum_{i\in S}\dot\varphi_i = \frac{-P_z + \langle p_{zk}\rangle }{\alpha}.
\end{equation}
Therefore, the synchronization frequency of the synchronized cluster is not zero as in the original system, but, on average, lies at
\begin{equation}
    \omega_\text{sync} \coloneqq \frac{-P_z + \langle  p_{zk} \rangle}{\alpha (N-1)}
    = -\frac{\alpha_z}{\alpha} \frac{\omega_s}{N-1},
    \label{eq:definition_omega_sync}
\end{equation}
with the second equality according to Eq.~(\ref{eq:self_consistency_implicit}). This linear relationship between $\omega_s$ and $\omega_\text{sync}$ will come in useful later.
Note that for constant ratios $m_i/\alpha_i$ for all nodes including $z$, the $\alpha_i$-weighted sum over all frequencies exponentially decays towards zero.
Therefore, if $m_z/\alpha_z = m/\alpha$, we have $\alpha_z \omega_s = -\alpha (N-1) \omega_\text{sync}$.

To be able to linearize the dynamics, we need the proper coordinate frame that eliminates the collective rotation of $S$.
Therefore, we define the corotating solitary frequency, $\omega_c \coloneqq \omega_s - \omega_\text{sync}$.
If we shift the phases in the synchronized cluster by $\omega_\text{sync} t$, the synchronous periodic orbit becomes a fixed point, which induces a steady-state graph Laplacian $L$.

We know from numerical experiments (Fig.~2 in main article) that solitary states occur when $L$ has at least one eigenvector $\vec v^{[r]}$ (corresponding eigenvalue $\lambda^{[r]}$) with a sufficiently high concentration at node $k$, which we assume from here on.
We will see later that the self-consistent equation, Eq.~(\ref{eq:self_consistency_implicit}), shows no solutions if the assumption is not fulfilled (see also the discussion in the main article).

We observe that this localization is typical at dense root nodes, which are neighbors of dense sprout nodes.
As we show in the preliminaries, the frequency of a highly localized mode is closely connected to the degree of the respective node. 
We remark that, if such a localized eigenmode does not exist, we can just use the largest eigenmode for the rescaling in the subsequent derivation.
Later on, we see that all modes contribute to $\langle p_{zk} \rangle$.
The magnitude of their contribution depends on their localization at node $k$ and their frequency.

\subsection{Averaging approach}
\label{sub:averaging_approach}
Here, we derive an explicit expression for Eq.~(\ref{eq:self_consistency_implicit}) using the averaging technique.
It is the proper technique to treat an autonomous system that becomes non-autonomous by inserting an explicitly time-dependent ansatz like Eq.~(\ref{eq:ansatz_solitary_node}).
The averaging theorem allows us to approximate a weakly non-autonomous system by an averaged, autonomous one.

The first step is to rewrite  system Eq.~(\ref{eq:swing_lossless}) so that the fast and slow variables are split with a suitable small parameter $\varepsilon$. 
We consider the case $\alpha_i=\alpha$ and $m_i=m$ for all $i\in S = \{1,\ldots,N\}\setminus z$ in the synchronized cluster, and rewrite Eq.~(\ref{eq:swing_lossless}) by separating the solitary node $z$ as follows,
\begin{align}
    %\label{eq:swing_lossless_z}
    m \ddot{\varphi}_{i} & =P_{i}-\alpha \dot{\varphi}_{i}-\sum_{j=1}^{N}\kappa_{ij}\sin(\varphi_{i}-\varphi_{j}), 
    \quad i\in S,\\
    %\label{eq:swing_lossless_z1}
    m_{z}\ddot{\varphi}_{z} & =P_{z}-\alpha_{z}\dot{\varphi}_{z}-\kappa_{zk}\sin(\varphi_{z}-\varphi_{k}), \quad k\in\mathcal{N}_z \quad \left(\text{here, $\mathcal{N}_z = \{k\}$}\right).
\end{align}

We first rescale parameters and the time variable\footnote{Note that this implies an inverse rescaling of frequencies.} using
\begin{equation}
    t = \frac{\sqrt{m}}{\nu} t',
    \qquad
    P_i = \frac{P'_i \alpha \nu}{\sqrt{m}},
    \qquad
    \kappa_{ij} = \frac{\kappa'_{ij}\alpha \nu}{\sqrt{m}},
    \label{eq:parameter_rescaling}
\end{equation}
and omitting the prime for the dimensionless quantities (e.g., $t' \to t$) thereafter. The scalar $\nu$ will be specified later.
We also introduce the small dimensionless parameter
\begin{equation}
    \label{eq:eps}
    \varepsilon \coloneqq \frac{\alpha}{\sqrt{m} \nu} \ll 1.
\end{equation}
 With the rescaling Eqs.~(\ref{eq:parameter_rescaling})--(\ref{eq:eps}), the system in Eq.~(\ref{eq:swing_lossless}) transforms into
\begin{align}
    \ddot\varphi_i =& \varepsilon \left[ P_i - \dot\varphi_i - \sum_{j=1}^N \kappa_{ij} \sin(\varphi_i - \varphi_j) \right],
    \quad  i \in S,
    \label{eq:rescaled_eom_reduced_network}
    \\
    \ddot\varphi_z =& \frac{\varepsilon m}{m_z} \left[P_z - \frac{\alpha_z}{\alpha} \dot\varphi_z - \kappa_{zk} \sin(\varphi_z - \varphi_k) \right].
    \label{eq:rescaled_eom_solitary_oscillator}
\end{align}

If $\varepsilon$ is small, and the terms multiplied by it on the right-hand side are of order one, then, to zeroth order, the system is approximated by freely rotating oscillators.
If we set $\nu = 1$, small $\varepsilon$ means we have a system with very high inertia and low damping.
For such a system, this is indeed a good approximate solution.
However, it does not describe any coupling between the oscillators.

To obtain the solitary and resonant states we want to describe, we want to consider the case where $\varepsilon$ is small in part due to $\nu$, and $\nu$ is related to the coupling. If we fix $\nu^2 \varepsilon \approx 1$ then $\kappa' \approx \kappa$, and the right-hand side in  \eqref{eq:rescaled_eom_solitary_oscillator} is indeed small.
However, we will see that there are values of $\nu^2$ such that the sum in the right-hand side of \eqref{eq:rescaled_eom_reduced_network} is not small.
We will see that these are related to the eigenvalues of the Laplacian of the synchronized cluster and can be averaged out in the proper coordinate frame. Thus, we proceed to linearize the system near the synchronized state.

First, we denote $\delta_{ij}$ as the Kronecker delta: $\delta_{ik} = 1$ if and only if $i=k$, and $\delta_{ik} = 0$ else.
Then, we define shifted power constants
\begin{equation}
    \hat P_i \coloneqq P_i - \alpha \omega_\text{sync} + \delta_{ik} \langle  p_{zk} \rangle
    = P_i + \frac{\alpha_z \omega_s}{N-1} + \delta_{ik} \langle  p_{zk} \rangle.
    \label{eq:shifted_power_constants}
\end{equation}
In contrast to the original power constants $P_i$ they are balanced inside $S$: $\sum_{i\in S} \hat P_i = 0$.

For the phase angles of the synchronized cluster, we make the ansatz
\begin{equation}
    \varphi_i(t) \coloneqq \omega_\text{sync} t + \vartheta_i^* + \vartheta_i(t),
    \label{eq:coordinate_shift_corotating}
\end{equation}
again, without loss of generality.
The steady-state of the synchronized cluster with only $\langle p_{zk} \rangle$ as input is given by $\vartheta_i(t) = 0$.
The power flow equations for the steady-state in the synchronized cluster are
\begin{equation}
    \hat P_i = \sum_{j\in S} \kappa_{ij} \sin\left(\vartheta_i^* - \vartheta_j^*\right).
    \label{eq:definition_corotating_power_flow_steady_state}
\end{equation}
Equation~(\ref{eq:coordinate_shift_corotating}) can be interpreted as a transformation into the average corotating frame of the synchronized cluster.
We define
\begin{equation}
    \vartheta_z(t) \coloneqq \varphi_z(t) - \omega_\text{sync} t - \vartheta_k^*
    = \omega_c t  + \varepsilon \psi_z(t) - \vartheta_k^*.
    \label{eq:def_theta_N}
\end{equation}
The function $\psi_z(t)$ is zero with only $\langle p_{zk} \rangle$ as coupling.
The constant phase shift $\vartheta_k^*$ in $\vartheta_z(t)$ can be chosen by applying a translation in time and turns out to be a convenient choice later.

Next, we consider also the oscillating part of the energy flow between the two subsystems, $p_{zk}^{\text{osc}}(t) = \kappa_{zk} \sin(\vartheta_z - \vartheta_k) - \langle p_{zk}\rangle$.
With Eq.~(\ref{eq:shifted_power_constants}), the dynamics Eqs.~(\ref{eq:rescaled_eom_reduced_network}) and (\ref{eq:rescaled_eom_solitary_oscillator}) for the corotating coordinates read
\begin{eqnarray}
    \ddot\vartheta_i &=& \varepsilon \left[\hat{P}_i - \dot\vartheta_i - \sum_{j=1}^N \kappa_{ij} \sin(\vartheta_i^* + \vartheta_i - \vartheta_j^* - \vartheta_j) \right],
    \qquad  \text{for $i \in S\setminus k$,}
    \label{eq:full_shifted_eom_reduced_network_i}
    \\
    \ddot\vartheta_k &=& \varepsilon \left[\hat{P}_k - \dot\vartheta_k - \sum_{j\in S} \kappa_{kj} \sin\left(\vartheta_k^* + \vartheta_k - \vartheta_j^* - \vartheta_j\right) 
    + \kappa_{zk} \sin(\vartheta_z - \vartheta_k ) - \langle p_{zk} \rangle \right],
    \label{eq:full_shifted_eom_reduced_network_adjacent}
    \\
    \ddot\vartheta_z &=& \frac{\varepsilon m}{m_z} \left[P_z - \frac{\alpha_z}{\alpha} (\dot\vartheta_z + \omega_\text{sync}) - \kappa_{zk} \sin(\vartheta_z - \vartheta_k) \right].
    \label{eq:full_shifted_eom_solitary_oscillator}
\end{eqnarray}
We remark that due to the definition of $\hat{P}_k$, the last two terms in Eq.~(\ref{eq:full_shifted_eom_reduced_network_adjacent}) together have zero mean, hence we have successfully eliminated any constant offset in the perturbation induced by the coupling $p_{zk}$, which would cause a rotation.

Next, we approximate the full energy flow by linearization with respect to $\vartheta_k$,
\begin{equation}
    p_{zk} \approx \kappa_{zk} \left[ \sin\vartheta_z - \vartheta_k \cos\vartheta_z\right].
    \label{eq:linearization_power_flow}
\end{equation}
We also linearize Eqs.~(\ref{eq:full_shifted_eom_reduced_network_i}) - (\ref{eq:full_shifted_eom_solitary_oscillator}) with respect to $\vartheta_i(t)$ for $i\in S$.
Using Eq.~(\ref{eq:definition_corotating_power_flow_steady_state}), the $\hat P_i$ for $i \in S$ are eliminated,
\begin{eqnarray}
    \ddot\vartheta_i &=& \varepsilon \left[- \dot\vartheta_i - \sum_{j\in S} \left( \vartheta_i - \vartheta_j \right) \kappa_{ij} \cos\left(\vartheta_i^* - \vartheta_j^*\right)  \right]
    \qquad  \text{for $i \in S\setminus k$,}
    \label{eq:lin_shifted_eom_reduced_network_not_adjacent}
    \\
    \ddot\vartheta_k &=& \varepsilon \left[- \dot\vartheta_k - \sum_{j\in S} \left( \vartheta_k - \vartheta_j \right)  \kappa_{kj} \cos\left(\vartheta_k^* - \vartheta_j^*\right) 
    + \kappa_{zk} \left( \sin\vartheta_z - \vartheta_k \cos\vartheta_z \right) - \langle p_{zk} \rangle \right],
    \label{eq:lin_shifted_eom_reduced_network_adjacent}
    \\
    \ddot\vartheta_z &=& \frac{\varepsilon m}{m_z} \left[P_z - \frac{\alpha_z}{\alpha} (\dot\vartheta_z + \omega_\text{sync}) - \left(\kappa_{zk} \sin\vartheta_z - \vartheta_k \cos\vartheta_z \right) \right].
    \label{eq:lin_shifted_eom_solitary_oscillator}
\end{eqnarray}
We observe that to first order in $\varepsilon$, $\vartheta_k$ is being harmonically driven by $\varepsilon \kappa_{zk} \sin\vartheta_z \approx \varepsilon \kappa_{zk} \sin\omega_c t$.
Due to the additional nonlinear perturbation $\varepsilon \kappa_{zk} \vartheta_k \sin\vartheta_z$, we can not simply apply linear response.
Approximating it to first order in $\varepsilon$ as $\varepsilon \kappa_{zk} \vartheta_k \sin(\omega_c t)$ fully linearizes the system, but still leaves us with time-varying coefficients.
Therefore, in the following, we apply the averaging theorem to treat this term and show that it averages out in Eq.~(\ref{eq:full_shifted_eom_reduced_network_adjacent}).
We show later on in section \ref{sub:linear_response} that one obtains the same result as with the averaging technique, if one neglects this term right away, but in Eq.~(\ref{eq:full_shifted_eom_reduced_network_adjacent}) only, where it is only a small contribution to the overall coupling of node $k$.
It should not be ignored in Eq.~(\ref{eq:full_shifted_eom_solitary_oscillator}), as it gives a relevant contribution to $\langle p_{zk}\rangle$ in Eq.~(\ref{eq:self_consistency_implicit}).

The steady state in the corotating frame of $S$ induces a weighted graph Laplacian for the synchronized cluster,
\begin{equation}
    L_{ij} \coloneqq
    \begin{cases}
        \begin{aligned}
            \sum_{j\in S\setminus i} \kappa_{ij} \cos\left(\vartheta_i^* - \vartheta_j^*\right) &\quad \text{iff $i = j$},\\
            - \kappa_{ij} \cos\left(\vartheta_i^* - \vartheta_j^*\right) &\quad \text{iff $i\neq j$.}
        \end{aligned}
    \end{cases}
    \label{eq:steady_state_Laplacian_corotating}
\end{equation}
for $i,j\in S$.
We remark that it is reminiscent of the Laplacian of the original system, $\mathcal{L}$, if one removes all entries associated to node $z$, i.e., row and column $z$, as well as the contribution to $\mathcal{L}_{kk}$.

Let $\vec\vartheta$ be the vector with entries $\vartheta_i$ for $i\in S$.
We can now write Eqs.~(\ref{eq:lin_shifted_eom_reduced_network_not_adjacent}) and (\ref{eq:lin_shifted_eom_reduced_network_adjacent}) in vectorized form,
\begin{eqnarray}
    \ddot{\vec\vartheta} &=& \varepsilon \left(-\dot{\vec\vartheta} - L \vec\vartheta + \vec f \right),
    \label{eq:vectorized_eom_reduced_network}
    \\
    \vec f &\coloneqq& \hat e_k \left( \kappa_{zk} \sin\vartheta_z 
     - \kappa_{zk} \vartheta_k \cos\vartheta_z - \langle p_{zk}\rangle \right),
     \\
     &\approx&
     \hat e_k \left( \kappa_{zk} \sin\omega_c t 
     - \kappa_{zk} \vartheta_k \cos\omega_c t - \langle p_{zk}\rangle \right),\label{eq:definition_perturbation_vector}
\end{eqnarray}
where $\hat e_k$ is the Cartesian basis vector in $k$-direction, which has components $\delta_{ik}$.
We have approximated $\varepsilon \vec f$ to first order in $\varepsilon$ for $\vartheta_z$, that is, we approximated $\vartheta_z \approx \omega_c t$ in $\vec f$.
For this, we assume that $\omega_c \gg \varepsilon \dot\psi_z(t)$, i.e., that $\psi(t)$ does not vary faster than $O(\omega_c \varepsilon^{-1}).$ This is supported by the numerical observations that in solitary state, the phase of the solitary node is dominated by its mean time-linear growth and not by the fluctuations around it.
With this approximation, we can proceed with Eq.~(\ref{eq:vectorized_eom_reduced_network}) and can set Eq.~(\ref{eq:lin_shifted_eom_solitary_oscillator}) aside.

As mentioned earlier, there is a network mode with associated eigenvalue $\lambda^{[r]}$ that is highly localized at node $k$.
We have shown above in section \ref{subsub:eigenvector_localization} that this implies that the associated frequency is closely connected to $d_k$, the degree of node $k$.
We set $\nu^2 = \lambda^{[r]}$ to measure the systems frequencies in units of $\sqrt{\lambda^{[r]}}$.
Therefore, the above approximations are most valid for typical root-sprout configurations with highly connected root nodes.

Since $\kappa'_{ij} = \frac{\kappa_{ij}}{\varepsilon \lambda^{[r]}}$ in Eq.~(\ref{eq:parameter_rescaling}), the Laplacian got rescaled like $L' = \frac{L}{\varepsilon \lambda^{[r]}}$ (renaming $L'$ to $L$ thereafter).
We now rescale the Laplacian again, such that we measure it in units of $\lambda^{[r]}$ and make the $\varepsilon$-scaling explicit, that is, $L' = \frac{L''}{\varepsilon}$, thus substituting $L = \frac{L''}{\varepsilon}$ in Eq.~(\ref{eq:vectorized_eom_reduced_network}), and renaming $L''$ to $L$ afterward,
\begin{equation}
     \ddot{\vec\vartheta} = \varepsilon \left(-\dot{\vec\vartheta} + \vec f \right) - L \vec\vartheta.
     \label{eq:vectorized_eom_reduced_network_rescaled_Laplacian}
\end{equation}

Now, we diagonalize the system by transforming into the eigenbasis $\left\{\vec{v}^{[\ell]} \right\}$ of $L$,
\begin{eqnarray}
    \vec \xi &\coloneqq & Q^T \vec\vartheta,
    \label{eq:def_mode_variables_trafo}
    \\
    Q_{ij} &\coloneqq & v^{[j]}_i,
    \\
    Q^T L Q &\coloneqq & \text{diag} (\lambda^{[\ell]}) \coloneqq \Gamma,
\end{eqnarray}
with the eigenvalues $\left\{\lambda^{[\ell]}\right\}$ ordered by magnitude from $\ell = 0$ to $N-1$.
The $\xi_\ell$ are the excitations of network modes corresponding to $\omega^{[\ell]} \approx \sqrt{\lambda^{[\ell]}}$ with their nodal excitations $\vec\vartheta$ distributed across the network according to $\vec v^{[\ell]}$.

Eq.~(\ref{eq:vectorized_eom_reduced_network_rescaled_Laplacian}) becomes
\begin{eqnarray}
    \ddot{\vec{\xi}} &=& \varepsilon\left( - \dot{\vec{\xi}} + \vec g \right) - \Gamma \vec\xi,
    \label{eq:diagonalized_eom_second_order}
\end{eqnarray}
where $\vec g \coloneqq Q^T \vec f$.
Equation~(\ref{eq:diagonalized_eom_second_order}) has the form of the perturbation of the Hamiltonian system $\ddot{\vec{\xi}} = -\Gamma \vec\xi$, where the perturbation is due to the friction term $\dot{\vec\xi}$ and due to $\vartheta_z$ in $\vec{g}$.
The appropriate perturbative approach in this case is averaging [30], where one should first write the system for the slowly varying amplitudes of the network modes, i.e., the amplitudes of the periodic solutions of the unperturbed system.
These zeroth order solutions in $\varepsilon$ of are
\begin{equation}
    \xi_\ell( t) =   x_\ell \cos \left(\sqrt{\lambda^{[\ell]}} \cdot  t\right) -  z_\ell \sin \left( \sqrt{\lambda^{[\ell]}} \cdot  t \right),
\end{equation}
where $\vec x$ and $\vec z$ are constants specified by initial conditions.
Since the zeroth order system is conservative, they describe a set of periodic orbits on tori in phase space.
For $\ell=1$, we have $\lambda^{[1]} = 0$ and an additional constant plus linear motion in time.
However, this uniform motion of the synchronized cluster is already accounted for in Eq.~(\ref{eq:ansatz_solitary_node}) via a global phase shift in the original system, hence we can set it to zero here.

Following reference [30] (theorem 4.1.1, example II), for the first-order solution, we can make the ansatz\begin{equation}
     \vec\xi(t) \coloneqq \vec x(t) \cos \omega_c t - \vec z(t) \sin \omega_c t,
    \label{eq:averaging_network_ansatz_modes}
\end{equation}
where $\vec x(t)$ and $\vec z(t)$ are now time-dependent amplitude variables.
As the ansatz above, this can also be considered as a change of coordinates without loss of generality.
As we see later, this change of coordinates is necessary to achieve the proper form to apply averaging.
Due to coupling, damping, and resulting resonance between the solitary rotation and the oscillation of the synchronized cluster, the $\xi_\ell$ slowly move on the solution manifold of the zeroth order system, and the frequency of their collective oscillation is shifted from the root of the Laplacian eigenvalue $\sqrt{\lambda^{[\ell]}}$ to $\omega_s$.

Since Eq.~(\ref{eq:diagonalized_eom_second_order}) is not ready to apply averaging, because the right-hand side is not completely of order $\varepsilon$, we need to consider the dynamics of the $\vec x(t)$ and $\vec z(t)$ instead.
To do this, we rewrite the equations of motion, Eqs.~(\ref{eq:diagonalized_eom_second_order}), as a system of first-order differential equations with dimension $2(N-1)$,
\begin{equation}
    \frac{\mathrm{d}}{\mathrm{d} t}
    \begin{pmatrix}
        \vec{\xi}\\
        \vec{\zeta}
        \end{pmatrix}
        \coloneqq \begin{pmatrix}
            \vec\zeta\\
            \varepsilon\left( - \vec{\zeta} + \vec g \right) - \Gamma \vec\xi
        \end{pmatrix},
    \end{equation}
    and transform into a corotating frame with the amplitudes as new variables,
    \begin{equation}
        \begin{pmatrix}
            \vec x\\
            \vec z
        \end{pmatrix}
        \coloneqq
        A( t)
        \begin{pmatrix}
            \vec\xi\\
            \vec\zeta
        \end{pmatrix}
        \coloneqq
        \begin{pmatrix}
            \mathds{1} \cos\omega_c t & -\frac{\mathds{1}}{\omega_c}\sin\omega_c t\\
            -\mathds{1} \sin\omega_c t & -\frac{\mathds{1}}{\omega_c}\cos\omega_c t
        \end{pmatrix}
        \begin{pmatrix}
        \vec\xi\\
        \vec\zeta
        \end{pmatrix},
    \end{equation}
using the invertible van der Pol transformation $A(t)$, see reference [30] (theorem 4.1.1, example II).
   The equations of motion for the amplitude variables are
    \begin{eqnarray}
    \frac{\mathrm{d}}{\mathrm{d} t}
    \begin{pmatrix}
    \vec x\\
    \vec z
    \end{pmatrix}
    &=&
    \frac{\mathds{1} \omega_c^2 - \Gamma }{\omega_c}
    \begin{pmatrix}
    \vec z \sin^2\omega_c t  - \vec x \cos\omega_c t \sin\omega_c t \\
     \vec z\cos\omega_c t \sin\omega_c t  - \vec x  \cos^2\omega_c t
    \end{pmatrix}
    \nonumber
    \\
    &&-
    \frac{\varepsilon}{\omega_c}
    \begin{pmatrix}
        \sin\omega_c t \left(\omega_c \vec x \sin\omega_c t + \omega_c \vec z \cos\omega_c t  + \vec g\right)\\
        \cos\omega_c t \left(\omega_c \vec x \sin\omega_c t + \omega_c \vec z \cos\omega_c t  + \vec g\right)
    \end{pmatrix}.
    \label{eq:averaging_amplitues_slow_system}
\end{eqnarray}
To apply the averaging theorem, we need the right-hand side to be of order $\varepsilon$.
For the minimal effective model that we discuss below in section \ref{sub:the_minimal_effective_model}, the synchronized cluster only has a single frequency ($\Gamma$ is scalar), and we can just assume a close resonance, $\omega_c^2 - \Gamma = O(\varepsilon)$.
For the networked case, we need to assume one of the following two possibilities for all modes.
First, if a mode is in resonance, we have $\omega_c^2 - \lambda^{[\ell]} = O(\varepsilon)$.
Second, if the driving frequency is not close to the eigenmode's intrinsic frequency $\omega^{[\ell]}$, i.e., $\omega_c^2 - \lambda^{[\ell]} = O(1)$, the mode $z_\ell$ and $x_\ell$ will be only weakly excited.

This can be understood heuristically by the observation that in the absence of a resonant excitation, the response will be of the order of the driving force. There are two independent reasons for the driving force to be small: first, the weak localization of most modes at node $k$, and second, the fact that the driving comes with a factor $\varepsilon$.

To make this more explicit, consider the transfer function of Eq.~(\ref{eq:diagonalized_eom_second_order}),
\begin{eqnarray}
    T(s) &=& (s^2 + \varepsilon s + \Gamma)^{-1},
    \\
    \tilde \xi(s) &=& \varepsilon  T(s) \tilde g(s),
\end{eqnarray}
where $\tilde\xi(s)$ and $\tilde g(s)$ are the Laplace transforms of $\vec\xi(t)$ and $\vec g(t)$, respectively.
Absence of resonance implies that $\tilde g(s)$ has no support near $\Gamma$, and thus $T(s)g(s)$ is never of order $\varepsilon^{-1}$.

We note that this is an additional assumption to example II of the averaging theorem as in reference [30], which does not treat the networked case $|S| > 1$.
Averaging over a range of network modes is a novelty in our approach.
As we show later on, this assumption is self-consistent with the solutions it produces, and with empirically observed behavior of the full system in simulations.

We define the diagonal matrix $\Delta$ by
\begin{equation}
    \mathds{1}\omega_c^2 - \Gamma 
    \coloneqq \varepsilon \Delta.
\end{equation}
Since the system Eq.~(\ref{eq:averaging_amplitues_slow_system}) is slow moving it is of the right form to apply the averaging theorem in reference [30] (theorem 4.1.1 therein).
It states that a system of this form, where the right-hand side is periodic in time and of order $\varepsilon$, can be approximated at order $O(\varepsilon)$ by an autonomous system which is obtained by time-averaging the right-hand side for fixed (average) $\vec x$ and $\vec z$.
The periodicity is maintained in a $O(\varepsilon^2)$ term, which we neglect here.
The solutions of both systems are $\varepsilon$-close for time scales up to $O(\varepsilon^{-1})$ (in dimensionless time).
Translating this back by inverting Eq.~(\ref{eq:parameter_rescaling}), this amounts to time scales of the order $m/\alpha$, which is typically much larger than $T_s \coloneqq 2\pi/\omega_s$, the approximate periodicity of the solitary state.
Further, hyperbolic fixed points of the averaged system possess a unique hyperbolic orbit in the original system $\varepsilon$-close to the fixed point for all $\varepsilon$ up to some $\varepsilon_0$.
We make use of the last statement to determine an approximation for the (average) orbit $\vartheta_k$ from the unique stable fixed point of the averaged system.

By averaging Eq.~(\ref{eq:averaging_amplitues_slow_system}) over time, we obtain the linear system
\begin{equation}
\frac{\mathrm{d}}{\mathrm{d} t}
    \begin{pmatrix}
        \vec x\\
        \vec z
    \end{pmatrix}
    =
    J \begin{pmatrix}
        \vec x\\
        \vec z
    \end{pmatrix}
    -\vec h
    \coloneqq
    -\frac{\varepsilon}{2\omega_c}
    \begin{pmatrix}
        \mathds{1}\omega_c & -\Delta\\
        \Delta & \mathds{1}\omega_c
    \end{pmatrix}
    \begin{pmatrix}
        \vec x\\
        \vec z
    \end{pmatrix}
    - \vec h,
\end{equation}
where the components of the inhomogeneity vector are given by
\begin{equation}
    h_\ell =\begin{cases}
        \begin{aligned}
            \frac{\varepsilon  \kappa_{zk}}{2\omega_c} v_k^{[\ell]} & \quad \text{for $\ell = 1,\ldots,N-1$}
            \\
            0 &\quad \text{for $\ell = N,\ldots,2N-2$}.
        \end{aligned}
    \end{cases}
    \label{eq:inhomogeneity_vector}
\end{equation}
We note that the nonlinear term in $\vec g$, that we approximated by a time-varying linear term, is averaged out. This explains why we can obtain the same result by neglecting the term from the outset and applying linear response, as shown below in section \ref{sub:linear_response}.
The eigenvalues of the Jacobian $J$ are given by
\begin{equation}
    j_\pm^{[\ell]}
    =-\frac{\varepsilon}{2} \pm i \frac{\varepsilon \Delta_\ell}{2\omega_c},
\end{equation}
where $\Delta_\ell$ are diagonal elements of $\Delta$.
Note that due to its block structure and the invertibility of the blocks, we can easily invert $J$.
The stable fixed point is located at
\begin{equation}
    \begin{pmatrix}
        \vec x^*\\
        \vec z^*
    \end{pmatrix}
    = J^{-1} \vec h
    = -\frac{2}{\varepsilon} \left(\mathds{1} + \frac{\Delta^2}{\omega_c^2}\right)^{-1}
    \otimes
    \begin{pmatrix}
        \mathds{1} & \frac{\Delta}{\omega_c}\\
        -\frac{\Delta}{\omega_c} & \mathds{1}
    \end{pmatrix}
    \vec h,
\end{equation}
where in the last expression, the first (diagonal) matrix is to be multiplied with each of the four blocks (also diagonal) in the second matrix, as indicated by the symbol $\otimes$.
The fixed point solution for the components of $\vec\xi$ lies at
\begin{equation}
    \xi_\ell = x_\ell^* \cos\omega_c t - z_\ell^* \sin\omega_c t
    = -\frac{ \kappa_{zk} \omega_c}{\omega_c^2 + \Delta_\ell^2} v_k^{[\ell]} \cos\omega_c t - \frac{ \kappa_{zk}\Delta_\ell}{\omega_c^2 + \Delta_\ell^2} v_k^{[\ell]}  \sin\omega_c t.
    \label{eq:solution_xi_ell}
\end{equation}
This is a periodic orbit that serves as an approximation for the trajectories in $S$ in the solitary state at frequency $\omega_c$.

To finally obtain $\omega_c$, we have to insert this trajectory into the self-consistent equation.
With Eqs.~(\ref{eq:linearization_power_flow}), (\ref{eq:def_mode_variables_trafo}) and (\ref{eq:solution_xi_ell}), and the zeroth order approximation of $\vartheta_z$, we can now evaluate
Eq.~(\ref{eq:self_consistency_implicit}),
\begin{eqnarray}
    p_{zk} &\approx& \kappa_{zk} \left[ \sin\omega_c t  - \vartheta_k \cos\omega_c t \right],
    \nonumber\\
    \langle p_{zk} \rangle &=& p_s \approx  - \langle \kappa_{zk} \vartheta_k \cos\omega_c t \rangle
    \nonumber\\
    &=& -\kappa_{zk} \Bigg\langle \sum_{\ell=1}^{N-1} v_k^{[\ell]} \xi_\ell \cos\omega_c t \Bigg\rangle
    \nonumber\\
    &=&
    -\kappa_{zk} \Bigg\langle \sum_{\ell=1}^{N-1} v_k^{[\ell]} \cos\omega_c t  \left( -\frac{ \kappa_{zk} \omega_c}{\omega_c^2 + \Delta_\ell^2} v_k^{[\ell]} \cos\omega_c t - \frac{ \kappa_{zk}\Delta_\ell}{\omega_c^2 + \Delta_\ell^2} v_k^{[\ell]}  \sin\omega_c t\right) \Bigg\rangle
    \nonumber\\
    &=&
    \kappa_{zk} \sum_{\ell=1}^{N-1} v_k^{[\ell]} \frac{1}{2}  \frac{ \kappa_{zk} \omega_c}{\omega_c^2 + \Delta_\ell^2} v_k^{[\ell]},
    \nonumber\\
    0 &=& Z(\omega_s) \coloneqq  P_{z}-\alpha_z \omega_s - \frac{ \kappa_{zk}^2 }{2} \sum_{\ell=1}^{N-1} \frac{\alpha \omega_c \left(v_k^{[\ell]}\right)^2}{\left(\lambda^{[\ell]} - m \omega_c^2\right)^2 + \alpha^2\omega_c^2}.
    \label{eq:self_consistency_explicit}
\end{eqnarray}
Note that we have already re-substituted $\Delta_\ell$ and the original parameters by inverting Eq.~(\ref{eq:parameter_rescaling}).
The sum over $\ell$ stems from the decomposition of $\vartheta_k$ into network modes $\xi_\ell$.\

After having derived an explicit form, we close this subsection with a few notes on closing the self-consistent equation, Eq.~(\ref{eq:self_consistency_explicit}) and on the evaluation.
A detailed evaluation can be found in Section~\ref{sec:evaluation_of_the_self-consistent-equation}.

First, we need to insert a relation between $\omega_s$ and $\omega_c$.
Following Eq.~(\ref{eq:definition_omega_sync}), the corotating solitary frequency $\omega_c$ can be expressed by
\begin{equation}
    \omega_c = \left( 1 + \frac{\alpha_z}{\alpha} \frac{1}{N-1} \right) \omega_s.
    \label{eq:corotating_and_static_phase_velocity}
\end{equation}
We assume that $\alpha_z$ and $\alpha$ are of similar magnitude, hence for very large $N$, one could approximate $\omega_c \approx \omega_s$, however,  we keep the more general form.

Second, we need an explicit expression for the Laplacian eigenvectors and eigenvalues that appear in the self-consistent equation.
Since the Laplacian given in Eq.~(\ref{eq:steady_state_Laplacian_corotating}) and its eigenvectors and eigenvalues depend on $\langle p_{zk} \rangle$, we need to approximate it, in order to evaluate Eq.~(\ref{eq:self_consistency_explicit}).
We assume that the dependence is weak, and choose to use the steady-state phase angles $\varphi_i^*$ determined by Eq.~(\ref{eq:power_flow_equations}) instead of $\vartheta_i^*$ (for $i\in S$).
Alternatively, one could omit $\langle p_{zk} \rangle$ in the definition of $\hat P_i$, and $\omega_\text{sync}$ when evaluating Eqs.~(\ref{eq:definition_omega_sync}), (\ref{eq:shifted_power_constants}) and (\ref{eq:steady_state_Laplacian_corotating}).

Given an approximation of the Laplacian, Eq.~(\ref{eq:corotating_and_static_phase_velocity}) closes Eq.~(\ref{eq:self_consistency_explicit}), which can be evaluated numerically to find solutions for $\omega_s$.
In principle, the solutions can be used to increase the accuracy of the approximation of the Laplacian, leading to an iterative solution algorithm.
Here, we evaluate the zeroth-order solutions of this cycle, as they are already accurate.
$L$ is then obtained by omitting column and line $z$ from the $N\times N$ Laplacian of the total system $\mathcal{L}$, and omitting the term in $\mathcal{L}_{kk}$ that stems from the coupling between nodes $k$ and $z$ as well.
This last steps conserves the Laplacian property.
Without this last step, due to the Poincar\'e separation theorem,
the eigenvalues of $L$ lie between those of $\mathcal{L}$ (so-called interlacent property).
We suppose that something similar holds, at least approximately, when including the correction of the diagonal entry $\mathcal{L}_{kk}$.

Lastly, we comment on the expected stability of solutions.
The zeros of Eq.~(\ref{eq:self_consistency_explicit}) give rise to approximate solutions of the dynamics in Eq.~(\ref{eq:swing_lossless}) with $\langle \dot\varphi_z \rangle = \omega_s$.
The linear stability of a solution $\omega_s$ of $Z(\omega_s)=0$ is given by $\mathrm{d} Z/\mathrm{d} \omega_s < 0$, which is equivalent to
\begin{eqnarray}
    \alpha_z &<& -\frac{\mathrm{d}\langle p_{zk} \rangle}{\mathrm{d} \omega_s}
    = -\frac{\mathrm{d}\langle p_{zk} \rangle}{\mathrm{d} \omega_c} \frac{\mathrm{d}\omega_c }{\mathrm{d} \omega_s}
    \nonumber
    \\
    &=&
    \sum_{\ell = 1}^{N-1}
    \kappa_{zk}^2 \alpha \left( v_k^{[\ell]} \right)^2
    \frac{ \left(\lambda^{[\ell]} - m \omega_c^2\right)^2 + \alpha^2 \omega_c^2 
    + 2 \omega_c^2 \left[2 m \left(\lambda^{[\ell]} - m \omega_c^2\right) - \alpha^2 \omega_c \right]}{2 \left[\left(\lambda^{[\ell]} - m \omega_c^2\right)^2 + \alpha^2 \omega_c^2\right]^2} 
    \quad \left( 1 + \frac{\alpha_z}{\alpha} \frac{1}{N-1} \right)
    .
    \label{eq:stability_of_self_consistency}
\end{eqnarray}
Graphically, this means that the curve $\langle p_{zk} \rangle(\omega_s)$ intersects the straight line $P_z - \alpha_z \omega_s$ with a slope greater than $-\alpha_z$, i.e., coming from below in positive $\omega_s$-direction.

\subsection{Linear response}
\label{sub:linear_response}
Here, we provide a second derivation of Eq.~(\ref{eq:self_consistency_explicit}) with a heuristic, but intuitive method that relies on linear response, cf. reference [28].
If we split the linearized energy flow in Eq.~(\ref{eq:linearization_power_flow}) in its mean and oscillating part, cf. Eq.~(\ref{eq:def_mean_and_time_dependent_power_flow_Nk}), we can approximate
\begin{equation}
    p_{zk}^{\text{osc}}(t) \approx \kappa_{zk} \sin\omega_c t,
\end{equation}
in Eq.~\ref{eq:full_shifted_eom_reduced_network_adjacent}) for small $\vartheta_k$.
This is valid, if the neglected nonlinear term $\vartheta_k\sin\vartheta_z$ is only a small contribution to the overall coupling of $\vartheta_k$.
As noted before, the term should not be neglected in Eq.~\ref{eq:full_shifted_eom_solitary_oscillator}), where it is the only coupling term and crucial for nonzero $\langle p_{zk}\rangle$ in Eq.~(\ref{eq:self_consistency_implicit}).

Now, with only harmonic forcing, we can apply linear response.
Defining $\chi_i \coloneqq \omega_i - \omega_\text{sync} = \dot\vartheta_i$, Eq.~(\ref{eq:swing_lossless_omega}) for the synchronized cluster transforms into
\begin{equation}
    m \dot\chi_i 
    \approx
     \hat P_i - \alpha \chi_i
    - \sum_{j=\in S}  \kappa_{ij} \sin\left(\vartheta_i - \vartheta_j\right)
     + \delta_{ik} \kappa_{kz}  \sin\left(\omega_c t\right).
     \label{eq:linear_response_network_xi_i_full_with_ansatz_omega_N_approximated_coupling}
\end{equation}
We linearize with respect to the $\vartheta_i$ around $\vartheta_i^*$, and solve for them using linear response with the ansatz
\begin{equation}
    \vartheta_i = \vartheta_i^* + \kappa_{kz} \theta_i + \kappa_{kz} \vartheta_i^{(k)}(t).
\end{equation}
Here, $\theta_i$ is the constant part, and $\vartheta_i^{(k)}(t)$ the time-varying part of the response.
The external perturbing signal is the sinusoid with amplitude $\kappa_{kz}$ and frequency $\omega_c$.
We write it as a vector $\vec F^{(k)}(t)$ with components $\delta_{ik} \kappa_{iz} \Im \left(e^{\imath \omega_c t}\right)$, where $\Im$ is the imaginary part.
The linearized system in vectorized form reads
\begin{equation}
    m \dot{\vec{\chi}} = \alpha \vec{\chi} - L \kappa_{kz} \vec \vartheta^{(k)} + \vec F^{(k)}(t).
    \label{eq:linear_response_network_xi_vectorized_linearized}
\end{equation}
Zhang et al. show that the linear response is valid for amplitudes of this magnitude, see reference [28].
Following their calculation, the constant parts of the response are the same, $\theta_i = \theta$ for all $i$, and fall out of the equation, while the time-varying part of the response is
\begin{equation}
    \kappa_{kz} \vec\vartheta^{(k)}(t) = \Im \left[ \left( L - m \omega_c^2 + \imath \alpha \omega_c \right)^{-1} \vec F^{(k)}(t)\right].
    \label{eq:linear_response_matrix_inverse}
\end{equation}
An analytical expression is obtained by projecting $\vec F^{(k)}(t)$ onto the orthonormal eigenbasis of $L$, i.e., inserting $\mathds{1}_{ij} = \sum_{\ell=1}^{N-1}  v_i^{[\ell]} \left( v_j^{[\ell]}\right)^T $,
\begin{equation}
    \kappa_{kz} \vec\vartheta^{(k)}(t) = \Im \left( \sum_{\ell=1}^{N-1} \frac{\vec{v}^{[\ell]} v_k^{[\ell]}}{\lambda^{[\ell]} -m \omega_c^2 + \imath \alpha \omega_c} \kappa_{kz} e^{\imath \omega_c t}\right).
\end{equation}
If we define the modal response amplitudes at node $i$ as
\begin{equation}
    a_i^{(k)[\ell]} \coloneqq \frac{ \kappa_{kz}  v_i^{[\ell]}  v_k^{[\ell]}  }{\sqrt{\left(\lambda^{[\ell]} - m \omega_c^2\right)^2 + \alpha^2 \omega_c^2 }},
    \label{eq:linear_response_resonance_amplitude}
\end{equation}
and the modal phase lags $\delta^{[\ell]}$ as
\begin{align}
    \tan\delta^{[\ell]} &\coloneqq -\frac{\alpha \omega_c}{\lambda^{[\ell]} - m \omega_c^2},
    \nonumber
    \\
    \sin\delta^{[\ell]} &= -\frac{\alpha \omega_c}{\sqrt{\left(\lambda^{[\ell]} - m \omega_c^2\right)^2 + \alpha^2 \omega_c^2 }},
    \nonumber
    \\
    \cos\delta^{[\ell]} &= \frac{\lambda^{[\ell]} - m \omega_c^2}{\sqrt{\left(\lambda^{[\ell]} - m \omega_c^2\right)^2 + \alpha^2 \omega_c^2 }},
    \label{eq:linear_response_resonance_phase_lag}
\end{align}
we can write the response as
\begin{equation}
    \kappa_{kz} \vartheta_i^{(k)}(t) = \Im \left( \sum_{\ell=1}^{N-1} a_i^{(k)[\ell]} e^{\imath (\omega_c t + \delta^{[\ell]} ) } \right).
    \label{eq:linear_response_solution_with_amplitudes_and_phases}
\end{equation}
This provides us with the following interpretation.
The solitary rotation with frequency $\omega_c$ excites network modes given by $\omega^{[\ell]}\approx \sqrt{\lambda^{[\ell]}}$.
The nodal amplitudes depend on the mismatch between the driving frequency $\omega_c$ and the eigenfrequencies of the network, as well as the localization of the corresponding eigenvectors and the damping.

To close the self-consistent equation, we need to determine how the excitation of the network acts back on $\omega_c$.
Averaging over $\dot\omega_z$, see Eq.~(\ref{eq:self_consistency_implicit}), using Eqs.~(\ref{eq:linearization_power_flow}) and (\ref{eq:linear_response_solution_with_amplitudes_and_phases}) and $\vartheta_z \approx \omega_c t$, yields the same self-consistent equation as the averaging approach, Eq.~(\ref{eq:self_consistency_explicit}).
In terms of the response amplitudes and phase lags, the self-consistency function takes the form
\begin{equation}
    Z(\omega_s) = P_z - \alpha_z \omega_s + \frac{\kappa_{kz}}{2} \sum_{\ell=1}^N a_k^{(k)[\ell]} \sin\delta^{[\ell]}.
\end{equation}

In fact, the terms in the denominators in the sum over all modes in Eq.~(\ref{eq:self_consistency_explicit}) resemble the resonance curve of linear oscillators.
It is known as Cauchy–Lorentz distribution, Lorentz(ian) function, or Breit–Wigner distribution.

\subsection{Solitary nodes with several neighbors}
\label{sub:solitary_nodes_with_several_neighbors}
In the previous derivations,
we assumed that the solitary node is of degree one with the neighbor $k$.
If the solitary node has a higher degree, we can generalize the self-consistent equation as follows.
Due to linearity, we can simply superpose the perturbations entering the network through the links from the solitary node to its neighbors $k \in \mathcal{N}_z$.
This transforms Eq.~(\ref{eq:swing_eq_solitary_node}) into
\begin{equation}
    m_z \ddot\varphi_z = P_z - \alpha_z \dot\varphi_z - \sum_{k \in \mathcal{N}_z} p_{zk},
    \label{eq:swing_eq_solitary_node_several_neighbors}
\end{equation}
i.e., replacing $p_{zk}$ by $p_z \coloneqq \sum_k p_{zk}$ from there on.
For the averaging approach, this corresponds to an additional sum over $k \in \mathcal{N}_z$ in Eq.~(\ref{eq:inhomogeneity_vector}) and thereafter.
For the linear response, we sum over several perturbation vectors in Eqs.~(\ref{eq:linear_response_network_xi_i_full_with_ansatz_omega_N_approximated_coupling}) to (\ref{eq:linear_response_network_xi_vectorized_linearized}).
The result is
\begin{equation}
    0 = Z(\omega_s) \coloneqq  P_{z}-\alpha_z \omega_s - \sum_{k \in \mathcal{N}_z} \frac{ \kappa_{zk}^2 }{2} \sum_{\ell=1}^{N-1} \frac{\alpha \omega_c \left(v_k^{[\ell]}\right)^2}{\left(\lambda^{[\ell]} - m \omega_c^2\right)^2 + \alpha^2\omega_c^2}.
    \label{eq:self_consistency_explicit_several_neighbors}
\end{equation}
Eq.~(\ref{eq:stability_of_self_consistency}) also gets an additional sum over $k\in\mathcal{N}_z$.

\subsection{Heterogeneous system parameters}
\label{sub:heterogeneous_system_parameters}
So far, we assumed that the synchronized cluster has homogeneous inertia $m_i = m$ and damping $\alpha_i = \alpha$ for $i\in S$,
while allowing for different parameters $m_z$ and $\alpha_z$ at the solitary node, as well as heterogeneity in the power constants $P_i$ and coupling matrix $\kappa_{ij}$ for the whole network.
If one allows generally heterogeneous inertia and damping in the whole network, the self-consistent equation can still be evaluated numerically,\footnote{Note that numerical computation of the eigenvectors and eigenvalues of the Laplacian is necessary for complex networks, anyway. Numerical evaluation of the zeros of the self-consistent equation is necessary for networks as complex as the star graph, already.} as outlined below in section \ref{sec:evaluation_of_the_self-consistent-equation}.

To see this, consider Eq.~(\ref{eq:linear_response_matrix_inverse}) and replace $m$ by $M \coloneqq \text{diag}(m_i)$ and $\alpha$ by $A \coloneqq \text{diag}(\alpha_i)$ for $i\in S$.
If $M$, $A$ and $L$ do not commute,\footnote{As Jörg Liesen pointed out in personal correspondence, for a general connected graph Laplacian, the only commuting diagonal matrix of the Laplacian is the identity up to a scalar multiplicative constant.} they can not be diagonalized simultaneously and the projection onto $L$'s eigenbasis of network modes can not be applied to invert the matrix
\begin{equation}
    R^{-1}(\omega_c) \coloneqq L - \omega_c^2 M + \imath \omega_c A.
\end{equation}
However, this can still be done numerically.
Multiplication of the \textit{response matrix} $R(\omega_c)$ with $\vec F^{(k)}(t)$, taking the imaginary part, projecting out component $k$ to get $\vartheta_k^{(k)}(t)$, and time-averaging Eq.(\ref{eq:linearization_power_flow}), which is a projection on the $\kappa_{zk} \cos\omega_c t$ function, all amount to taking the matrix element $-\frac{\kappa_{kz}^2}{2}\Im\left(R_{kk} \right)$ for each solitary neighbor $k \in \mathcal{N}_z$.
Hence, the self-consistent equation can be written as
\begin{equation}
    0 = Z(\omega_s) \coloneqq  P_{z}-\alpha_z \omega_s + \sum_{k \in \mathcal{N}_z} \frac{ \kappa_{zk}^2 }{2} \Im\left[R_{kk}(\omega_c) \right].
    \label{eq:self_consistency_explicit_heterogeneous}
\end{equation}
Note that, here, $\omega_\text{sync}$ does not obey Eq.~(\ref{eq:definition_omega_sync}), but the following generalized version.
Averaging over Eq.~(\ref{eq:reduced_network_sum_swing_eqs}) with the ansatz $\langle \dot \varphi_i \rangle = \omega_\text{sync}$ for $i\in S$ yields
\begin{equation}
    \omega_\text{sync} = \frac{-P_z + \langle  p_{z} \rangle}{\sum_{i\in S}\alpha_i}
    =  \frac{-\omega_s \alpha_z}{\sum_{i\in S} \alpha_i}
    = -\frac{\alpha_z}{\alpha^*} \frac{\omega_s}{N-1},
    \label{eq:definition_omega_sync_heterogeneous}
\end{equation}
where $\alpha^*$ is the average $\alpha_i$ in the synchronized cluster, and
\begin{equation}
    \omega_c \coloneqq \omega_s - \omega_\text{sync}
    = \left(1 + \frac{\alpha_z}{\sum_{i\in S} \alpha_i}  \right) \omega_s
    = \left( 1 + \frac{\alpha_z}{\alpha^*} \frac{1}{N-1} \right) \omega_s.
\end{equation}

We expect that the mechanism behind the solitary frequency tuning in heterogeneous networks is still resonance with network modes, as was the case for homogeneous networks.
The modes are still the imaginary parts of the Jacobian's eigenvalues in Eq.~(\ref{eq:linearized_dynamics}), however, a simple relationship to the eigenvalues of the Laplacian like Eq.~(\ref{eq:jacobian_eigenvalues}) does not hold anymore.
We reserve it for future work to test Eq.~(\ref{eq:self_consistency_explicit_heterogeneous}) for heterogeneous $M$ and $A$.

\subsection{Several solitary nodes}
\label{sub:several_solitary_nodes}
The self-consistency framework presented in this work can be generalized from 1-solitary states to $l$-solitary states with $l$ nodes that oscillate around desynchronized frequencies.
If all $l$ nodes have different frequencies, they do not interact with each other directly, because their coupling cancels out.
However, they all excite network modes, and we can superpose the respective responses of the synchronized cluster, which has size $N-l$ now.
Analogous to the considerations in this work, we can get a set of $l$ self-consistent equations that are only weakly coupled through $\omega_\text{sync}$, because averaging over the equations for $\ddot\varphi_l$ averages out the responses to frequencies other than the respective $\omega_{c;l}$.

If solitary nodes have identical frequencies, we enter the regime of cluster synchronization.
When averaging over the equations for $\ddot\varphi_l$, the response of the network to all identical frequencies add up in every respective equation.
One has to take care of additional constant phase shifts between the members of the cluster.\footnote{Note that the system is of $l$ equations determines $\omega_s$ and $l-1$ relative phases, as we can fix one.}
In this work, only concerned with 1-solitaries, this could be ignored due to the averaging process.

We expect that there are limits to the amount of perturbations and responses that can be superposed, because the assumption that linearization works might break down at some point.
It is a task for future work to explore the capabilities and limits of extensions of the theory of solitary states presented here.

\section{Evaluation of the self-consistent equation}
\label{sec:evaluation_of_the_self-consistent-equation}
Starting from the self-consistent equation, Eq.~(\ref{eq:self_consistency_explicit}), one can apply model reduction by simply replacing the Laplacian eigenvalues and eigenvectors with the ones from effective models.
In the following, we evaluate several effective models and a complex network without any reduction as examples.

\subsection{The minimal effective model}
\label{sub:the_minimal_effective_model}
The minimal effective model is the simplest model that shows tristability between the synchronous state and solitary states at several frequencies.
The graph of the minimal effective model can be seen in the inset in Fig.~3 (right) in the main article.
Starting from a network with dense a sprout node, $z$, and root node, $k$, we rewrite the dynamics using the local order parameter of node $k$,
\begin{equation}
       \rho_k e^{\imath \Psi_k} \coloneqq \frac{1}{\kappa_{kz}^* n} \sum_{j\in \mathcal{N}_k} \kappa_{kj}  e^{\imath \varphi_j},
       \label{eq:order_parameter_root}
\end{equation}
where 
\begin{equation}
    \kappa_{kz}^* n \coloneqq \sum_{j \in \mathcal{N}_k} \kappa_{kj},
    \label{eq:complex_network_mean_coupling_strength}
\end{equation}
so that $\kappa_{kz}^*$ is the mean coupling strength of the links adjacent to node $k$ except the link $(k,z)$.
This is inspired by the classical Kuramoto order parameter, see reference [7].
Now we can rewrite Eq.~(\ref{eq:swing_lossless_omega}) for node $k$,
\begin{equation}
        m\dot\omega_k = P_k - \alpha \omega_k 
        - \kappa_{kz} \sin(\varphi_k - \omega_s t)
        - \rho_k \kappa_{kz}^* n \sin(\varphi_k - \Psi_k).
\end{equation}
The complex network is then reduced to a slack node\footnote{A slack node has infinite mass and fixed phase. It is also called infinite-grid node. It can absorb any amount of power instantaneously and is widely used in engineering to balance systems.} and the two nodes $z$ and $k$ by setting the inertia of the rest of the grid to infinity.
This is especially accurate for large, well-synchronized networks.
The frequencies of the nodes in the synchronized cluster are zero and their phases constant.
Without loss of generality, we set $\Psi_k=0$.
We approximate $\rho_k$ by its asymptotic temporal mean value $\langle\rho_k\rangle$.
Between the root and the slack node, there are $n$ lines of coupling strength $K$.
Due to incoherence of the nodes in $\mathcal{N}_k$, we know that $\rho_k < 1$.
This leads to an effective coupling strength $\langle\rho_k\rangle  \kappa_{kz}^* n$, which is smaller than in the idealized case with $\kappa_{kz}^* n$.
Therefore, we expect resonances at smaller $\omega_s$ in complex networks.
For the analysis of the minimal effective model, we label the sprout node $z=1$, the root node $k=2$, and define $K_1 \coloneqq \kappa_{12}$, $K_2 \coloneqq \kappa_{21}$, and set and $K_3 n \coloneqq \kappa_{kz}^* n \langle \rho_k\rangle$.

The equations of motion of the minimal effective model are
\begin{eqnarray}
    m_1 \ddot \varphi_1 &=& P_1 - \alpha_1 \dot\varphi_1 - K_1 \sin(\varphi_1 - \varphi_2),
    \label{eq:toy_model_eom_node_1}
    \\
    m_2 \ddot \varphi_2 &=& P_2 - \alpha_2 \dot\varphi_2 - K_2 \sin(\varphi_2 - \varphi_1) - K_3 n \sin\varphi_2.
    \label{eq:toy_model_eom_node_2}
\end{eqnarray}
We set $m_1 = m_2 = 1$, $P_1 = 1 = -P_2$, $\alpha_1 = \alpha_2 = 0.1$, and $K = 6$.
In a power grid context, $K_1 = K_2 = K$; furthermore, $n$ gives us the freedom to set $K_3 = K$, too.
However, this system can be analyzed in a more general form.
For the slack node, we just have $\varphi_3 = 0$ at all times.
Node 1 can have different solitary frequencies, while node 2 and 3 form the synchronized cluster.\footnote{Node 2 can also be solitary at its natural frequency, but we focus on node 1, for which this toy model is designed.}
Due to the presence of a slack node in the synchronized cluster, we have $\omega_\text{sync} = 0$, hence $\omega_c = \omega_s$.
Effectively, $\varphi_3$ does not appear in the equations of motion, and the synchronized cluster only contains node 2.
The steady state is given by $\varphi_2 = 0$.
The Laplacian and its eigenvectors and eigenvalues are scalars,
$L = K_3 n$, $\vec v = 1$, $\lambda = L$.
The self-consistent equation reads
\begin{equation}
    0 = P_1 - \alpha_1 \omega_s \cdot \left(  1
    +\frac{\alpha_2}{2 \alpha_1}\frac{K_1 K_2\alpha_2}
        {\left(K_3 n - m_2\omega_s^2\right)^2 + \alpha_2^2 \omega_s^2} \right).
        \label{eq:self_consistency_two_node_toy_model_explicit}
\end{equation}
We see that $0 < |\omega_s| < |\Omega_1|$.
Furthermore, the self-consistency condition is symmetric under $\omega_s \to \omega_s$ while $P_1 \to -P_1$.
The conditions can be solved for $n$ explicitly, revealing two branches that are horizontally centered around $\omega_s = \text{sign}(P_1) \sqrt{\lambda/m_2}$ in the $\omega_s$-$n$-plane,
\begin{equation}
    n_\pm = \frac{m_2 \omega_s^2}{K_3}
    \pm \sqrt{\frac{\alpha_2 K_1 K_2}{2\alpha_1 K_3^2}\frac{\omega_s}{\Omega_1 - \omega_s} - \frac{\alpha_2^2 \omega_s^2}{K_2^2}}.
    \label{eq:n_of_omega_1}
\end{equation}
The $\omega_+ = \omega_s(n_+)$ branch is stable, while the $\omega_-$ branch has stable and unstable parts, see Fig.~3 in the main article.

\subsection{The star graph}
\begin{figure*}%[h]
    \centering
    \includegraphics[width=\textwidth]{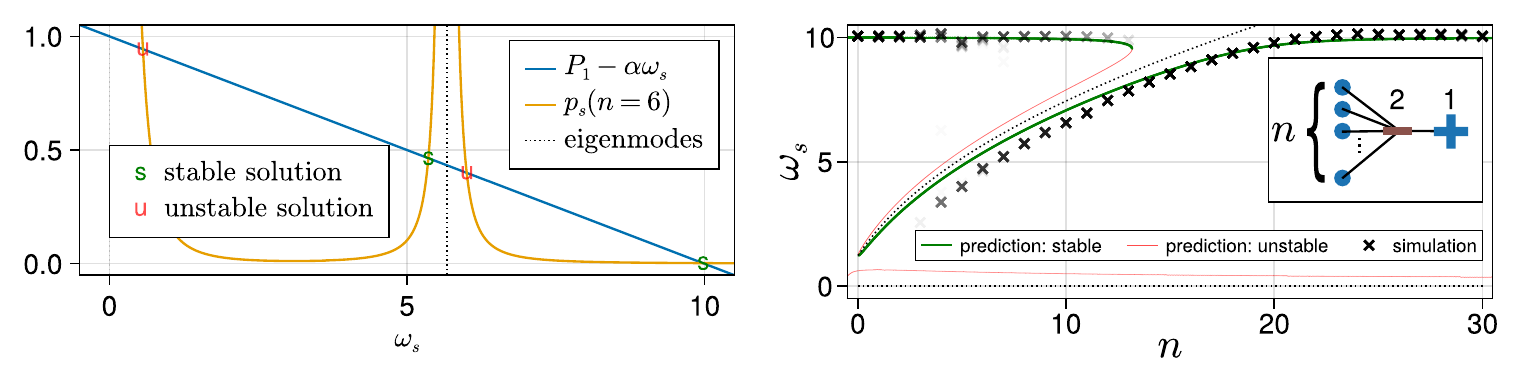}
    \caption{How network modes shape the landscape of solitary attractor states for a star-shaped graph, cf. Fig.~3 in the main article.
    (left) Intersections between the straight line $P_1 - \alpha \omega_s$ and the mean energy flow $p_s(n=6)$ generate stable and unstable solutions according to Eq.~(\ref{eq:self_consistency_explicit}).
    We use $P_1 = 1$, $\alpha=0.1$.
    The single non-zero peak of $p_s$ is centered around the network mode close to $\sqrt{K (n+1)}$.
    (right) Bifurcation diagram with bifurcation parameter $n$.
    The inset shows the graph.
    Round nodes have $P_i = 0$.
    We plot the asymptotic frequencies obtained from simulations of random initial conditions.}
    \label{fig:star_graph_intersection_and_bifurcation_diagram}
\end{figure*}

A more detailed model replaces the slack node in the minimal effective model by $n$ identical nodes with zero power generation to mimic a synchronized power-balanced grid connected that is well-connected to the root node.
The resulting star-shaped graph with $2+n$ nodes is shown in the inset in Fig.~\ref{fig:star_graph_intersection_and_bifurcation_diagram} (right).
The equations of motion are
\begin{eqnarray}
    m_1 \dot{\omega}_1 &=& P_1 - \alpha_1 \omega_1 - \kappa_{12} \sin\left( \varphi_1 - \varphi_2 \right),
    \label{eq:2+n_node_toy_omega_1}
    \\
    m_2 \dot{\omega}_2 &=& P_2 - \alpha_2 \omega_2 - \kappa_{21} \sin\left( \varphi_2 - \varphi_1 \right)
    - \sum_{j=3}^{2+n} \kappa_{2j} \sin\left(\varphi_2 - \varphi_j\right),
    \label{eq:2+n_node_toy_omega_2}
    \\
    m_i \dot{\omega}_i &=& P_i - \alpha_i \omega_i - \kappa_{i2} \sin\left( \varphi_i - \varphi_2 \right)
    \qquad \text{for } i\in\left\{3,\ldots, 2+n\right\}.
    \label{eq:2+n_node_toy_omega_i}
\end{eqnarray}
For identical $\varphi_{i\ge3}$, the latter two equations of motion can be reduced to
\begin{align}
    m_2 \dot{\omega}_2 &= P_2 - \alpha_2 \omega_2 - \kappa_{21} \sin\left( \varphi_2 - \varphi_1 \right)
    - \kappa_{23} n \sin\left(\varphi_2 - \varphi_3\right),
    \label{eq:3_node_toy_omega_2}
    \\
    m_3 \dot{\omega}_3 &= P_3 - \alpha_3 \omega_3 - \kappa_{32} \sin\left( \varphi_3 - \varphi_2 \right).
    \label{eq:3_node_toy_omega_3}
\end{align}
Again, we use identical $m_i = m = 1$, $\alpha_i=\alpha = 0.1$ and $\kappa_{ij} = 6$ (if connected).
The power constants are given by $P_1 = -P_2 = 1$, and $P_3 = 0$.
We observe that node 3 does not have solitary states, because $P_3 = 0$.
Any initial difference in the $\varphi_{i\ge3}$ are dissipated over time, which makes any additional links between these nodes obsolete.
For example, simulations of a network with additional all-to-all coupling except for node $1$, show the same asymptotic behavior.

The steady state of the original network, including node 1, is given by $P_1 = K \sin\left(\varphi_1^* - \varphi_2^*\right)$ and $0 = P_3 = K \sin\left(\varphi_3^* - \varphi_2^*\right)$.
Using the approximation discussed at the end of section \ref{sub:averaging_approach}, i.e, using the known $\varphi_i^*$ instead of the unknown $\vartheta_i^*$, the synchronized cluster has the $(1+n) \times (1+n)$ steady-state Laplacian
\begin{equation}
    L = K 
    %\gamma_3
        \begin{bmatrix}
        n & -1 & -1 & \ldots & -1\\
        -1 & 1 & 0 & \ldots & 0\\
        -1 & 0 & 1 & \ddots & \vdots\\
        \vdots & \vdots & \ddots & \ddots & 0\\
        -1 & 0 & \ldots & 0 & 1
    \end{bmatrix}.
\end{equation}
Its eigenvalues and eigenvectors are
\begin{eqnarray}
    \left\{\lambda^{[\ell]} \right\} &=&  K %\gamma_3 
    \cdot \{0, 1, \ldots,1, n+1 \},
    \\
    v_i^{[1]} &=& \frac{1}{\sqrt{n+1}},
    \quad
    v_i^{[n+1]} = \frac{1 - \delta_{i1} (n+1)}{\sqrt{n^2 + n}}.
\end{eqnarray}
The eigenvectors to the $n-1$ eigenvalues $K$ do not contribute to the self-consistent equation for solitary node 1, because they are strictly zero at the neighbor, node $2$.
They can be obtained by orthogonalization of the vectors with
\begin{equation}
    \tilde{v}_i^{[\ell]} = \frac{\delta_{i2} - \delta_{i(\ell+1)}}{\sqrt{2}}
    \quad \text{for $\ell = 2,\ldots,n$}.
\end{equation}
Note that component $i$ of the eigenvectors corresponds to $\varphi_{i+1}$ in the original graph.

Eq.~(\ref{eq:self_consistency_explicit}) explicitly evaluates to
\begin{align}
    Z(\omega_s) &= P_1 - \alpha_1 \omega_s 
    - \frac{K^2 \alpha \omega_c}{2}
    \left(  \frac{\frac{1}{n+1}}{ \left( m \omega_c^2 \right)^2 + \alpha^2 \omega_c^2} +  \frac{\frac{n}{n+1}}{\left(K %\gamma_3
    (n+1)  - m\omega_c^2 \right)^2 + \alpha^2 \omega_c^2 } \right),
\end{align}
where $\omega_c$ is given by Eq.~(\ref{eq:corotating_and_static_phase_velocity}).
The solutions and their linear stability are shown for $n=6$ in Fig.~\ref{fig:star_graph_intersection_and_bifurcation_diagram} (left).
The bifurcation diagram is shown in the right panel.
In comparison to the minimal effective model in Fig.~3 in the main article, we see here that Eq.~(\ref{eq:self_consistency_explicit}) for any network with a synchronized cluster of size larger than one (not counting infinite-grid nodes) has an additional unstable solution due to the bulk mode at $\lambda^{[1]}=0$. Other than that, the features are similar to the minimal effective model. However, the only contributing non-zero network mode lies close to $\lambda^{[n+1]} = K (n+1)$.

Note that the network modes that are shown are translated into the static frame via
\begin{equation}
    \mu^{[\ell]} \coloneqq \text{sign}(P_1) \cdot \left( 1 + \frac{\alpha_z}{\alpha} \frac{1}{N-1} \right)^{-1} \sqrt{ \frac{\lambda^{[\ell]}} {m}},
\end{equation}
cf. Eq.~(\ref{eq:corotating_and_static_phase_velocity}).
The reason is that $\omega_c$ is in resonance with $\sqrt{\lambda^{[\ell]}}$, such that $\lambda^{[\ell]} - m\omega_c^2$ becomes small and $\text{sign}(\omega_s) = \text{sign} (\omega_c) = \text{sign}(P_1)$.
However, we plot $\omega_s$, for which the original network modes $\pm \sqrt{\lambda^{[\ell]} / m}$ appear shifted away from the peaks.

We remark that the minimal effective model and the star graph show a maximum of degree heterogeneity between node $k$ (degree $n$ or $N-1$) and all other nodes (degree one), which results in high localization of an eigenmode at node $k$, see also the discussion in the main article.

\subsection{Complex synthetic power grids}

\begin{figure*}
    \includegraphics[width=0.5\textwidth]{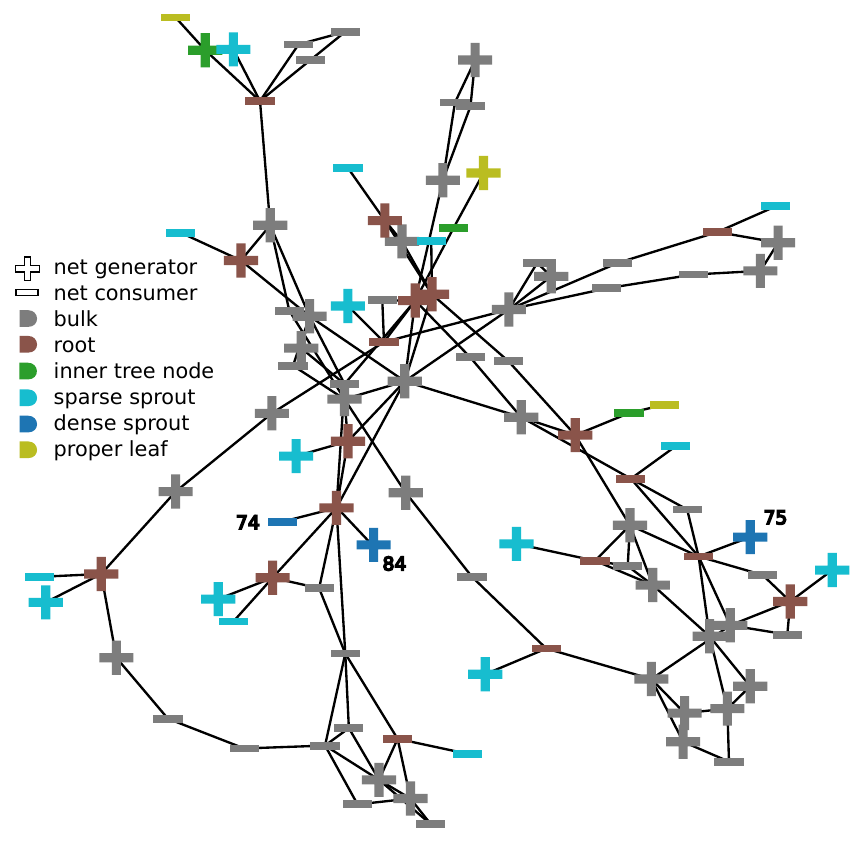}
    \caption{A synthetic power grid of $N=100$ nodes and their topological classes according to reference [25].
    Generated using a random growth model, see reference [37].
    Nodes marked with a "$+$" ("$-$") have $P_i=\pm 1$.
    The grid and its three dense sprout nodes, highlighted with node numbers 74, 75 and 84, are used as examples throughout the figures of the main article and the supplement.}
    \label{fig:grid_1}
\end{figure*}

Fig.~\ref{fig:grid_1} shows the graph of a spatially embedded synthetic power grid.

\begin{figure*}
    \centering
    \includegraphics[width=.29\textwidth]{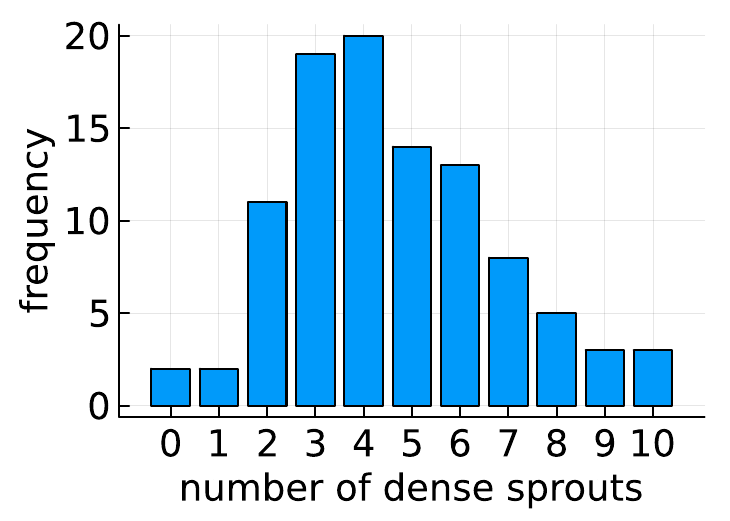}
    \includegraphics[width=.29\textwidth]{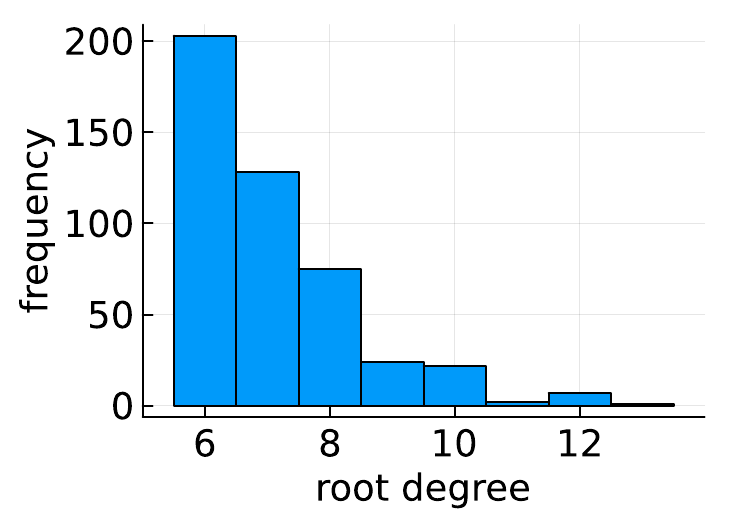}
    \caption{Topological properties of the ensemble of 100 synthetic networks.
    (left) Histogram of the number of dense sprout nodes per network.
    (right) Histogram of the degrees of the respective root nodes.
    Each root node appears as many times as it has dense sprout nodes as neighbors.
    }
    \label{fig:ensemble_properties}
\end{figure*}

Fig.~\ref{fig:ensemble_properties} shows properties of the ensemble of 100 synthetic power grids with $N=100$ nodes each.

\begin{figure*}
    \centering
    \includegraphics[width=.65\textwidth]{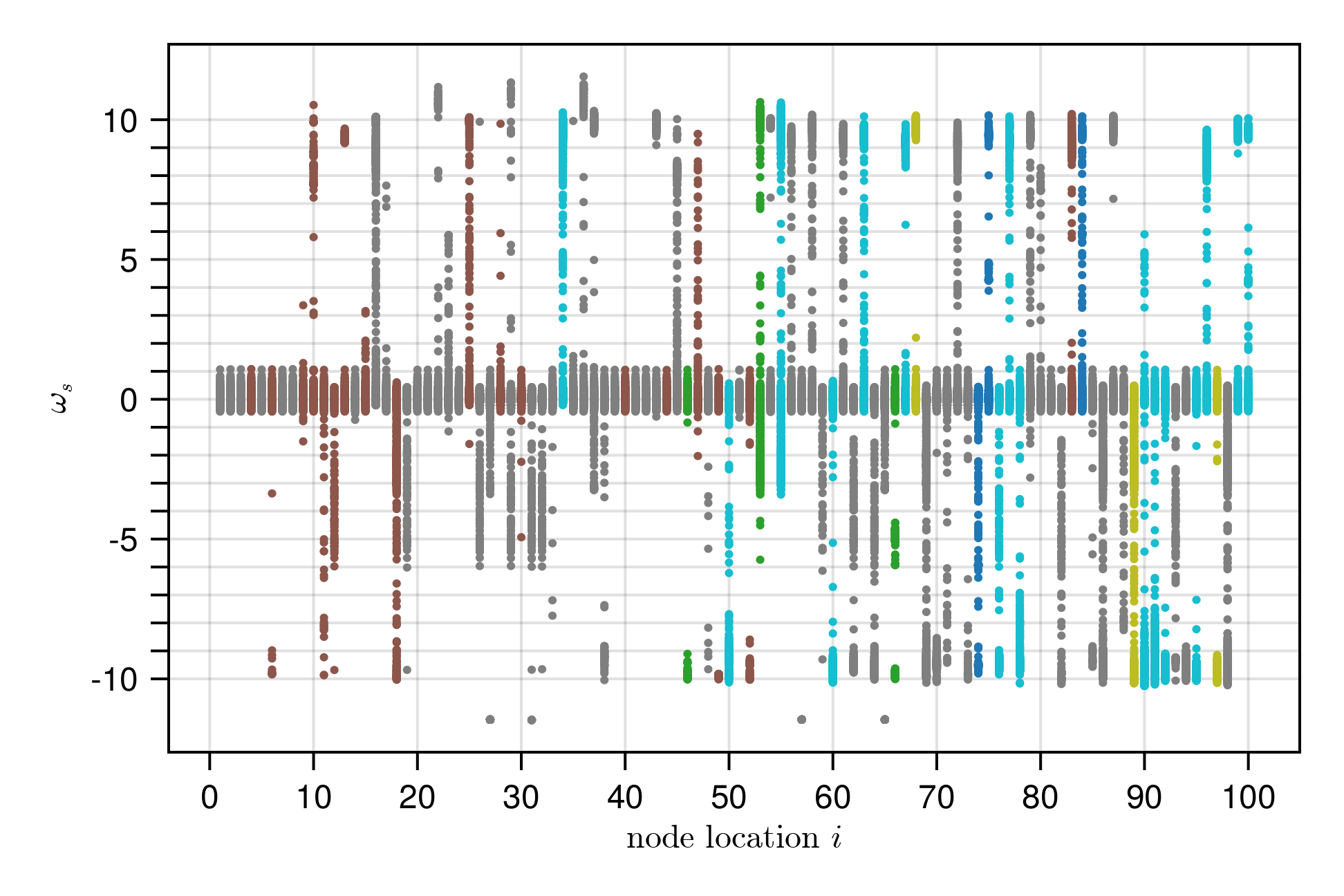}
    \includegraphics[width=.65\textwidth]{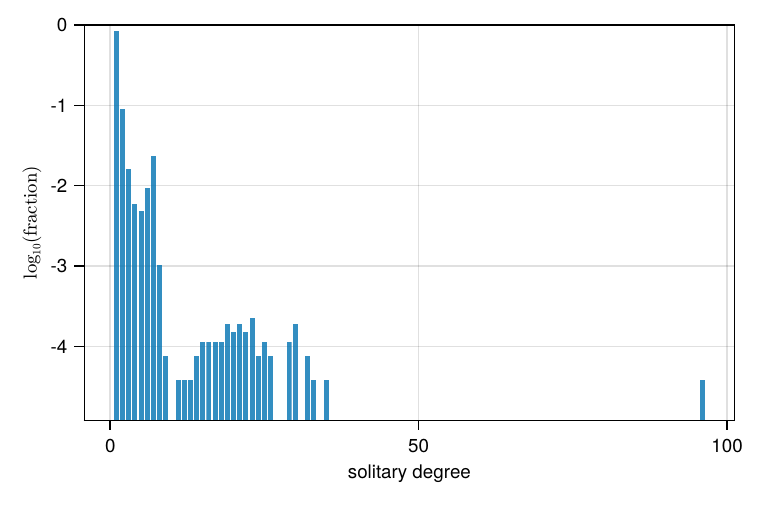}
    \caption{Asymptotic frequencies in a synthetic power grid.
    We apply a fine mesh of 1,200 evenly spaced single-node perturbations to each of the 100 nodes in a synthetic power grid, see Fig.~\ref{fig:grid_1}.
    About 22\% of the perturbations lead to  desynchronized asymptotic states.
    (top) The wide spectrum of possible asymptotic mean frequencies in this highly multistable system is scattered over the node locations.
    We see that the natural frequencies place approximate bounds on the asymptotic mean frequencies.
    The difference to Fig.~2 in the main article is that here, all asymptotic states are shown (not only 1-solitaries).
    (bottom) Histogram of the relative abundance of solitary degrees (numbers of desynchronized nodes, not degrees of solitary nodes) of all the desynchronized asymptotic states.
    In a single occasion, a single-node perturbation desynchronized the whole network.
    }
    \label{fig:ensemble_scan_solitary_freqs}
\end{figure*}

Fig.~\ref{fig:ensemble_scan_solitary_freqs} shows all asymptotic frequencies and a histogram of the number of desynchronized nodes for grid 1.

\begin{figure}%[h]
    \centering
    \includegraphics[width=0.7\columnwidth]{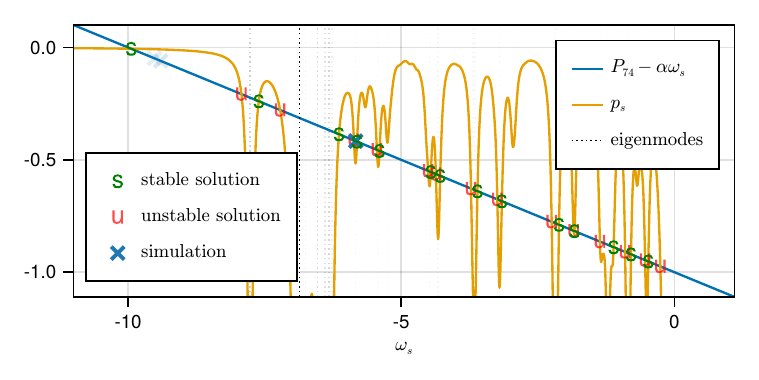}
    \caption{Solutions of Eq.~(\ref{eq:self_consistency_explicit}) for a dense sprout node ($i=74)$ in a synthetic power grid, see Fig.~\ref{fig:grid_1}, with standard parameters.
    The number of intersections is much higher, but the simulations show qualitatively similar results.}
    \label{fig:self_consistency_intersection_grid_1_node_74}
\end{figure}

Fig.~\ref{fig:self_consistency_intersection_grid_1_node_74} shows the solutions of Eq.~(\ref{eq:self_consistency_explicit}) for $z=74$ and homogeneous parameters $P_i = \pm 1$ and $\alpha_i = 0.1$ for the solitary oscillator ($P_{74}=-1$) and all other oscillators.
Fig.~4 in the main article shows the same, but for $P_{74} = -1 + \Delta P$ with $\Delta P = -1.1$ and $\alpha_{74} = 0.18$. The other $P_i$ are shifted from their standard values $\pm1$ by $-\Delta P/N$ to maintain power balance.
This slight adjustment makes the analytical predictions in the figure much clearer.
The simulations confirm that the dynamics, especially the locations and stability of the solitary states, are virtually unchanged.

\begin{figure}%[h]
    \centering
    \includegraphics[width=0.7\columnwidth]{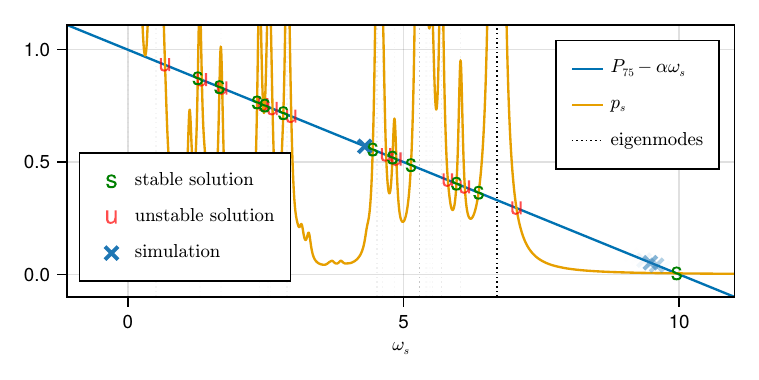}
    \caption{Solutions of Eq.~(\ref{eq:self_consistency_explicit}) for a dense sprout node ($i=75)$ in a synthetic power grid, see Fig.~\ref{fig:grid_1}, with standard parameters.
    The landscape of network modes can shift the main solution far from the most localized network mode.}
    \label{fig:self_consistency_intersection_grid_1_node_75}
\end{figure}

Fig.~\ref{fig:self_consistency_intersection_grid_1_node_75} shows the solutions of Eq.~(\ref{eq:self_consistency_explicit}) for node $75$, which is a generator.\footnote{Note that the visibilities (alpha values) of the simulation data points are not comparable between plots due to normalization and manual increase if necessary.}
This illustrates that these intersections plots are point-symmetric with respect to the origin if consumers and generators are exchanged.
This corresponds to the symmetry of Eq.~(\ref{eq:self_consistency_explicit}) under $P_z \to -P_z$ while $\omega_s \to -\omega_s$, which implies $\omega_c \to -\omega_c$.
We remark that due to other modes, the solution can be shifted significantly from the network mode that is most localized at the root node, cf. Fig~\ref{fig:self_consistency_intersection_grid_1_node_75}.

\begin{figure}
    \centering
    \includegraphics[width=0.7\textwidth]{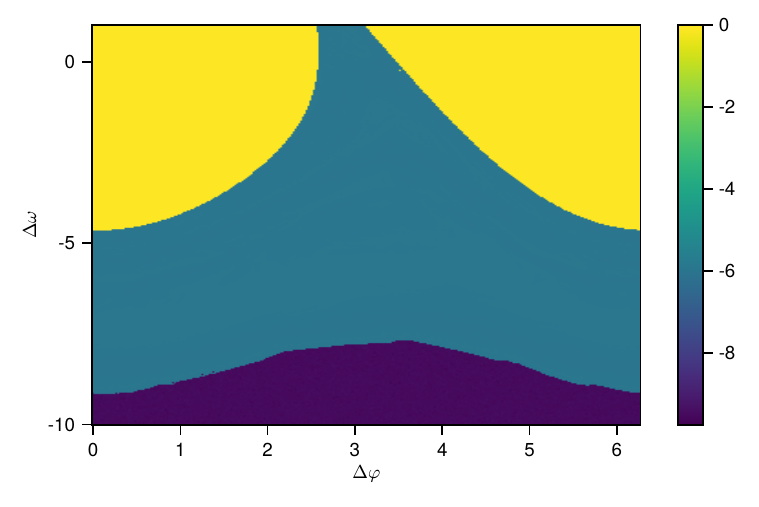}
    \caption{Phase portrait of node 74 of grid 1. Color indicates the mean frequency after the transient following perturbations to the synchronous state with $\Delta\varphi$ and $\Delta\omega$ to $\varphi_{74}$ and $\omega_{74}$ respectively. Yellow is the synchronous state, turquoise a resonant solitary state, purple the natural solitary state.}
    \label{fig:phase_portrait}
\end{figure}

Fig.~\ref{fig:phase_portrait} shows a phase portrait of node 74 of grid 1, the grid in Fig.~\ref{fig:grid_1}.

\begin{figure*}
    \centering
    \includegraphics[width=0.7\textwidth]{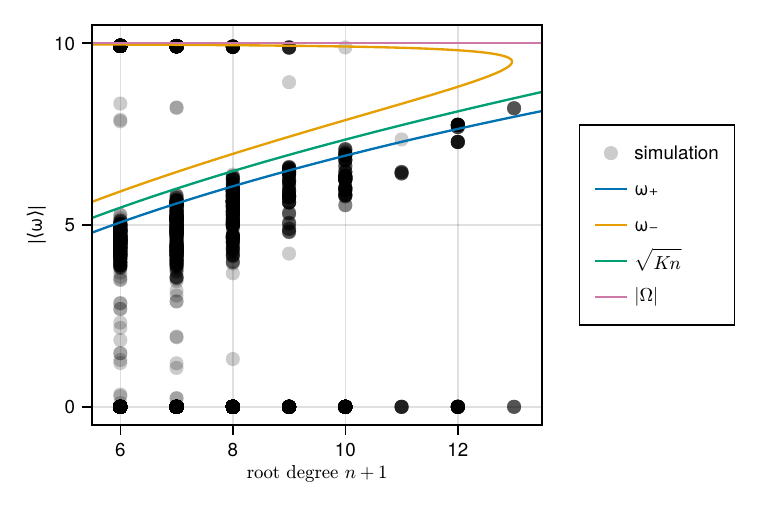}
    \caption{The minimal effective model provides an upper bound for the mean frequencies of resonant solitary states, as seen by comparison with the simulation results from the ensemble of synthetic power grids.
    A mesh of single-node perturbations is applied to all dense sprout nodes in the ensemble.
    All asymptotic mean frequencies at the dense sprout nodes are then scattered and overlaid by the analytical solution for the effective toy model. The branches $\omega_\pm$ correspond to $\omega_s(n_\pm)$ in Eq.~(\ref{eq:n_of_omega_1}).
    $\sqrt{K n}$ is the network mode at the center between the branches. $\Omega=\pm10$ is the natural frequency of generators or consumers, respectively, and the asymptotic bound for the solitary frequencies.}
    \label{fig:toy_vs_ensemble}
\end{figure*}

Fig.~\ref{fig:toy_vs_ensemble} compares the analytical prediction of the minimal effective model with simulation data from all 462 dense sprouts in the ensemble. Our prediction from the model reduction that the minimal model provides an upper bound are confirmed.

\subsection{Validity}
In this subsection, we comment on the range of validity of the derivation of the self-consistent equation, and when we expect to see solutions.
It is not the scope of this work to make extensive numerical studies of the phase space of different networks to work out the exact basins and how they change with the system's parameters, which is a highly complex question due to the involvement of complex network topologies.
In the following, we provide a few insights, and starting points for further investigations.

\subsubsection{Validity of rescaling}
To begin with, the derivation of the self-consistent equation is based on system parameters and topologies that result in small $\varepsilon = \alpha/(\sqrt{m \lambda})$ and localized eigenmodes, which is always given in our examples.
Concerning the parameters, (resonant) solitary states are more likely to be seen in the presence of small damping $\alpha$ when compared to inertia $m$, and comparatively strong coupling strength $K_{ij}$.

In addition to parameter ratios, the network topology plays a vital role in enabling solitary states in networks.
The solitary states are observed at leaf nodes with a high degree neighbor, because the neighbors tend to have highly localized network modes.
Such localization is present in networks with heterogeneous degree distributions, see reference [39].

\subsubsection{Validity by self-consistency}
One can see from the self-consistent equation, Eq.~(\ref{eq:self_consistency_explicit}) that it will not admit solutions if the localization of the eigenmodes, captured by the $v_k^{[\ell]}$, is too weak. This becomes clear in the examples provided in section~\ref{sec:evaluation_of_the_self-consistent-equation} and in the main article.
There, we observe that the self-consistent equation does not show solutions, when it is not valid.
Hence, when given an example network, the self-consistent equation can be used to assess the validity of its derivation self-consistently.

\subsubsection{Analytical validity of the ansatz}
To be sure that the solutions one finds to the self-consistent equation are viable, one can always double-check, if they fulfill all assumptions.
If that is the case, we expect to observe the predictions of the self-consistent equation numerically.
Given a specific system, one could test the accuracy of the approximations we made, by comparing the approximation of the trajectories from Eqs.~(\ref{eq:linearization_power_flow}), (\ref{eq:def_mode_variables_trafo}) and (\ref{eq:solution_xi_ell}) or statistics obtained from those with simulations.
The same goes for Eq.~(\ref{eq:linear_response_solution_with_amplitudes_and_phases}).
Note that an approximation of $\vartheta_z$ as in \ref{eq:def_theta_N} can be obtained by approximation and integration of \ref{eq:lin_shifted_eom_solitary_oscillator}.
See reference [38] for some examples.

In reference [22], which inspired our work, an approximation for the natural solitary state is derived.
Menck et al. start with a single (solitary) node of mass $m=1$ that is connected to an infinite grid with fixed phase and zero frequency.\footnote{In reference [36], a parameter study of the existence regime of the natural solitary state in this system can be found.}
Therefore, their effective model is given by
\begin{eqnarray}
    \dot\varphi &=& \omega,
    \\
    \dot\omega &=& P -\alpha \omega - K \sin\varphi.
\end{eqnarray}
They make the ansatz $\omega = \Omega + f(t)$, where $\Omega = P/\alpha$, and $f(t)$ is to be determined.
Here, the ansatz contains only a known frequency $\Omega$, hence no self-consistent equation is needed.
Approximating the coupling term as $K \sin(\Omega t)$, similar to what we did, they arrive at the expression
\begin{equation}
    f(t) = -\frac{\alpha K}{\Omega^2 + \alpha^2} \left(\sin(\Omega t + \varphi_0) - \frac{\Omega}{\alpha} \cos(\Omega t + \varphi_0) \right),
\end{equation}
where $\varphi_0 = \varphi(0)$ is the initial condition for the phase.
For $|\Omega|/\alpha = |P|/\alpha^2 \gg 1$, this is approximated by
\begin{equation}
    f(t) \approx \frac{K}{\Omega} \cos(\Omega t + \varphi_0).
\end{equation}
It follows that
\begin{equation}
    \varphi(t) \approx \Omega t + \frac{K}{\Omega^2} \sin(\Omega t + \varphi_0) + \varphi_0.
\end{equation}
Therefore, $\varphi(t) \approx \Omega t + \varphi_0$ will be observed if $\Omega^2 \gg K$ in addition to $|\Omega|/\alpha \gg 1$.

A similar statement can be made for the effective model in section \ref{sub:the_minimal_effective_model}, see reference [38].
For simplicity, we show this here for the most conceptual system that shows tristability with two distinct solitary states. This model is a partially linearized version of Eqs.~(\ref{eq:toy_model_eom_node_1}-\ref{eq:toy_model_eom_node_2}),
\begin{eqnarray}
    m_1 \dot{\omega}_1 &=& P_1 - \alpha \omega_1 - K_1 \sin\varphi_1 + K_1 \varphi_2 \cos\varphi_1,
    \label{eq:2_node_toy_omega_1_lin_simpler}
    \\
    m_2 \dot{\omega}_2 &\approx& P_2 - \alpha \omega_2 + K_2 \sin\varphi_1 - K_3 n \varphi_2.
    \label{eq:2_node_toy_omega_2_lin_simpler}
\end{eqnarray}
A similar calculation as in [22] shows that for $m_1^2 \omega_s^2 \gg K_1$ and $m_1 |\omega_s| /\alpha \gg 1$ we expect to see solitary states at node 1 with frequency $\omega_s$, see reference [38].

\subsubsection{Numerical validity of the ansatz}
\begin{figure}
    \centering
    \begin{subfigure}[b]{0.49\textwidth}
    \begin{overpic}[width=\textwidth]{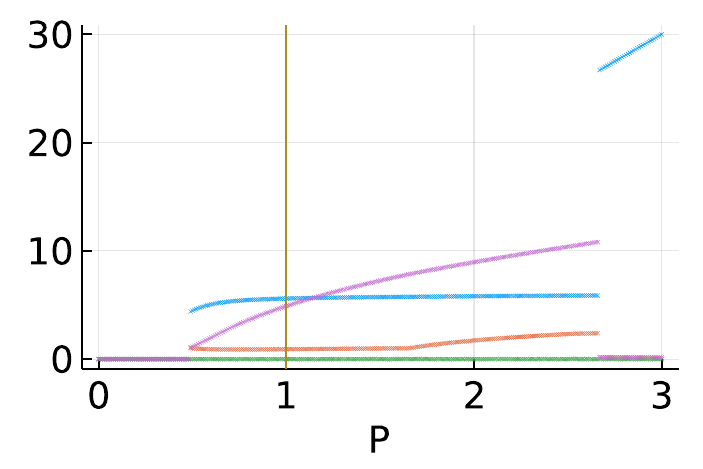}
    \put(0,70){\bf a}
    \end{overpic}
    \end{subfigure}
    \begin{subfigure}[b]{0.49\textwidth}
        \begin{overpic}[width=\textwidth]{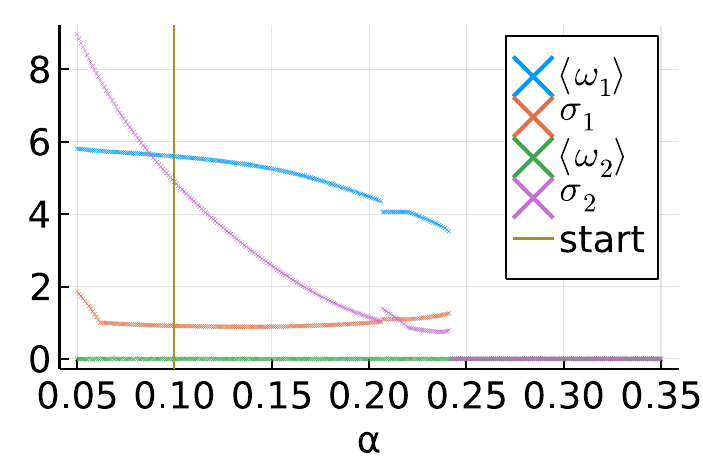}
        \put(0,70){\bf b}
        \end{overpic}
    \end{subfigure}
    \begin{subfigure}[b]{0.49\textwidth}
        \begin{overpic}[width=\textwidth]{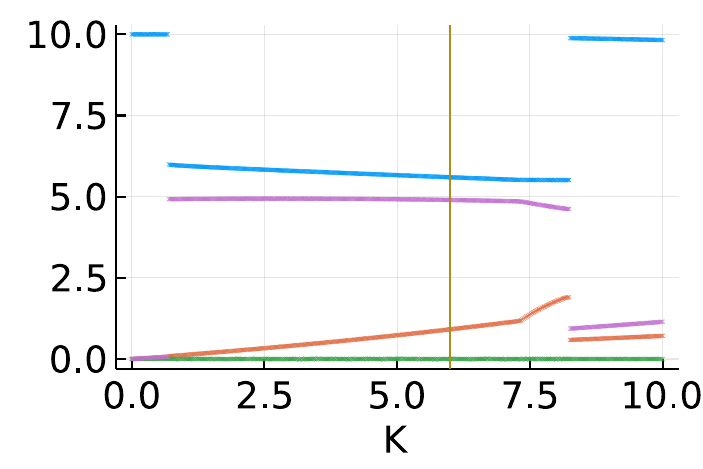}
        \put(0,70){\bf c}
        \end{overpic}
    \end{subfigure}
    \begin{subfigure}[b]{0.49\textwidth}
        \begin{overpic}[width=\textwidth]{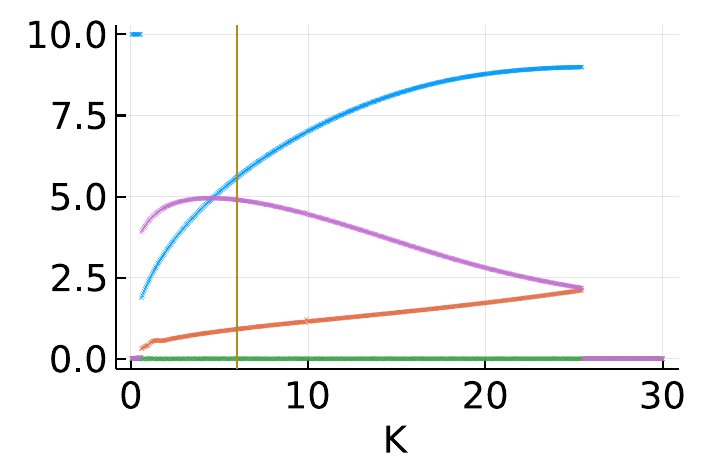}
        \put(0,70){\bf d}
        \end{overpic}
    \end{subfigure}
    \caption{Parameter study of the minimal toy model.
    The system Eqs.~(\ref{eq:2_node_toy_omega_1_lin_simpler})-(\ref{eq:2_node_toy_omega_2_lin_simpler}) is initialized with $(\varphi_1, \varphi_2, \omega_1, \omega_2) = (0,0,5,0)$ and the parameters $(P, \alpha, K, n) = (1, 0.1, 6, 10)$, indicated by the vertical line ``start".
    Note that $P_1 = P = -P_2$, and $\alpha_i = \alpha$, and $K_i = K$.
    The parameter under investigation is then tuned adiabatically in small steps, one parameter at a time.
    Starting from the standard values, we explore both positive and negative direction separately and show them in one plot.
    For each step, the system is integrated numerically for 1000 time units.
    To avoid transient artifacts, the statistics of the tails are computed for the last 300 time units.
    Then, the parameter is tuned instantaneously, i.e.,
    the final state of the system is used as initial condition in the next step.
    The horizontal axis shows the time-averaged statistics: the mean velocities $\langle \omega_i \rangle$ and respective standard deviations $\sigma_i$ defined in Eq.~(\ref{eq:def_standard_deviation_phase_velocity}).
    The minimum and maximum parameter value are indicated by the subscript min and max.
    The step size is indicated by a prefix $\Delta$.
    (a) $P_\text{min} = 0$, $P_\text{max} = 3$, $\Delta P = 0.01$;
    (b) $\alpha_\text{min} = 0.05$, $\alpha_\text{max} = 0.35$, $\Delta \alpha = 0.001$;
    (c) $K_\text{min} = 0$, $K_\text{max} = 10$, $\Delta K = 0.02$, while $n$ is adjusted to $n=60/K$ at each step to fulfill the constraint $K n = 6 \cdot 10$ at all times;
    (d) $K_\text{min} = 0.05$, $K_\text{max} = 30$, $\Delta K = 0.05$.}
    \label{fig:Kuramoto_adi}
\end{figure}

To evaluate the parameter space, in which the resonant solitary states are present, we conducted a parameter study with the minimal effective model in Eqs~(\ref{eq:2_node_toy_omega_1_lin_simpler})-(\ref{eq:2_node_toy_omega_2_lin_simpler}), where we used $m_i = 1$, and homogeneous parameters: $P_1 = P = -P_2$, $\alpha_i = \alpha$, and $K_i = K$.
Let $\sigma_i$ denote the standard deviation of frequency $\omega_i$ with respect to its temporal mean, i.e,
\begin{equation}
    \sigma_i \coloneqq \sqrt{\langle\left(\Delta\omega_i\right)^2\rangle},
    \label{eq:def_standard_deviation_phase_velocity}
\end{equation}
where $\Delta \omega_i = \omega_i - \langle\omega_i\rangle$.
The asymptotic statistics, $\langle\omega_i\rangle$ and $\sigma_i$ are shown for ranges of single system parameters in Fig.~\ref{fig:Kuramoto_adi}.
We see that all parameters have to be in an intermediate range, except for $n$, that can be arbitrarily high, cf. Fig.~3 in the main article.
If one parameter is tuned out of range, the resonant (intermediate) solitary state breaks down and the system finds another attractor.
Which attractor, generally depends on the exact timing of the perturbation, i.e., position in phase space, and the parameters values and the step size of parameter change.
It is intuitive that for small $P$ or large $\alpha$ we find a synchronous state and for large $P$ or small $\alpha$ an asynchronous state.
Fig.~\ref{fig:Kuramoto_adi} and Fig.~3 in the main article show that $\langle\omega_1\rangle$ depends mostly on the product $K n$, which is confirmed by the calculations in \ref{sub:the_minimal_effective_model}.

\section{Numerical Simulations}
Here, we provide some details about the numerical experiments. They were conducted with the \texttt{julia} programming language and the \texttt{PowerDynamics.jl} package, see reference [40].
package version 1.6.3. The code is openly accessible in reference [38].

The ensemble of 100 synthetic power grids was generated using the algorithm in reference [37] in the form of the software package \texttt{SyntheticNetworks.jl} \footnote{\url{https://github.com/PIK-ICoNe/SyntheticNetworks.jl/blob/main/src/SyntheticNetworks.jl}} with the parameters $N_0=1$, $p=1/5$, $q=3/10$, $r=1/3$, $s=1/10$, $u=0.0$ ($u$ is currently not implemented).

For Fig.~2 in the main article and Fig.~\ref{fig:ensemble_scan_solitary_freqs}, we used a mesh of evenly spaced initial conditions. The perturbations of the phases were sampled in 10 steps of $\Delta\varphi = 0.2 \pi$ between 0 and $1.8\pi$, i.e., 10 steps in $[0,2\pi)$. The perturbations of the frequencies were sampled in 120 evenly spaced steps in $[-5,6) +  0.5 \Omega_i$ for each node $i$.
These initial perturbations were added to the state $(\varphi_i,\omega_i)$ of a single node $i$ in the synchronous state $(\varphi_i^*, 0)$, cf. Eq.~(\ref{eq:power_flow_equations}).
This is called a single-node perturbation, as the other nodes were left in the synchronous state. From these initial conditions, we simulated the dynamics Eqs.~(\ref{eq:swing_lossless_phi}-\ref{eq:swing_lossless_omega}) for 200 time units.
We used the solver \texttt{Tsit5} with absolute tolerance $10^{-3}$ and relative tolerance $10^{-3}$.
The statistics of the trajectories, specifically the asymptotic mean frequencies, were calculated from the last 50 time units of the trajectories of all nodes. For this, the tails of the solutions of the (adaptive) solver were evaluated (by interpolation) at 10001 evenly spaced points in time.
This procedure was repeated with perturbing every node in the network individually.
\begin{itemize}
    \item We used the same solver settings for the trajectories shown in Fig.~1 and in the animation ``Supplementary Video~1''.
    \item For the intersection plots in Figs.~3, 4,
    \ref{fig:star_graph_intersection_and_bifurcation_diagram},
    \ref{fig:self_consistency_intersection_grid_1_node_74},
    \ref{fig:self_consistency_intersection_grid_1_node_75} 
    we obtained the asymptotic frequencies the same way.
    \item For the bifurcation diagrams in Figs.~3 and \ref{fig:star_graph_intersection_and_bifurcation_diagram} we proceeded the same way, but with 12 steps in the frequency perturbations $\Delta\omega$.

    \item For the phase portrait in Fig.~\ref{fig:phase_portrait}, we proceeded identically, but with 300 steps of initial perturbations to both the phase and the frequency of (only) node 74 of grid 1 of the ensemble.

    \item For Fig.~\ref{fig:toy_vs_ensemble}, we only perturbed the frequency of the nodes. We used steps of $\Delta\omega_i = \text{sgn}(P_i) = \pm 1 $ in $[0,11] \cdot\text{sgn}(P_i)$.
    The dynamics were integrated using \texttt{Tsit5} or \texttt{Rodas4}.
    The statistics were calculated over the last 20 time units of the 200 time units long trajectories. For this, the tails of the trajectories were evaluated at 998 evenly spaced points in time.
    For more details, we refer to the thesis that can be found in the repository reference [38].

    \item The procedure for Fig.~\ref{fig:Kuramoto_adi} is described in the caption and the according section.

\end{itemize}